\documentclass[11pt,a4paper]{article} 
\usepackage{jheppub}
\pdfoutput=1

\usepackage{amsmath,amsfonts}
\usepackage{latexsym}
\usepackage{amssymb}
\usepackage{hhline}
\usepackage{stmaryrd}
\usepackage{color}
\usepackage{verbatim}
\usepackage{graphicx}
\usepackage{epstopdf}

\makeatletter
\@addtoreset{equation}{section}
\makeatother

\newcommand{\be}{\begin{equation}}
\newcommand{\ee}{\end{equation}}
\newcommand{\bea}{\begin{eqnarray}}
\newcommand{\eea}{\end{eqnarray}}

\preprint{
\begin{flushright}
CP3-16-50\\
DESY 16/193\\
IFT-UAM/CSIC-16-096\\
FTUAM 16/36\\
BONN-TH-2016-07
\end{flushright}}

\title{\centering 
Confronting SUSY models with LHC data\\ via electroweakino production}

\author[1]{Chiara Arina}
\author[2]{, Mikael Chala}
\author[3,4]{, V\'ictor Mart\'in-Lozano}
\author[5]{and Germano Nardini}

\affiliation[1]{Centre for Cosmology, Particle Physics and Phenomenology (CP3), Universit\'e catholique de Louvain, B-1348 Louvain-la-Neuve, Belgium}
\affiliation[2]{Deutsches Elektronen Synchrotron, Notkestrasse 85, D-22603, Hamburg, Germany}
\affiliation[3]{Departamento de F\'isica Te\'orica \& Instituto de F\'isica Te\'orica UAM/CSIC, Universidad Aut\'onoma de Madrid, E-28049, Madrid, Spain}
\affiliation[4]{Bethe Center for Theoretical Physics \& Physikalisches Institut der Universit\"{a}t Bonn, Nu{\ss}allee 12, 53115, Bonn, Germany}
\affiliation[5]{Institute for Theoretical Physics, Albert Einstein Center, University of Bern, Sidlerstrasse 5, CH-3012 Bern, Switzerland}

\abstract{
We investigate multi-lepton signals produced by ElectroWeakino
(EWino) decays in the MSSM and the TMSSM scenarios with sfermions, gluinos and
non Standard Model Higgses at the TeV scale, being the Bino electroweak-scale dark
matter. We recast the present LHC constraints on EWinos for these
models and we find that wide MSSM and TMSSM parameter regions prove to be allowed. We forecast the number of events expected in the
signal regions of the experimental multi-lepton analyses in the next LHC runs. The
correlations among these numbers will help to determine whether future
deviations in multi-lepton data are ascribable to the EWinos, as well
as the supersymmetric model they originate from.
}


\begin{document}

\maketitle
\flushbottom


\section{Introduction} \label{sec:intro}
The first run of the LHC (Run~1) led to unexpected results. It was a
common perception that SUperSYmmetry (SUSY), if related to stabilizing
the ElectroWeak (EW) scale, would have been discovered quite quickly
while Higgs physics would have needed to wait for higher
statistics. However a Higgs boson was
found~\cite{Aad:2012tfa,Chatrchyan:2012ufa} but there is still no sign
of physics beyond the Standard Model (SM).

In fact, the collected LHC data impose quite strong bounds on
SUSY. First and second generation of squarks and gluinos need to be
well above 1\,TeV~\cite{atlas:susy:twiki,cms:susy:twiki}. Sbottoms and
stops lighter than about 800\,GeV are difficult to accommodate in view
of direct search constraints and of the 125 GeV Higgs mass
observation, at least in the Minimal
Supersymmetric SM (MSSM)~\cite{Arbey:2011ab,Carena:2011aa,Kang:2012bv,Fan:2014txa}.
Bounds pushing the chargino and neutralino sector well above the
EW scale exist as well~\cite{atlas:susy:twiki,cms:susy:twiki}.

Even though these constraints are based on strong model dependent
assumptions, there is a general feeling that perhaps the hierarchy
problem should be given up. In this spirit, MSSM scenarios where most
of the new physics is far away from the reach of the LHC ({\it e.g.}~in
high-scale~\cite{Djouadi:2013vqa}, spread~\cite{Hall:2011jd} or
split~\cite{Giudice:2004tc,Bhattacharyya:2012ct,Benakli:2013msa} SUSY)
are gaining popularity. Nevertheless, before departing towards these
drastically fine-tuned scenarios, it is wise to understand better the
model dependence of the experimental bounds. In particular, among
several plausible options, it seems sensible to generalize these
bounds in frameworks where charginos and neutralinos, somehow
protected by the chiral symmetry, feature EW-scale masses, while the
beyond-the-SM scalars are in the TeV range and do not interfere with
the neutralino and chargino production and subsequent decay.

In SUSY frameworks with only light charginos and neutralinos, dubbed
ElectroWeakinos (EWinos) hereafter, the lightest neutralino is the
Lightest Supersymmetric Particle (LSP). This particle is an excellent
EW scale Dark Matter (DM) candidate. Its relic density reproduces the observed
DM abundance in the parameter regions of the Higgs/$Z$-boson funnel
and the well-tempered neutralino~\cite{Masiero:2004ft, Pierce:2004mk,
  Baer:2005zc,ArkaniHamed:2006mb,Baer:2016ucr}. In the former case the LSP is
Bino-like, with mass close to half of the Higgs/$Z$-boson mass. In the
latter the LSP is a tuned mixture of gaugino and Higgsinos that
achieves the correct relic density away from resonances and
coannihilations with non-EWino particles.

At colliders it is the mass gap between the LSP and the other EWinos
what differentiates the
well-tempered region from the Higgs/$Z$-boson funnel one. Whereas in the former region a compressed
spectrum close in mass to the LSP is unavoidable and very hard to
probe~\cite{Cirelli:2014dsa, Golling:2016gvc}~\footnote{Recently CMS has published a new analysis with the data at 13 TeV to look specifically for soft leptons and set the first constraints on compressed spectra~\cite{CMS:2016zvj}.}, in the latter a gap of at least
40\,GeV is guaranteed (by {\it e.g.}~the LEP chargino mass bound
$m_{\widetilde{\chi}^\pm}\gtrsim 104\,$GeV~\cite{PDG}). Therefore, provided
that EWinos other than the LSP are sufficiently produced, the
Higgs/$Z$-boson funnel case exhibits a rich LHC phenomenology with
energetic EWino decay products that are easily tagged. In order to
characterize this rich phenomenology, in the present paper we
determine in detail the multi-lepton plus Missing Transverse Energy
(MET) signatures coming from EWinos.

For concreteness we focus on two SUSY models: the MSSM and the TMSSM,
{\it i.e.} the MSSM supplemented by one hyperchargeless $SU(2)_L$-triplet
chiral superfield~\cite{Espinosa:1991wt,DiChiara:2008rg}~\footnote{Among its
  appealing features (see {\it e.g.}~Refs.~\cite{Delgado:2012sm,
    Delgado:2013zfa, Bandyopadhyay:2013lca, Arina:2014xya}), the TMSSM
  provides a reduction of the little hierarchy problem with respect to
  the MSSM. In a bottom-up approach, the tuning of the model is
  similar to the one of the MSSM singlet
  extension~\cite{Ellwanger:2012ke} but it can be actually much
  smaller in appropriate ultraviolet
  embeddings~\cite{Delgado:2016vib}.}. In both cases we fix the LSP at
the Higgs funnel region while the other EWinos, consisting of Winos,
Higgsinos and, for the TMSSM, the fermionic components of the triplet
(dubbed Triplinos), are above the chargino mass bound $m_{\widetilde
  \chi^{\pm}}\gtrsim 104\,$GeV~\cite{PDG}. Sfermions and non-SM Higgses are assumed decoupled 
from the EWino LHC phenomenology but not very heavy in order not to
exacerbate the little hierarchy problem. This implies for instance
that within the MSSM the 125 GeV Higgs mass is possible only in the
large $\tan\beta$ regime, while such a regime is not required in the TMSSM
due to the additional $F$ terms increasing the tree level Higgs
mass~\cite{Espinosa:1991wt,DiChiara:2008rg}. In these MSSM and TMSSM
scenarios we determine the following:
\begin{enumerate}
\item {\it Present bounds on EWinos:} We recast the experimental
  analyses~\cite{Aad:2014vma,Aad:2014nua,ATLAS-CONF-2013-036,Aad:2014iza} constraining the anomalous production of two or
  more charged leptons in final states with MET, as it occurs in the
  production and subsequent decay of EWinos. We then generalize the
  ATLAS and CMS simplified-model constraints to the above MSSM and
  TMSSM scenarios. This analysis takes an approach similar to the one
  in Ref.~\cite{Martin:2014qra}.
  \item {\it EWino signatures in future data:} In the parameter space
    compatible with the present EWino bounds, we produce forecasts for
    multi-lepton searches. For the whole MSSM and TMSSM EWino
    parameter regions that we consider, we highlight the number of events
    that are expected in each Signal Region (SR) of the above multi-lepton
    analyses. We display results for a luminosity of $100$\,fb$^{-1}$
    at center-of-mass energy of 13\,TeV, nonetheless forecasts for the high luminosity phase are also discussed.
    \item {\it Disentangling the MSSM from other models:} We prove
      that the correlations among the events commented above are
      sensitive to the details of the EWino sector. In particular,
      SUSY models with an extended EWino sector can produce signals
      whose correlations are not produced by the EWino sectors of
      other models. Specifically, we prove that there is a small
      parameter region where the above MSSM and TMSSM scenarios can be
      disentangled already with $100$\,fb$^{-1}$. The region where
      this disentanglement is possible becomes wide for the luminosity of $3000$\,fb$^{-1}$ expected in the high luminosity LHC run.
\end{enumerate}

The rest of the paper is organized as follows. In the next section,
Sec.~\ref{sec:models}, we detail the EW sector of the MSSM and the TMSSM
and introduce the parameters and assumptions relevant for our
study. Section~\ref{sec:analysis} deals with the technical details of
the analyses and experimental searches we use. In
Sec.~\ref{sec:excreg} we present the most stringent constraints on the
MSSM and the TMSSM EWino parameter space. This sets the basis for our
multi-lepton forecasts at the present and future LHC runs, provided in
Sec.~\ref{sec:res}. In this section the possibility of disentangling
SUSY models by means of the correlation among multi-lepton signals is
also presented. We then come to our conclusions in
Sec.~\ref{sec:Concl}. We provide further details in Appendix~\ref{sec:appA}.

\section{The SUSY models}\label{sec:models}

In Sec.~\ref{sec:mssm} and Sec.~\ref{sec:tmssm} we briefly describe
the EWino sectors of the MSSM and the TMSSM as well as the SM-like Higgs boson
emerging in these models. Since we work under the premise that all
sfermions, non-SM Higgses and gluinos are at the TeV scale, SM particles and EWinos are the only
particles accessible by the LHC. This
assumption is supported by the experimental constraints of ATLAS and
CMS after Run~1~\cite{atlas:susy:twiki,cms:susy:twiki} that tend to push all the scalar SUSY sector to the TeV scale. We also
outline the free parameters relevant for our numerical analysis and
give technical details about the model implementation in
Sec.~\ref{sec:setup}.

\subsection{The MSSM model}\label{sec:mssm}

The EW sector of the MSSM is constituted by the neutralinos
$\widetilde\chi^0_i$, with $i=1,\dots,4$, and the charginos
$\widetilde\chi^\pm_j$, with $j=1,2$. In the limit considered in this work, the MSSM is
described by four free parameters, $\{M_1, M_2, \mu, \tan\beta \}$,
respectively the two gaugino masses, the bilinear term for the Higgs
sector in the superpotential and $\tan \beta=v_2/v_1$ with $v^2=
v_1^2+v_2^2=(174\,{\rm GeV})^2$, and $v_1$ and $v_2$ being the Higgs
Vacuum Expectation Values (VEVs). The mass matrices for the
neutralinos and charginos in terms of these four parameters are

\begin{equation} \label{eq:mnmassmssm}
{\mathcal M}^{tree}_{\widetilde\chi^0} = \left( 
\begin{array}{cccc}
M_1 &0 &-\frac{1}{2} g_1 v_1  &\frac{1}{2} g_1 v_2 \\ 
0 &M_2 &\frac{1}{2} g_2 v_1  &-\frac{1}{2} g_2 v_2 \\ 
-\frac{1}{2} g_1 v_1  &\frac{1}{2} g_2 v_1  &0 &- \mu \\ 
\frac{1}{2} g_1 v_1  &-\frac{1}{2} g_2 v_2  &- \mu &0\end{array} 
\right)\, ,
\end{equation} 
and
\be
\mathcal
M_{\widetilde\chi^\pm}^{tree}=
\left(
\begin{array}{cc} M_2& g _{2}v\sin\beta\\ g _{2}v\cos\beta&\mu\end{array}
\right)\,.
\label{charginosmssm}
\ee

All the sfermions as well as the pseudo scalar and charged
Higgses are much heavier, say in the
TeV range. They are then decoupled from the EWino LHC phenomenology.
For concreteness we set all the masses of the gluinos, non-SM Higgses,
sleptons and 1st/2nd generation of squarks at 2\,TeV. All trilinear
terms are assumed vanishing. This choice of parameters is in agreement
with the current LHC observations.

When the CP-odd Higgs mass $m_A$ is large, the SM-like Higgs mass is given
at tree level by
\begin{equation}
m^2_{h,tree} = m^2_Z \cos2\beta\,,
\end{equation}
where $m_Z$ is the mass of the $Z$ boson. At loop level, the dominant
correction comes from the top squark masses.  For stops at the TeV
scale, the large $\tan\beta$ regime is the only viable option to
achieve $m_h \simeq 125$\,GeV. We thus fix $\tan\beta=10$, as we have
checked that within the large $\tan\beta$ regime the EWino production
and decay are rather insensitive to the specific value of
$\tan\beta$.

\subsection{The TMSSM model}\label{sec:tmssm}

The TMSSM is an extension of the MSSM in which a hyperchargeless
$SU(2)_L$-triplet superfield is added. If we express the triplet
superfield as
\begin{equation}
\Sigma=\left(
\begin{array}{cc}
\xi^0/\sqrt{2} & \xi^+_2 \\
\xi_1^-        & -\xi^0/\sqrt{2}
\end{array}
\right)\,,
\end{equation}
the superpotential reads
\begin{equation}\label{eq:superpotential}
W_{\rm TMSSM} = W_{\rm MSSM} + \lambda H_1\cdot \Sigma H_2 +
\frac{1}{2}\mu_\Sigma{\rm Tr}\,\Sigma^2\, ,
\end{equation}
where the dot $\cdot$ is the $SU(2)_L$ antisymmetric product. The soft
breaking Lagrangian can be written as
\begin{equation}\label{eq:sblagrangian}
\mathcal{L}_{\rm TMSSM_{SB}}=\mathcal{L}_{\rm MSSM_{SB}} + m_4^2 {\rm Tr}(\Sigma^{\dagger}\Sigma)+ [B_\Sigma{\rm Tr}(\Sigma^2)+\lambda A_{\lambda}H_1\cdot \Sigma H_2 + {\rm h.c.}]\, .
\end{equation}
We consider no $CP$ violation so the parameters appearing in
Eqs.~(\ref{eq:superpotential}) and (\ref{eq:sblagrangian}) are taken
as real.

The neutral scalar component $\xi^0$ acquires a VEV $\langle
\xi^0 \rangle$. This VEV is very constrained by the EW
precision observables that impose $\langle \xi^0 \rangle \lesssim 4$
GeV at 95\% C.L.~\cite{Delgado:2013zfa,PDG}. This limit is naturally satisfied
in the parameter region~\cite{Delgado:2013zfa}
\begin{equation}\label{eq:hierarchy}
|A_\lambda|,|\mu|,|\mu_{\Sigma}|\lesssim 10^{-2}\frac{m_\Sigma^2+\lambda^2 v^2 /2}{\lambda v}\, ,
\end{equation}
where $m_\Sigma^2\equiv m_4^2 +\mu_\Sigma + B_{\Sigma}\mu_\Sigma$ is
the squared mass term of $ \xi^0$. To fulfill this relation, we choose
$m_\Sigma=5\,$TeV and $A_\lambda=0$, while $\mu$ and $\mu_\Sigma$,
which we deal as varying parameters, are never taken larger than
1\,TeV. Of course, the EWino phenomenology is independent of the
specific values of $m_\Sigma$ and $A_\lambda$ we adopt.

\paragraph{The Higgs sector}

The mixing between the MSSM-like Higgs fields, $H_1$ and $H_2$, and
the scalar triplet is negligible for $m_{\Sigma}\gtrsim 5$\,TeV
\cite{Delgado:2012sm}. In this case the EW minimization conditions can
be obtained from the scalar potential of $H_1$ and $H_2$. These
imply~\cite{Delgado:2013zfa}
\begin{eqnarray}
m_3^2 & = & m_A^2\sin\beta\cos\beta\, , \label{eq:min1}\\
 m^2_Z & = & \frac{m_2^2-m_1^2}{\cos 2\beta} - m_A^2 + \lambda^2v^2/2\,,\label{eq:min2} \\
 m_A^2 & = & m_1^2 + m_2^2 + 2|\mu|^2 + \lambda^2v^2/2\,,\label{eq:min3} \\
m_{H^{\pm}} & = & m_A^2+m_W^2+\lambda^2v^2/2\,,
\end{eqnarray}
where $m_W$ is the mass of the $W$ vector boson,
and $m_1^2$, $m_2^2$ and $m_3^2$ are the MSSM soft parameters of the
Higgs fields $H_{1,2}$. In the EW minimum, moreover, the squared mass
matrix of the $H_{1,2}$ CP-even Higgs components reads
\begin{equation} \label{eq:higgsmass}
{\mathcal M}^{2}_{h,H} = \left( 
\begin{array}{cc}
m_A^2\cos^2\beta + m_Z^2\sin^2\beta & (\lambda^2v^2 -m^2_A-m_Z^2)\sin\beta\cos\beta\\
(\lambda^2v^2-m_A^2-m_Z^2)\sin\beta\cos\beta & m_A^2\sin^2\beta + m_Z^2\cos^2\beta
\end{array} 
\right)\,.
\end{equation} 
Like in the MSSM, for $m_A$ at the TeV scale, all the non-SM Higgses
are heavy and the lighter CP-even Higgs, $h$, is aligned to the SM
Higgs. Nevertheless, the $h$ tree-level mass
\begin{equation}\label{eq:mhtmssm}
m_{h,tree}^2=m_Z^2\cos^2 2\beta + \frac{\lambda^2}{2}v^2\sin^2 2\beta\,
\end{equation}
can be larger than in the MSSM. Thus, in the TMSSM the little
hierarchy problem can be less severe than in the MSSM and can be
ameliorated with
$\lambda$ of order one and $\tan\beta$ low (in this respect the TMSSM
is similar to the singlet extension of the
MSSM~\cite{Ellwanger:2012ke}).

Motivated by the above features, in our analysis we consider TMSSM
reference scenarios with $\tan\beta=3$. We prefer however not to fully
exploit the boost in the tree-level Higgs mass~\footnote{Otherwise
  stop masses would turn out to be upper bounded by the 125 GeV Higgs
  mass constraint and possibly too light to be decoupled from the
  EWino phenomenology.}, consequently we fix $\lambda=0.65$ and work
in a regime where the sfermions, gluino and non-SM Higgses are as
heavy as in the MSSM case described in Sec.~\ref{sec:mssm}.

\paragraph{The EWino sector}\label{subsec:ewk}
As the TMSSM is an extension of the MSSM with a triplet superfield,
not only the scalar sector is enlarged but also the fermionic one. The
fermionic components of the triplet augment the number of neutralino
states to five and the number of chargino states to three. The
Triplinos mix with the MSSM EWinos. The tree-level mass matrices of
the neutralino and chargino sector are given by
\begin{equation} \label{eq:mnmass}
{\mathcal M}^{tree}_{\widetilde\chi^0} = \left( 
\begin{array}{ccccc}
M_1 &0 &-\frac{1}{2} g_1 v_1  &\frac{1}{2} g_1 v_2  &0\\ 
0 &M_2 &\frac{1}{2} g_2 v_1  &-\frac{1}{2} g_2 v_2  &0\\ 
-\frac{1}{2} g_1 v_1  &\frac{1}{2} g_2 v_1  &0 &- \mu  &-\frac{1}{2} v_2 \lambda \\ 
\frac{1}{2} g_1 v_1  &-\frac{1}{2} g_2 v_2  &- \mu  &0 &-\frac{1}{2} v_1 \lambda \\ 
0 &0 &-\frac{1}{2} v_2 \lambda  &-\frac{1}{2} v_1 \lambda  & \mu_\Sigma \end{array} 
\right)\,,
\end{equation} 
and
\be
\mathcal
M_{\widetilde\chi^\pm}^{tree}=
\left(
\begin{array}{ccc} M_2& g _{2}v\sin\beta& 0\\ g _{2}v\cos\beta&\mu& -\lambda v\sin\beta\\0&\lambda v\cos\beta& \mu_\Sigma\end{array}
\right)\,.
\label{charginos}
\ee
The presence of the Triplino increases the number of the EWino
parameters to six, which are now $\{M_1, M_2, \mu, \mu_\Sigma,
\tan\beta, \lambda\}$. Among them, only the first four parameters are
free once we fix $\tan\beta$ and $\lambda$ as previously described.

The enlarged EW sector can induce deviations in the $h\to \gamma
\gamma$ and $h\to \gamma Z$ decay channels while keeping all the
lightest-Higgs tree-level couplings
SM-like~\cite{Delgado:2012sm}. These deviations are sizeable only if $m_\Sigma\sim
100\,$GeV (when $\tan\beta$ is small and $\lambda$ is large), while
for $m_\Sigma\gtrsim 300\,$GeV they are generally
negligible~\cite{Arina:2014xya}. We use this lower bound on $m_\Sigma$
in our analysis.

The rest of the paper investigates the imprints of the additional
EWinos on the multi-lepton searches.  The reader interested in other
phenomelogical aspects of the TMSSM is referred to
Refs.~\cite{Delgado:2012sm,Delgado:2013zfa, Bandyopadhyay:2013lca, Arina:2014xya, Delgado:2016vib}.

\subsection{Numerical implementation of the (T)MSSM models}\label{sec:setup}

The multi-lepton signatures arise from EWino production and their
subsequent decay into the LSP. The charginos and the neutralinos
typically decay into $\widetilde{\chi}^0_1$ and a $W,Z$ or $h$ boson,
which subsequently can decay leptonically.  It should be noted that
the current LHC limits strongly depend on the presence of particular
decay modes, and are considerably weakened in case of compressed
\cite{Dreiner:2012gx} or stealth \cite{Fan:2011yu} spectra. For
instance the strongest bounds on EWino parameters are obtained with a
very light (LSP) Bino mass, $M_1 \sim 10$ GeV, on-shell $W,Z$ and $h$
bosons, and high $p_T$ final state leptons. On the other hand, for
$M_1$ above 100 GeV, the Run~1 LHC constraints are actually not much
stronger that the LEP bound $m_{\chi^\pm}\gtrsim 104\,$GeV~\cite{PDG}
(see~{\it e.g.}~Ref.~\cite{Martin:2014qra}).

In our analysis we consider a fixed LSP one-loop mass, namely
$m_{\widetilde{\chi}_1^0} = 63$ GeV. This choice of mass implies $M_1$
varying in between approximately 50 and 80 GeV, depending on the
particular values of the other EWino parameters. As previously explained, the choice of a 63 GeV pure Bino neutralino
is dictated by DM requirements. Indeed in both the MSSM and the TMSSM the
Higgs pole is a region where the LSP achieves the correct relic
density and is compatible with DM direct detection
searches~\cite{Bagnaschi:2015eha,Arina:2014xya}. At the same time the
invisible decay channel $h \to \widetilde{\chi}_1^0
\widetilde{\chi}_1^0$ is closed. This, together with a sensible
parameter choice suppressing deviations in the Higgs loop-induced
decays ({\it e.g.}~$m_\Sigma\gtrsim 300\,$GeV), guarantees full
agreement with the experimental Higgs measurements~\cite{Aad:2015zhl}.  Notice
that a slightly different LSP mass would not alter significantly our
multi-lepton results, while for a significantly lower value of $M_1$ the constraints will be tighter~\cite{Martin:2014qra}. Nevertheless in this latter case, for such lower masses, the EWino
signals should be also cross correlated to the Higgs invisible width
and the spin-independent DM-nucleon exclusion limits.

Both MSSM and TMSSM models we consider are implemented numerically in the following way:
\begin{itemize}
\item The models are generated by means of {\sc
  SARAH}~v4~\cite{Staub:2009bi,Staub:2010jh,Staub:2012pb}, which
  produces the model files for {\sc
    SPheno}~v3~\cite{Porod:2003um,Porod:2011nf} and the UFO
  files for {\sc MadGraph5\_aMC@NLO}~\cite{Alwall:2014hca};
\item The particle mass spectrum is computed in {\sc SPheno}: all
  EWino masses are computed at one loop level while the Higgs mass is
  computed at two loop level. Practically, at each parameter point we
  consider, we adjust the stop soft masses, which are in the TeV
  range, to obtain the observed Higgs mass. Concerning the TMSSM we
  refer to Ref.~\cite{Arina:2014xya} for the detailed description on
  how we compute the Higgs mass including the most relevant two-loop
  contributions;
\item The Branching Ratios (BRs) as well as the total decay widths of
  all particles are computed by means of {\sc SPheno};
\item Regarding the TMSSM, we keep $\lambda$ fixed at 0.65 and
  consider two values for the Triplino mass: $\mu_\Sigma = 300\,$GeV
  in one reference scenario (called TMSSM\_1 hereafter) and
  $\mu_\Sigma = 350\,$GeV in another scenario (called TMSSM\_2 hereafter). We
  will comment on the impact of changing $\lambda$ in Sec.~\ref{sec:excreg}.
\end{itemize}

\section{Searches for EWinos at the LHC: multi-lepton signals}\label{sec:analysis}

There are several SUSY searches implemented by the experimental collaborations. Relevant to our analysis are mainly those searches involving the direct production of charginos and neutralinos, as these are the only particles in the reach of LHC in our setup. More specifically we consider the searches that have only leptons in the final state, cleaner signatures with respect to jets + MET. We concentrate on 8 TeV data. As a matter of fact, most of dedicated analyses at 13 TeV are either preliminary~\cite{ATLAS-CONF-2016-075, CMS-PAS-SUS-16-024, CMS-PAS-SUS-16-022} or do not provide stronger constraints in general due to the still small luminosity~\cite{Aad:2016tuk}. At any rate, our results are not expected to be sensibly modified in the short term.

The multi-lepton searches look for departures in particular leptonic final states + MET  with respect to the SM predictions. If a deviation is seen, that could be interpreted as the production and subsequent decay to the LSP of EW particles, depending on the specific final state under investigation. The observed number of events in a specific search is typically studied in terms of Simplified Model Spectra (SMS). The SMS rely on the assumptions that only $\widetilde{\chi}^0_2$ and $\widetilde{\chi}^\pm_1$ are produced, that they both have the same mass and they decay 100\% into the LSP plus leptons via specific decay channels (some of them can include a decay mediated by a slepton).  Within the SMS interpretation, for instance, the three-lepton searches constrain the chargino mass up to 700 GeV for massless LSP~\cite{ATLAS-CONF-2013-035,Chatrchyan:2014aea}.

However if the EW particle content is richer than in SMS approach or the topologies leading to multi-leptons + MET are different, the exclusion bounds from Run~1 on $\widetilde{\chi}^\pm_1$ and $\widetilde{\chi}^0_1$ cannot be naively applied to the model to constrain chargino/neutralino masses. The correct approach in this case is to produce full event simulations at the detector level for all processes leading to a certain final state. In the (T)MSSM, for example, decays into the LSP can produce the following signals:
\begin{itemize}
\item $p p  \to \widetilde{\chi}^\pm_i  \widetilde{\chi}^\pm_j$ (with $i,j=1,2,(3)$) and $\widetilde{\chi}^\pm_i \to W^\pm \widetilde{\chi}_1^0$ give rise to two Opposite Sign (OS) leptons when both $W$s decay leptonically;
\item $p p  \to \widetilde{\chi}_i^0  \widetilde{\chi}^\pm_j$ (with $i=2,3,4,(5)$ and $j=1,2,(3)$) and $\widetilde{\chi}^\pm_j \to W^\pm \widetilde{\chi}_1^0$, $\widetilde{\chi}_i^0 \to Z \widetilde{\chi}_1^0$ lead to {three-lepton} final {states} when both $W$ and $Z$ bosons decay leptonically, with two leptons being of the Same Flavour and OS (SFOS);
\item $p p  \to \widetilde{\chi}_i^0  \widetilde{\chi}_j^0$ (with $i,j=2,3,4,(5)$) with $\widetilde{\chi}_i^0 \to Z \widetilde{\chi}_1^0$ gives rise to four-lepton final states, with two pairs of SFOS;
\item Also decay chains can contribute to the multi-lepton final states. For instance $p p  \to \widetilde{\chi}_3^0  \widetilde{\chi}_3^0$ with $\widetilde{\chi}_3^0 \to Z \widetilde{\chi}^0_1$ and $\widetilde{\chi}^0_3 \to W^- \widetilde{\chi}^+_1 \to W^- W^+ \widetilde{\chi}_1^0$ produce different number of leptons plus MET signatures depending on whether the $W$ and $Z$ bosons decay leptonically.
\end{itemize}
All these possibilities will be taken into account in our analysis as described below. The relative weight of these decay chains with respect to the two or three body decays into the LSP will depend on the composition of the neutralinos and charginos that are produced as well as on the mass spectrum.

\subsection{Recasting of the experimental searches}

In the following we describe the experimental searches we consider and how they are implemented in our analysis. Among the leptonic searches available we take into account the di-lepton search plus MET, the three-lepton search plus MET and the four-lepton search with MET. The latter is based on the data that ATLAS collected with a luminosity of 20.7 fb$^{-1}$, while the former rely on the selected events collected by the ATLAS detector with a luminosity of 20.3 fb$^{-1}$, all during Run~1 with $\sqrt{s} = 8$ TeV center-of-mass energy.
We do not consider the one-lepton and two $b$-jets + MET~\cite{Aad:2015jqa} search, as its sensitivity is not yet competitive with the other leptonic searches. It should be noted however that this search might probe a parameter space which is poorly covered by other SUSY analyses, namely at large $M_2$ and $\mu$~\cite{Martin:2014qra}.

\paragraph{Two-lepton + MET search~\cite{Aad:2014vma}} This search looks for a pair of OS leptons and MET and has a veto on $\tau^\pm$ leptons in the selected events. In the SMS approach, this search is sensitive to chargino pair production, followed by the decay of the charginos either directly into the LSP via $W$ boson emission either mediated by sleptons, or direct slepton pair production, depending on the assumption on the SUSY mass spectrum. It is also designed to be sensitive to chargino and next-to-lightest neutralino production, decaying into the LSP via a $W$ and a $Z$ boson, with the $W$ decaying hadronically and $Z$ boson leptonically.

Irrespectively of the flavour, all OS lepton pairs must satisfy the following criteria:
\begin{itemize}
\item $p_T > 35$ GeV for the higher-$p_T$ lepton;
\item $p_T > 20$ GeV for the other lepton;
\item an invariant mass of the di-lepton pair $m_{ll} > 20$ GeV.
\end{itemize}
Besides $m_{ll}$, another variable to tag the selected events that suppresses the main backgrounds, namely $WW$, $ZV$ (with $V$ a vector boson) and top ($t$) production, is the stransverse mass $m_{T2}$, defined as
\begin{equation}\label{eq:mt2}
m_{T2} = \rm{min}_{\mathbf{q}_T} \left\{ \rm{max}( m_T(\mathbf{p}_T^{l1},\mathbf{q}_T), m_T(\mathbf{p}_T^{l2},\mathbf{p}_T^{\rm miss}-\mathbf{q}_T)) \right\}
\end{equation}
with $\mathbf{p}_T^{l1}$ and $\mathbf{p}_T^{l2}$ being the transverse momentum of each of the two leptons, $\mathbf{p}_T^{\rm miss}$ the MET vector and $\mathbf{q}_T$ the transverse vector that minimises the larger of the two transverse masses defined as
\begin{equation}\label{eq:mt}
 m_T= \sqrt{2 ( p_T^l p_T^{\rm miss} - \mathbf{p}_T^l \cdot \mathbf{q}_T)}\,.
 \end{equation}
The end point of $m_{T2}$ is correlated with the difference in mass between the produced particles and the LSP. 

In some SRs, the signal events are also selected based on the variable $E^{\rm miss, rel}_T$, defined as
\begin{equation}
E^{\rm miss,rel}_T = \left\{ \begin{array}{ll}
         p_T^{\rm miss} & \mbox{if $\sin\Delta\Phi_{l,j}  \geq \pi/2$}\,,\\
         p_T^{\rm miss} \times \sin\Delta\Phi_{l,j} & \mbox{if $\sin\Delta\Phi_{l,j}  < \pi/2$}\,,\end{array} \right.
\end{equation}
where $\sin\Delta\Phi_{l,j}$ stands for the azimuthal angle between the direction of $\mathbf{p}_T^{\rm miss}$ and, depending on the case, that of the nearest electron, muon, central $b$-jet or central light flavour jet.

The flavour of the lepton pair depends on the SR, in some case both Same Flavour (SF, $e^+ e^-$ and $\mu^+ \mu^-$) and Different Flavour (DF, $e^\pm \mu^\mp$) are selected, while in other SRs only SF di-leptons are considered. There are in total seven SRs, each one with specific cuts as follows:
\begin{itemize}
\item{\bfseries{SRm$_{T2,90}$}}: both SF and DF leptons, $Z$ veto, namely $m_{ll}$ must be at least 10 GeV from the $Z$ boson mass, $m_{T2} > 90$ GeV and $E^{\rm miss,rel}_T> 40 $ GeV, zero central and forward light and $b$-jets. The main background is given by $WW$, $ZV$ and $t$ production and is estimated to be $61.5$ events;
\item{\bfseries{SRm$_{T2,110}$}}: same as above with however $m_{T2} > 110$ GeV and an expected of background events of 12.5;
\item{\bfseries{SRm$_{T2,150}$}}: same as  SR-m$_{T2,90}$ with $m_{T2} > 150$ GeV and an expected of background events of 4.2;
\item{\bfseries{SRWWa}}: both SF and DF leptons, $Z$ veto, $\mathbf{p}_T^{ll} > 80$ GeV, $m_{ll} < 120$ GeV and $E^{\rm miss,rel}_T > 80$ GeV, zero central and forward light and $b$-jets. This region is designed to look for production of on-shell $W$ bosons, however close to threshold. The main background is given by $WW$, $ZV$ and $t$ production, giving rise to $160.1$ expected events;
\item{\bfseries{SRWWb}}: both SF and DF leptons, $Z$ veto, $\mathbf{p}_T^{ll} > 80$ GeV, $m_{ll} < 170$ GeV and $m_{T2}> 90$ GeV, zero central and forward light and $b$-jets and expected background events of $48.3$. This region, together with SRWWc, is designed for charginos with masses larger than 120 GeV and boosted $W$ bosons;
\item{\bfseries{SRWWc}}: same as SRWWb however with  no cut on $m_{ll}$ and $m_{T2}> 100$ GeV and expected background events of $29.3$;
\item{\bfseries{SRZjets}}: SF leptons only, two central light jets (from the $W$ decaying hadronically), the di-lepton invariant mass should be within 10 GeV of the $Z$ boson mass, $E^{\rm miss,rel}_T > 80$ GeV, $\mathbf{p}_T^{ll} > 80$ GeV, the separation between the two leptons should be $0.3~<~\Delta R_{ll}~<~1.5$, the two highest-$p_T$ jets are required to satisfy $p_T> 45$ GeV and the invariant mass of the jet pair should be $50 \, {\rm GeV}< m_{jj} < 100$ GeV. Here the expected number of background events is $1.4$.
\end{itemize}

In our models this search will constrain mostly chargino production decaying via $W$ bosons into the lightest neutralino plus MET, hence the most relevant SRs are SRWWa, SRWWb and SRWWc.

\paragraph{Three-lepton + MET search~\cite{Aad:2014nua}} This search looks for the production of chargino and neutralinos, which decay further into three-leptons plus MET in the form of two LSPs and neutrinos. In the SMS approach this search constrains $\widetilde{\chi}^\pm_1$ and $\widetilde{\chi}^0_2$, which decay into $\widetilde{\chi}_1^0$  via off or on-shell $W$ and $Z/h$ bosons respectively. 

The selected events must contain exactly three leptons, two of them are required to be of OS, while the flavour can be different depending on the definition of the SR (signal electrons or muons are labelled with $l$  and $l'$, where the flavours of $l$ and $l'$ are different). The leptons should be separated from each other by $\Delta R > 0.3$. The selected events should not contain any $b$-jet, however there is no requirement on the number of non $b$-jets. The trigger for the signal leptons depends  on their flavour:
\begin{itemize}
\item single isolated $e$ or $\mu$: $p_T > 25$ GeV;
\item $e^+e^-$: $p_T >  14$ GeV and $p_T >  14$ GeV or  $p_T >  25$ GeV and $p_T >  10$ GeV;
\item $\mu^+\mu^-$: $p_T >  14$ GeV and $p_T >  14$ GeV or  $p_T >  18$ GeV and $p_T >  10$ GeV;
\item $e^\pm \mu^\mp$: $p_T^e >14 $ and $p_T^\mu > 10$  GeV or $p_T^\mu > 18$ GeV and $p_T^e > 10$ GeV.
\end{itemize}

To further suppress the background (mainly given by $WZ$, $ZZ$, $VVV$, $t\bar{t} V$ and $tZ$), relevant kinematic variables are $p^{\rm miss}_T$, $m_{T2}$ and $m_T$ defined in Eqs.~(\ref{eq:mt2}) and (\ref{eq:mt}).

In total there are five SRs, defined by the flavour and the charge of the leptons and sometimes requiring the $Z$ veto (no SFOS lepton invariant mass within 10 GeV of the $Z$ boson mass). In each SR the cuts on the kinematic variables are:
\begin{itemize}
\item{\bfseries{SR0$\tau$a}}: $l^\pm l^\mp l', l^\pm l^\mp l$, $\tau$ flavour veto. This is a complicated SR, with all kinematic variables separated in 20 bins, as detailed in Tab.~\ref{tab:SR0taua}.
\end{itemize}
{We do not consider other SRs reported by the experimental collaboration, as they tag more $\tau$ leptons and hence reduce the sensitivity with respect to SR0$\tau$a.} In our model this search will constrain the production of a chargino and a neutralino (not only $\widetilde{\chi}^\pm_1$ and $\widetilde{\chi}^0_2$) decaying directly or via a decay chain into the LSP plus three leptons, depending on the mass spectrum. Only $Z,W$ and the {$h$} bosons are considered in the decays, as all non-SM scalars are much heavier.
\begin{table}[t!]
\begin{center}
\caption{Definition of the bins for  {SR0$\tau$a} belonging to the three-lepton plus MET experimental search.}
\begin{tabular}{c|c|c|c|c|c}
\hline
 Bin & $m_{\rm SFOS}$ [GeV] & $m_T$ [GeV] & $E_T^{\rm miss}$ [GeV] & Z veto & SM background [\# events]\\
\hline
1 & 12-40 & 0-80& 50-90& no & 23\\
2 & 12-40 & 0-80& $>90$ & no& 4.2\\
3 & 12-40 & $> 80$& $50-75$ & no&10.6 \\
4 & 12-40 & $> 80$ &  $> 75$ & no &8.5\\
\hline
5 &40-60 &0-80 & 50-75 &  yes &12.9 \\
6 &40-60 &0-80 & $>75$& no & 6.6\\
7 & 40-60 & $ > 80$& 50-135 & no & 14.1\\
8 & 40-60 & $>80$ & $>135$ & no & 1.1\\
\hline
9 & 60-81.2 & 0-80 & 50-75 & yes & 22.4\\
10 & 60-81.2 & $> 80$ & 50-75 & no & 16.4\\
11 & 60-81.2 & 0-110 & $> 75$ & no & 27\\
12 & 60-81.2 & $>110$ & $> 75$ & no & 5.5\\
\hline
13 & 81.2 - 101.2 & 0-110&50-90 & yes &  715\\
14 & 81.2 - 101.2  &0-110 &$>90$ & no & 219\\
15 & 81.2 - 101.2  & $>110$ & 50-135 & no & 65\\
16 & 81.2 - 101.2  &$> 110$ &$>135$ &no &  4.6\\
\hline
17 & $>101.2$& 0-180& 50-210& no & 69\\
18 &  $>101.2$ & $>180$& 50-210& no & 3.4\\
19 &  $>101.2$ & 0-120& $>210$ & no & 1.2\\
20 &  $>101.2$ & $> 120$ & $>210$& no & 0.29\\
\hline
\end{tabular}
\label{tab:SR0taua}
\end{center}
\end{table}

\paragraph{Four-lepton + MET search~\cite{ATLAS-CONF-2013-036,Aad:2014iza}} This search looks for four or more isolated leptons in the final state plus {MET}. In the SMS approach this search can constrain the production of a pair of heavy neutralinos, decaying for instance into two $Z$ bosons and two LSPs. In the selected events {at least three `light leptons' are required}, where the term `light lepton' refers to electrons and muons only, including those from leptonic decay of the tau. The term `lepton' refers to electrons, muons and taus. Tau leptons that decay hadronically are reconstructed by requiring {the jets to} have $p_T > 10$ GeV and pseudorapidity $|\eta| < 2.5$.

The invariant mass of all possible SFOS light lepton {pairs} must be larger than 12 GeV to suppress the background from low energy resonances. The signal is discriminated over the background using the effective mass variable
\begin{equation}
m_{\rm eff} = {p_T^{\rm miss}} + \sum_\mu p_T^{\mu} + \sum_e p_T^e + \sum_\tau p_T^\tau +   \sum_j p_T^j\,,
\end{equation}
where $p_T^j$, the transverse momentum of the jets, must be at least 40 GeV. The SM processes that {originate} the main background are $ZZ$, $ZWW$, $t\bar{t}Z$ and Higgs production. There are in total {five} SRs, three requiring an extended $Z$ veto and two requiring $Z$ candidates (the updated version \cite{Aad:2014iza} contains additional SRs). The extended $Z$ veto is defined as each possible {pair, triplet and quadruplet} of light leptons in the selected events with an invariant mass between 81.2 GeV and 101.2 GeV to be discarded. The cuts in each region are:
\begin{itemize}
\item{\bfseries{SR0Z}}: at least {four} light leptons and no $\tau$ lepton, $p^{\rm miss}_T  > 75$ GeV, SFOS light leptons with an invariant mass between 81.2 GeV and 101.2 GeV. The number of SM background expected events is 1.7;
\item{\bfseries{SR1Z}}: {one} $\tau$ lepton and {three} light leptons, $p^{\rm miss}_T  > 100$ GeV, SFOS light leptons with an invariant mass between 81.2 GeV and 101.2 GeV. The number of SM background expected events is 1.6;
\item{\bfseries{SR0noZa}}: at least {four} light leptons and no $\tau$ lepton, $p^{\rm miss}_T  > 50$ GeV, extended veto. The number of SM background expected events is 2;
\item{\bfseries{SR0noZb}}: at least {four} light leptons and no $\tau$ lepton, $p^{\rm miss}_T  > 75$ GeV, $m_{\rm eff} > 600$ GeV, extended veto. The number of SM background expected events is 4.8;
\item{\bfseries{SR1noZ}}: {one} $\tau$ lepton and three light leptons, $p^{\rm miss}_T  > 100$ GeV, $m_{\rm eff} > 400$ GeV, extended veto. The number of SM background expected events is 1.3.
\end{itemize}
In our analysis this search will constrain heavy neutralino and heavy chargino production decaying two body as well as via long decay chains into the LSP.

\subsection{Numerical implementation of the analyses}\label{sec:numdet}

The details on the experimental analysis implementation and the generation of the event simulations for the MSSM and the TMSSM models are the following:
\begin{itemize} 
\item The event simulations at parton level are produced by {\sc MadGraph5\_aMC@NLO}. To be specific, we simulate the production cross section of all possible EWinos with 500k events, {\it i.e.}: $p p \to Z \to \widetilde{\chi}^0_i  \widetilde{\chi}^0_j$, $p p \to Z/\gamma \to \widetilde{\chi}^\pm_k \widetilde{\chi}^\pm_l$  and $p p \to W^{\pm} \to \widetilde{\chi}^0_i \widetilde{\chi}^\pm_k $ with $i,j=1,...,4, (5)$ and $k,l=1,2, (3)$;
\item The decays of charginos and neutralinos producing the leptonic final states are computed with {\sc Pythia}~v6~\cite{Sjostrand:2006za}, as well as the showering and hadronization;
\item We use {\sc Delphes}~v3~\cite{deFavereau:2013fsa} to simulate the detector response, with the default detector card and modified lepton efficiencies, in order to have a better matching with the experimental searches;
\item The experimental analyses are implemented in {\sc MadAnalysis}~v5~\cite{Conte:2012fm,Conte:2014zja}. Each analysis has been validated by considering few clearly defined benchmark points. More specifically the two-lepton + MET search is validated by reproducing figure 5 as well as the benchmark points in tables 5 and 6 of Ref.~\cite{Aad:2014vma}. The three-lepton + MET and four-lepton + MET searches are validated using the outflow of benchmark points provided by {\sc CheckMate}~v1~\cite{Drees:2013wra}~\footnote{{\sc CheckMate} uses {\sc FastJet}~\cite{Cacciari:2005hq,Cacciari:2011ma}, the anti-$k_t$ jet algorithm~\cite{Cacciari:2008gp}, the CL$_{\rm s}$ prescription~\cite{Read:2002hq} and the $m_{T2}$ algorithm~\cite{Lester:1999tx,Barr:2003rg,Cheng:2008hk}.} and {\sc Seer}~\cite{Martin:2015hra}. In all cases we find agreement within 20\%;
\item The exclusion bounds are computed {at the 95\% confidence level (CL)} with the CL$_{\rm s}$ method~\cite{Read:2002hq}.
In the case of the two-lepton search, our analysis takes into account only one SR, SRWWa. The SRs are indeed not independent, they highly and non trivially overlap, hence they cannot be easily combined together with SRWWa in a statistically meaningful way. The constraint coming from the {three-lepton} analysis considers all the 20 bins of SR0$\tau$a, since they are all statistically independent. In the {four-lepton} analysis, the first two SRs are not independent, hence we do not consider {SR1Z} for the constraints. We have checked that SR1Z has no impact on the shape of the exclusion region, as SR0Z contains many more events due to its looser cuts. 
\end{itemize}

\section{MSSM and TMSSM: excluded regions after the LHC Run~1}\label{sec:excreg}

\begin{figure}[ht]
\begin{minipage}[t]{0.32\textwidth}
\centering
\includegraphics[width=1.\columnwidth,trim=1mm 0mm 1mm 1mm, clip]{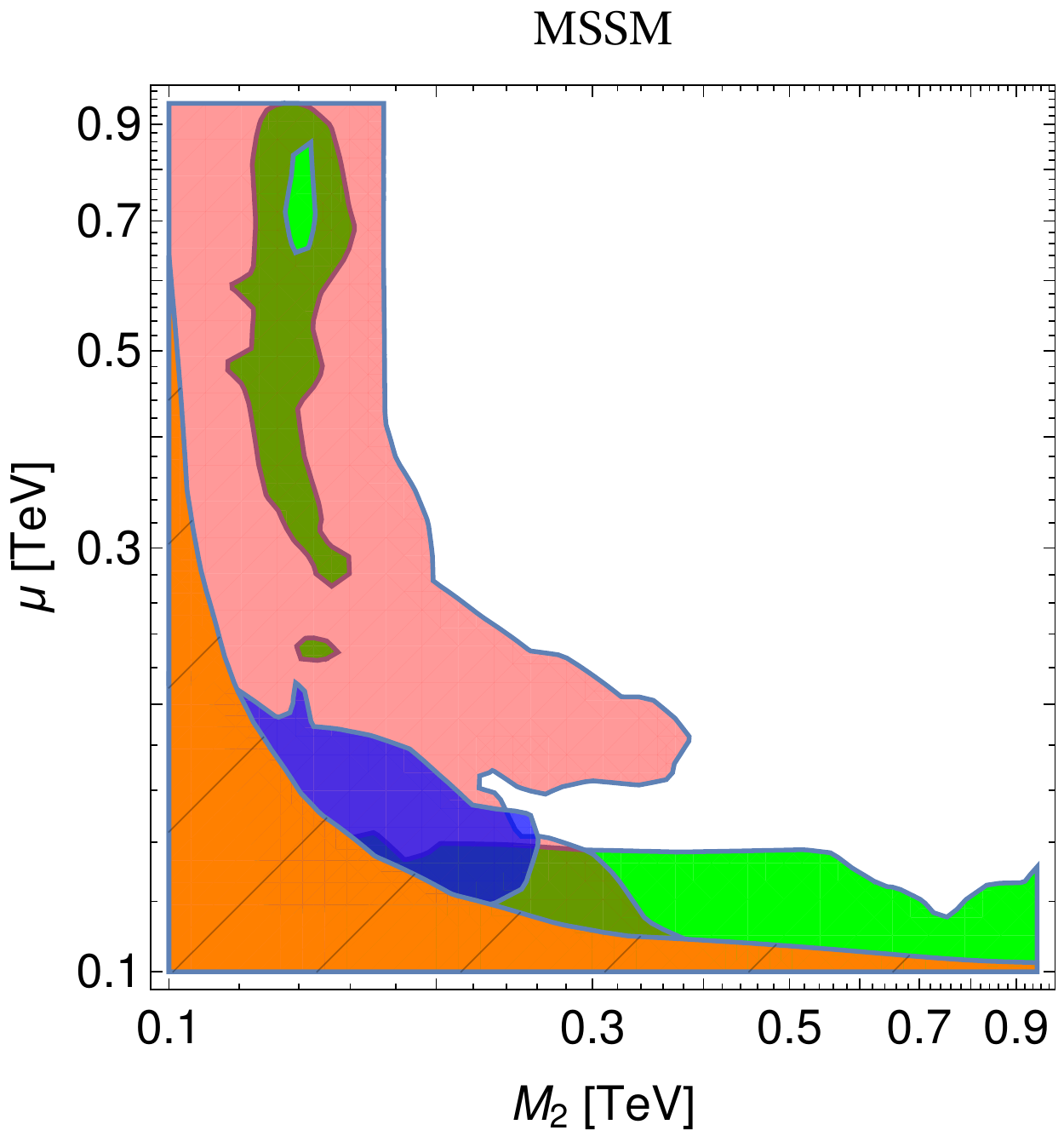}
\end{minipage}
\begin{minipage}[t]{0.32\textwidth}
\centering
\includegraphics[width=1.\columnwidth,trim=1mm 0mm 1mm 1mm, clip]{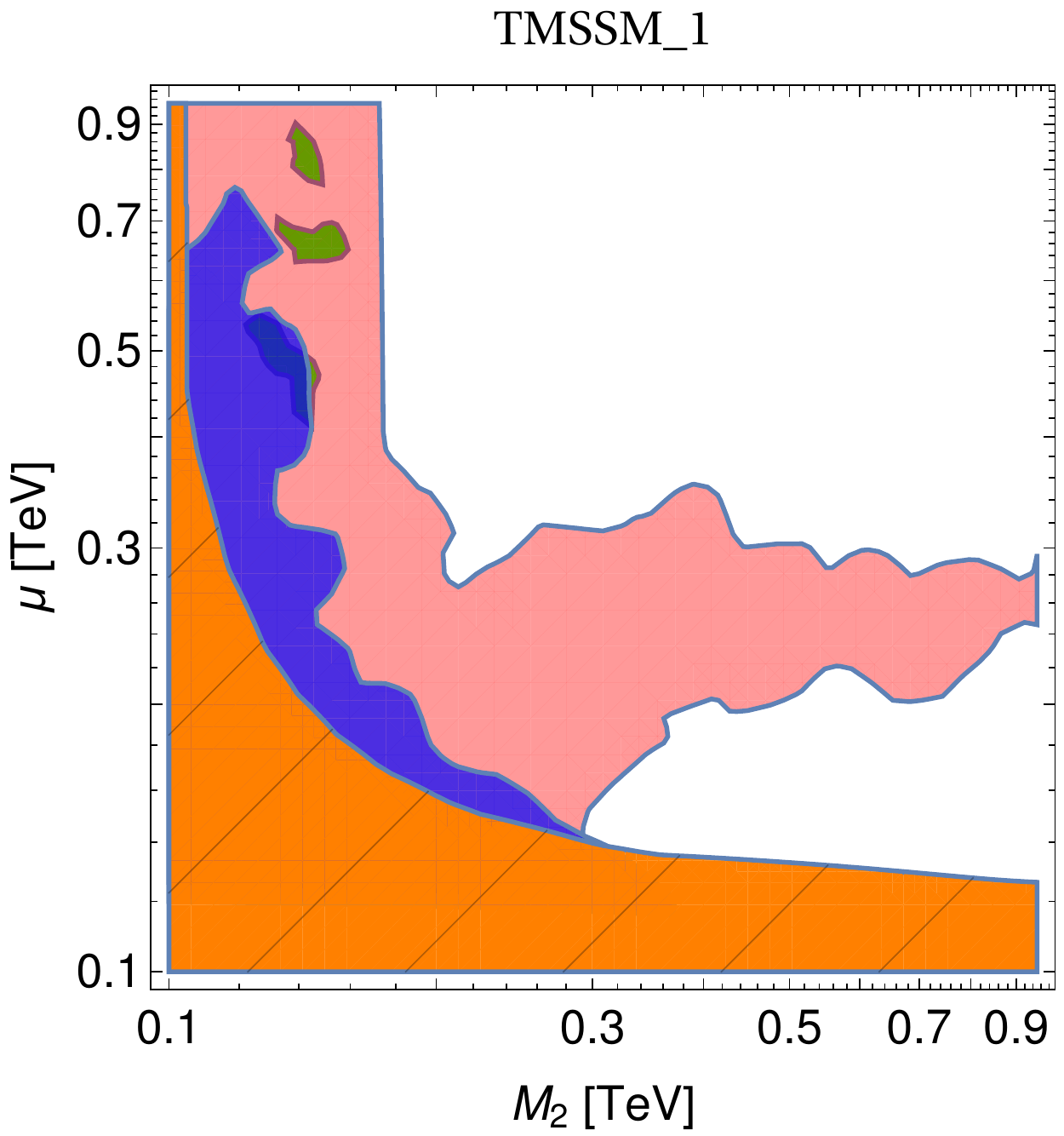}
\end{minipage}
\begin{minipage}[t]{0.32\textwidth}
\centering
\includegraphics[width=1.\columnwidth,trim=1mm 0mm 1mm 1mm, clip]{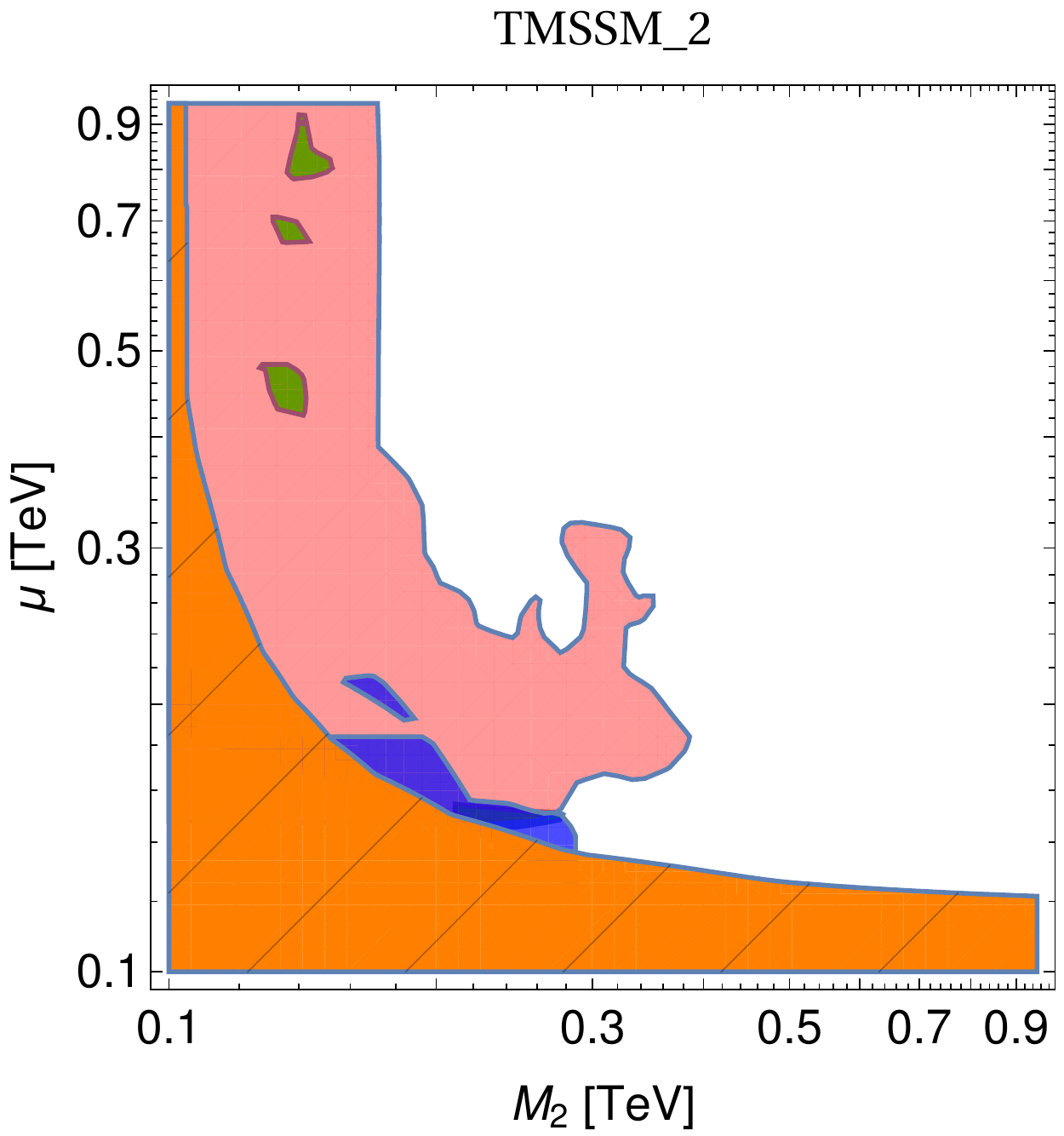}
\end{minipage}
\caption{Left: Excluded regions at 95\% CL  after the LHC Run~1 for the MSSM scenario with $\tan\beta=10$. The green, pink and blue shaded regions are the excluded regions from the two-lepton, three-lepton and four-lepton + MET search respectively, while the shaded orange region denotes the LEP bound on the chargino mass at 95\% CL. Other panels: Same as left for the TMSSM\_1 ($\lambda=0.65$, $\mu_\Sigma=300$ GeV, $\tan\beta=3$) and TMSSM\_2 ($\lambda=0.65$, $\mu_\Sigma=350$ GeV, $\tan\beta=3$) scenarios in the central and right panel respectively. In all panels the lightest neutralino is fixed at $m_{\widetilde{\chi}_1^0} = 63$ GeV.}
\label{fig:exclusion_plots}
\end{figure}

Figure~\ref{fig:exclusion_plots} illustrates the status of the MSSM and the TMSSM models we consider in the light of the multi-lepton searches of LHC Run~1. In all panels we show the exclusion contours derived from the two, three and four lepton + MET searches in green, pink and blue respectively, in the $\{\mu, M_2\}$-plane. The shaded orange region denotes the exclusion limit at 95\% CL on the chargino mass, $m_{\widetilde{\chi}^\pm} >103.5$ GeV~\cite{PDG}. This bound comes from the search for direct production of charginos at LEP via a $Z$ boson and holds for a generic MSSM scenario. The exclusion limit breaks down if the lightest chargino and neutralino are compressed in mass, in models  where the LSP is not the neutralino and in models with R parity violation~\cite{Batell:2013bka}. In the TMSSM models we consider the bound still holds, as the LSP neutralino is never close in mass with the lightest chargino and $\mu_\Sigma$ is sizeable~\cite{Batell:2013bka}.
\begin{table}[t!]
\begin{center}
\caption{Details of three MSSM points belonging to the excluded regions at 8 TeV (MSSM$\_$p1: $M_1 = 80$ GeV, $M_2 = 675$ GeV and $\mu = 110$ GeV; MSSM$\_$p2: $M_1 = 78$ GeV, $M_2 = 279$ GeV and $\mu = 122$ GeV; MSSM$\_$p3: $M_1 = 62$ GeV, $M_2 = 141$ GeV and $\mu = 900$ GeV). All benchmarks have $m_{\widetilde{\chi}_1^0}$ =  63 GeV ($\sim$ Bino) and $\tan\beta=10$. We indicate only the channels with a production cross sections $\sigma > 0.1$ pb and only the BRs larger than 0.01.}
\begin{tabular}{c|c|c|c}
\hline
 Model & Mass [GeV]  & Cross section [pb] & Branching ratios [\%]\\
 \hline
MSSM$\_$p1 &  $m_{\widetilde{\chi}_2^0}$ =  120  & $\sigma(p p \to \widetilde{\chi}_2^0 \widetilde{\chi}_1^0) = 0.52 $   & $\rm BR(\widetilde{\chi}_2^0 \to q \bar{q} \widetilde{\chi}_1^0) = 0.68 $\\
  & ($\sim$ Higgsino) & & $\rm BR(\widetilde{\chi}_2^0 \to l^+ l^- \widetilde{\chi}_1^0) =  0.11$ \\
    & & & $\rm BR(\widetilde{\chi}_2^0 \to \nu \bar{\nu} \widetilde{\chi}_1^0) =  0.21$ \\
\hline
 & $m_{\widetilde{\chi}_3^0}$ = 132  & $\sigma(p p \to \widetilde{\chi}_3^0 \widetilde{\chi}_2^0) =  0.31$   & $\rm BR(\widetilde{\chi}_3^0 \to q \bar{q} \widetilde{\chi}_1^0) = 0.57 $ \\
 &($\sim$ Higgsino) &  & $\rm BR(\widetilde{\chi}_3^0 \to l \bar{l} \widetilde{\chi}_1^0) = 0.25 $\\
 & & & $\rm BR(\widetilde{\chi}_3^0 \to q q' \widetilde{\chi}_1^\pm) = 0.12 $\\
 & &&  $\rm BR(\widetilde{\chi}_3^0 \to l l' \widetilde{\chi}_1^\pm) = 0.058 $\\
 \hline
&  $m_{\widetilde{\chi}_1^+}$ = 111  & $\sigma(p p \to \widetilde{\chi}_1^+ \widetilde{\chi}_1^-) = 0.78$ & $\rm BR(\widetilde{\chi}_1^\pm \to q \bar{q'} \widetilde{\chi}_1^0) = 0.67 $ \\
 & ($\sim$ Higgsino)  & $\sigma(p p \to \widetilde{\chi}_1^\pm \widetilde{\chi}_1^0) = 0.59 $ & $\rm BR(\widetilde{\chi}_1^\pm \to \nu l' \widetilde{\chi}_1^0) = 0.33 $\\
 & & $\sigma(p p \to \widetilde{\chi}_1^\pm \widetilde{\chi}_2^0) =  0.59$ & \\
\hline
MSSM$\_$p2  & $m_{\widetilde{\chi}_3^0}$ = 136   & $\sigma(p p \to \widetilde{\chi}_3^0 \widetilde{\chi}_1^0) =  0.30$ & $\rm BR(\widetilde{\chi}_3^0 \to q \bar{q} \widetilde{\chi}_1^0) = 0.68 $ \\
 & ($\sim$ Higgsino) &$\sigma(p p \to \widetilde{\chi}_3^0 \widetilde{\chi}_2^0) = 0.24$& $\rm BR(\widetilde{\chi}_3^0 \to l^+ l^- \widetilde{\chi}_1^0) = 0.11 $\\
  & & & $\rm BR(\widetilde{\chi}_3^0 \to \nu \bar{\nu} \widetilde{\chi}_1^0) = 0.21 $ \\
  \hline
&  $m_{\widetilde{\chi}_1^+}$ = 114  & $\sigma(p p \to \widetilde{\chi}_1^+ \widetilde{\chi}_1^-) = 0.84$ & $\rm BR(\widetilde{\chi}_1^\pm \to q \bar{q'} \widetilde{\chi}_1^0) = 0.67 $ \\
 & ($\sim$ Higgsino) &$\sigma(p p \to \widetilde{\chi}_1^\pm \widetilde{\chi}_1^0) =  0.55$ & $\rm BR(\widetilde{\chi}_1^\pm \to l^\pm \nu_{l'} \widetilde{\chi}_1^0) = 0.33 $\\
 & & $\sigma(p p \to \widetilde{\chi}_1^\pm \widetilde{\chi}_2^0) = 0.42 $ & \\
 & & $\sigma(p p \to \widetilde{\chi}_1^\pm \widetilde{\chi}_3^0) = 0.42 $ & \\
\hline
MSSM$\_$p3 &  $m_{\widetilde{\chi}_1^+}$ = 150  & $\sigma(p p \to \widetilde{\chi}_1^+ \widetilde{\chi}_1^-) = 0.86$ & $\rm BR(\widetilde{\chi}_1^\pm \to W^{\pm} \widetilde{\chi}_1^0) = 0.99 $\\
  & ($\sim$ Wino)  &$\sigma(p p \to \widetilde{\chi}_1^\pm \widetilde{\chi}_2^0) =  0.88$ & \\
 \hline
\end{tabular}
\label{tab:mssm_bench_8tev}
\end{center}
\end{table}

The left panel of Fig.~\ref{fig:exclusion_plots} illustrates the
disfavoured regions for the MSSM scenario with a fixed LSP mass of 63
GeV. The most constraining search is the three-lepton + MET search (pink
region), which excludes $\mu$ up to TeV for $M_2
    \lesssim 180$ GeV and $M_2 < 380$ GeV for $\mu
\lesssim 250$ GeV. The latter region has both $M_2$ and $\mu$ light, which lead
to a light and quite compressed mass spectrum for the lightest chargino and for the
neutralinos $\widetilde{\chi}^0_2$ and
$\widetilde{\chi}^0_3$. Consequently these particles are produced with
sizeable cross sections and further decay directly into the LSP via
on/off shell two/three body decay. As exemplified by the benchmark
point MSSM\_p2 in Tab.~\ref{tab:mssm_bench_8tev}, various
processes ($p p \to \widetilde{\chi}_1^\pm \widetilde{\chi}_2^0$, $p p
\to \widetilde{\chi}_1^\pm \widetilde{\chi}_3^0$) contribute to the three-lepton
signal and produce a large number of events. A similar
argument can be applied to understand the exclusion coming from the two-lepton
+ MET search (green region) in the same ballpark of values of $\mu$ and
$M_2$.  Again several processes ($p p \to \widetilde{\chi}_2^0
\widetilde{\chi}_1^0$, $p p \to \widetilde{\chi}_3^0
\widetilde{\chi}_1^0$, $p p \to \widetilde{\chi}_3^0
\widetilde{\chi}_2^0$, $p p \to \widetilde{\chi}_1^+
\widetilde{\chi}_1^-$) contribute to the two-lepton + MET signal, producing
numerous events. The most sensitive SR of this latter search is
SRWWa. The same corner of the $\{\mu, M_2\}$-plane is constrained by the
four-lepton search (blue region). The process giving rise to four-lepton + MET
in the final state is $p p \to \widetilde{\chi}_3^0
\widetilde{\chi}_2^0$, which is very sensitive to the value of $M_2$
and $\mu$. As soon as one of these parameters increases,
$\widetilde{\chi}_3^0$ becomes heavy and its production cross section
drops down, reducing drastically the signal. The most sensitive SRs of
this search are SR0Z and SR0noZb, which have the largest number of
signal events.  

Moving to the top left part of the plot (left panel of Fig.~\ref{fig:exclusion_plots}), the excluded
region with light $M_2$ is basically insensitive to the value of
$\mu$, as soon as $\mu > 300$ GeV: here only $\widetilde{\chi}_1^\pm$
and $\widetilde{\chi}_2^0$ (which are mostly Wino) can be produced with a significant cross section as they are always light no matter the value of $\mu$. Hence there is only one relevant process that contributes to the three-lepton signal ($p p \to \widetilde{\chi}_1^\pm \widetilde{\chi}_2^0$),
see {\it e.g.}~the benchmark point MSSM\_p3 in
Tab.~\ref{tab:mssm_bench_8tev}. Similarly,
only lightest chargino production is responsible for the region
excluded by the two-lepton + MET search in the parameter space with low
$M_2$ and large $\mu$. 

The two-lepton + MET search is complementary to the three-lepton + MET search as it is sensitive to the region with low $\mu \gtrsim 150$ GeV and large $M_2 \gtrsim 300$ GeV, which is basically independent of the value of the Wino mass. This case is exemplified by the benchmark point
MSSM\_p1 in Tab.~\ref{tab:mssm_bench_8tev}.  As shown in the table, there are
several processes that can produce two-lepton + MET signals while only one
able to produce three-lepton + MET  final states ($p p \to
\widetilde{\chi}_1^\pm \widetilde{\chi}_2^0$). Again the most
constraining SR of the two-lepton + MET search is SRWWa.

The central and right panels of Fig.~\ref{fig:exclusion_plots} show the excluded regions for the TMSSM cases under analysis. In both scenarios the most constraining search is the three-lepton + MET search. For the TMSSM\_1 (central panel), the three-lepton + MET search excludes the parameter space  $200\,  \rm GeV \lesssim \mu \lesssim 300$ GeV, quite independently of the value of the Wino mass. The rough argument to understand this exclusion is as follow. In this scenario the Triplino mass term is fixed at $\mu_\Sigma = 300$ GeV and a Triplino behaves similarly to a Wino. It is then reasonable to merely exchange $M_2$ with $\mu_\Sigma$ in the MSSM exclusion plot: it is clear that for  $\mu \simeq 200$ GeV a value of $\mu_\Sigma$ of 300 GeV is excluded, while $\mu_\Sigma \simeq 350$ GeV is still allowed by current searches. Hence we can also argue that the exclusion contours for the TMSSM\_2 (right panel), which has $\mu_\Sigma = 350$ GeV, should be closer to the case of the MSSM. This excluded region is indeed only slightly wider than the MSSM case as the EWinos mass spectrum of the TMSSM has a richer content, accordingly there are more processes contributing to the three-lepton + MET final state. In general the most sensitive SRs are those with the largest statistics: for instance bin 14 of the SR0$\tau$a of the three-lepton + MET search seems to be the most sensitive one to look for a BSM signal, no matter what is the SUSY model under investigation. We will discuss the sensitivity of the bins and SRs of the multi-lepton searches to the LHC run at 13 TeV center-of-mass energy in Sec.~\ref{sec:dis}. We provide further details in Appendix~\ref{sec:appA}.

We have performed the Monte Carlo simulations also for an additional scenario of the TMSSM, with $\mu_\Sigma = 350$ GeV and increased $\lambda=0.85$ to assess the impact of changing this parameter. We find that the excluded region is very similar to the TMSSM\_2 case, meaning that leptonic final states are rather insensitive to the value of $\lambda$ (namely they most likely arise from $Z$ and $W$ vector boson decays). The relevance of $\lambda$, which is partially responsible of the coupling between EWinos and the Higgs boson, could be however studied by the search for one-lepton and $b$-jets + MET, designed to tag the production of a Higgs boson in the decay of the SUSY particles into the LSP.  

For the TMSSM cases, the two-lepton + MET search looses sensitivity with respect to the case of the MSSM in the region with low $\mu$. The TMSSM has an enriched spectrum, with one additional neutralino and one additional chargino. For low $\mu$ there can be a greater number of light EWino states with respect to the MSSM case. This has the effect of adding new processes to the three-lepton + MET signal, while keeping constant the number of processes contributing to the two-lepton + MET final state, as shown in Tab.~\ref{tab:bench_tmssm_p1a}, hence reducing the sensitivity of the latter search. Still the most relevant SR for the two-lepton + MET search is SRWWa.

\begin{table}[t!]
\begin{center}
\caption{Details for a TMSSM benchmark point ($M_1 = 78$ GeV, $M_2 = 675$ GeV, $\mu = 135$ GeV, $\mu_\Sigma = 300$ GeV and $\lambda = 0.65$) as compared to MSSM$\_$p1. The LSP mass is fixed at 63 GeV and $\tan\beta=3$. Only BRs above $10^{-2}$  and production cross sections larger than 0.1 pb are reported.}
\begin{tabular}{c|c|c|c}
\hline
 Mass [GeV] & Composition & Cross section [pb] & Branching ratios [\%]\\
 \hline
 $m_{\widetilde{\chi}_2^0}$ =  123  & $\sim$ Higgsino & $\sigma(p p \to \widetilde{\chi}_2^0 \widetilde{\chi}_1^0) = 0.28 $&  $\rm BR(\widetilde{\chi}_2^0 \to q \bar{q} \widetilde{\chi}_1^0) = 0.68 $ \\
 & &  & $\rm BR(\widetilde{\chi}_2^0 \to l^+ l^- \widetilde{\chi}_1^0) = 0.11 $\\
 & & & $\rm BR(\widetilde{\chi}_2^0 \to \nu\bar{\nu} \widetilde{\chi}_1^0) = 0.21 $ \\
 \hline
 $m_{\widetilde{\chi}_3^0}$ = 156 & $\sim$ Higgsino & $\sigma(p p \to \widetilde{\chi}_3^0 \widetilde{\chi}_2^0) =  0.23$ & $\rm BR(\widetilde{\chi}_3^0 \to Z \widetilde{\chi}_1^0) = 0.96 $ \\
   \hline   
 $m_{\widetilde{\chi}_1^\pm}$ = 117 & $\sim$ Higgsino  & $\sigma(p p \to \widetilde{\chi}_1^+ \widetilde{\chi}_1^-) = 0.92$ &  $\rm BR(\widetilde{\chi}_1^\pm \to q \bar{q}' \widetilde{\chi}_1^0) = 0.67 $  \\
 & & $\sigma(p p \to \widetilde{\chi}_1^\pm \widetilde{\chi}_1^0) =  0.32$ & $\rm BR(\widetilde{\chi}_1^\pm \to \nu l' \widetilde{\chi}_1^0) = 0.33 $ \\
 & &$\sigma(p p \to \widetilde{\chi}_1^\pm \widetilde{\chi}_2^0) =  0.74$ & \\
 & & $\sigma(p p \to \widetilde{\chi}_1^\pm \widetilde{\chi}_3^0) = 0.24$ & \\
\hline
\end{tabular}
\label{tab:bench_tmssm_p1a}
\end{center}
\end{table}

To summarize, the most stringent constraints on the EWino parameter space are set by the three-lepton + MET search, in both the MSSM and the TMSSM models. In the MSSM the two-lepton + MET search is complementary to the three-lepton final state in the region with low $\mu$ and $M_2 \gtrsim 300$ GeV. This complementarity is lost in the TMSSM models, where only the three-lepton + MET search is able to exclude a significant part of the parameter space. Our findings for the excluded regions of the MSSM are compatible with
the analysis done in Ref.~\cite{Martin:2014qra}.

\section{Forecasts for multi-lepton signals at the LHC Run~2 and future runs}\label{sec:res}

Searches at the current LHC energy, $\sqrt{s} = 13$ TeV, will probe a
much larger region of the parameter space of both the MSSM and the
TMSSM, especially for large luminosities. To determine the capabilities of the
LHC in the near future, we consider the same searches as in
Sec.~\ref{sec:analysis} and estimate the excess of events over the SM
expectations in each SR. The complete results are provided
in Appendix~\ref{sec:appA}. Here we highlight our major findings.
\begin{figure}[t]
\begin{center}
\includegraphics[width=0.325\columnwidth]{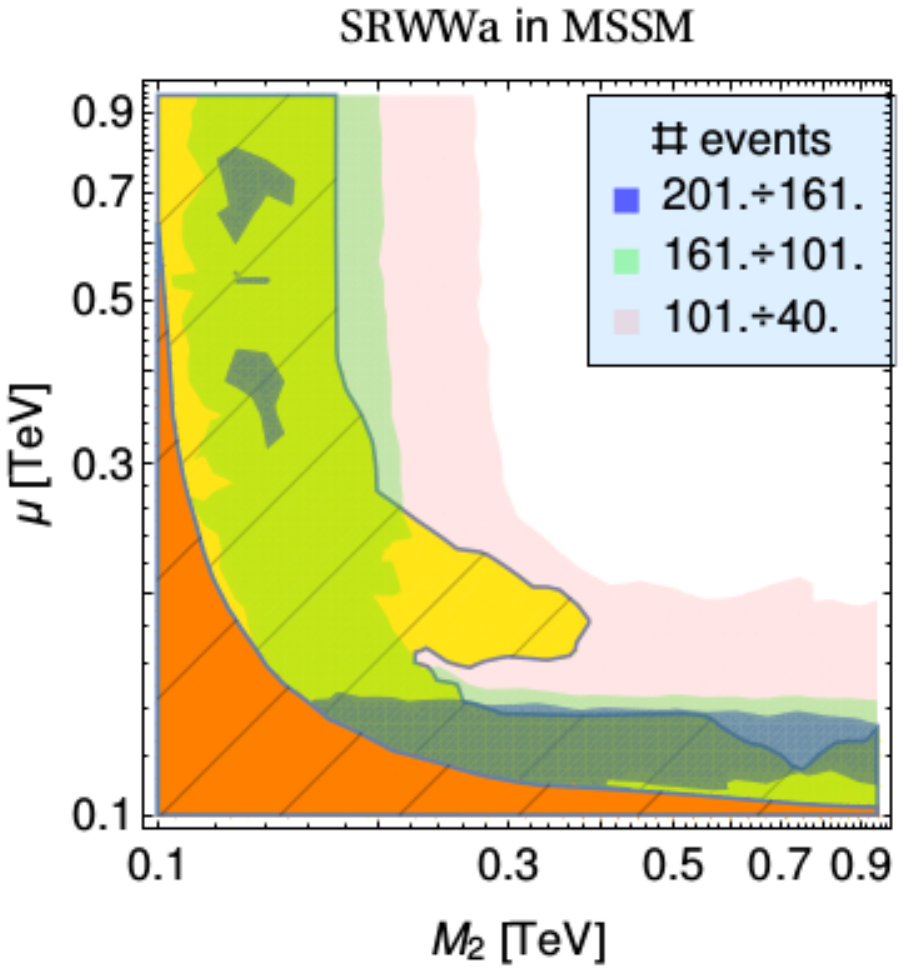}
\includegraphics[width=0.325\columnwidth]{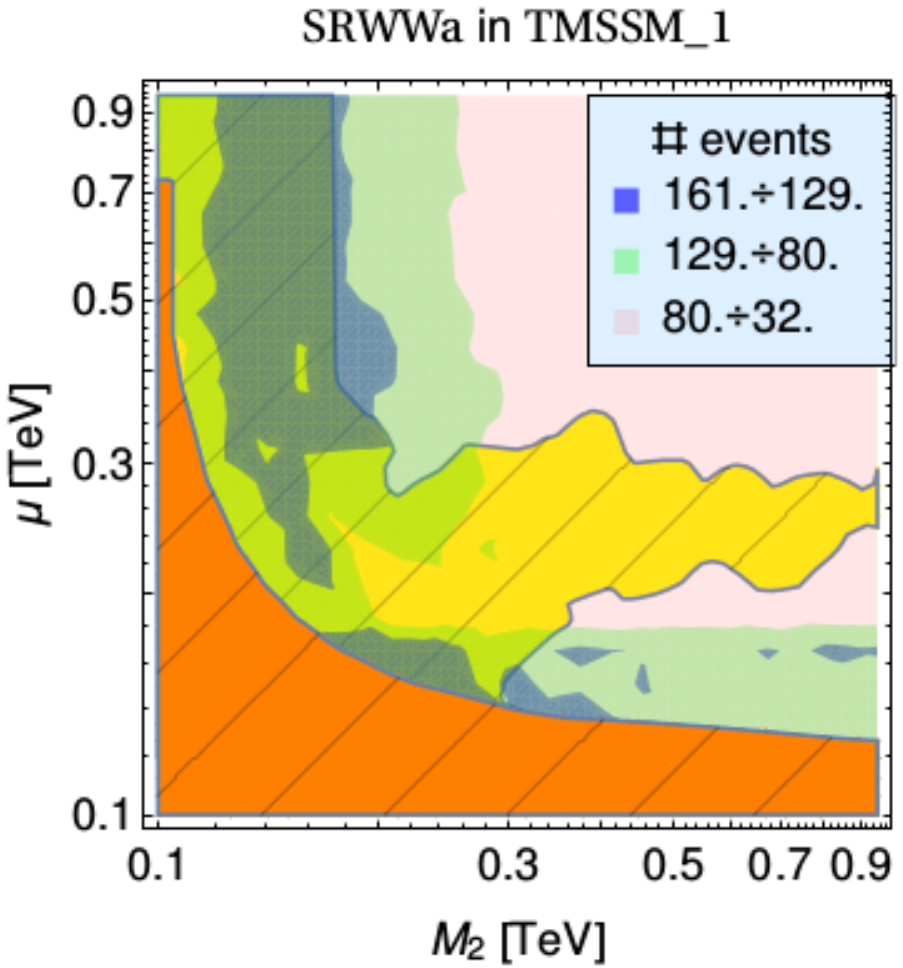}
\includegraphics[width=0.325\columnwidth]{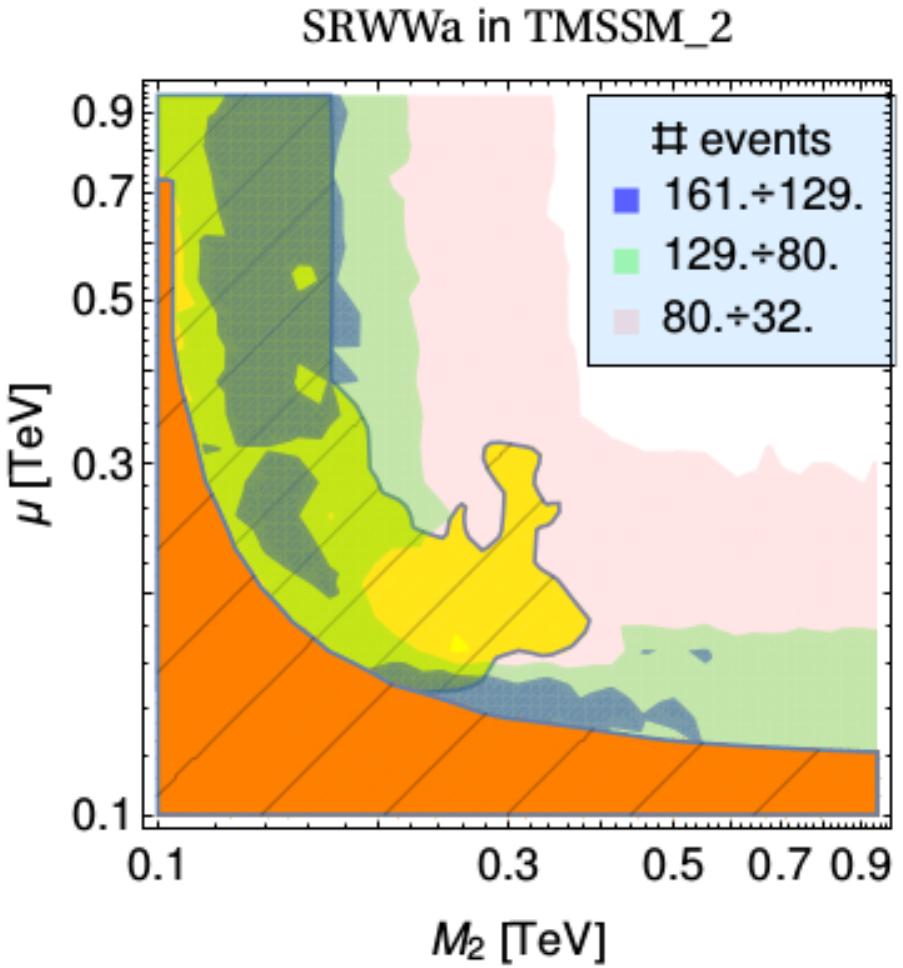}
\end{center}
\caption{Number of expected signal events, as labelled, in the MSSM (left panel), the TMSSM\_1 (center panel) and the TMSSM\_2 (right panel) in the SRWWa region of the two-lepton analysis. The excluded regions (yellow and orange for the LHC Run~1 bound and chargino mass bound respectively), surrounded by the solid line, are over-imposed.}\label{fig:excesses}
\end{figure}

\subsection{Ascribing a multi-lepton excess to the EWinos}

An illustrative example of the 13 TeV forecasts is shown in
Fig.~\ref{fig:excesses}.  From left to right, the expected number of
signal events in the SRWWa region of the two-lepton analysis is
plotted for a luminosity of $100$ fb$^{-1}$ in the MSSM,
the TMSSM\_1 and the TMSSM\_2, respectively.  Orange and yellow regions
are ruled out by the LEP chargino lower limit and the Run 1 bound
obtained before.  In all the three cases, in the regions not yet
excluded the expected number of signal events goes up to $\sim$150
events. This is about 10 times larger than the expectation in the
first LHC run, a factor of five being due to the luminosity, while the
remaining factor of two comes from the enhancement in the production
cross section as a result of the energy increasing and the PDFs. The
SM background, which is quoted in Sec.~\ref{sec:analysis}, is expected
to scale similarly. This is relevant for the significance of the
signal excess, whose estimate is described in the next section. For
the time, we disregard such a quantity and base our discussion on
order-of-magnitude arguments.

\begin{figure}[t]
\begin{center}
\includegraphics[width=0.325\columnwidth]{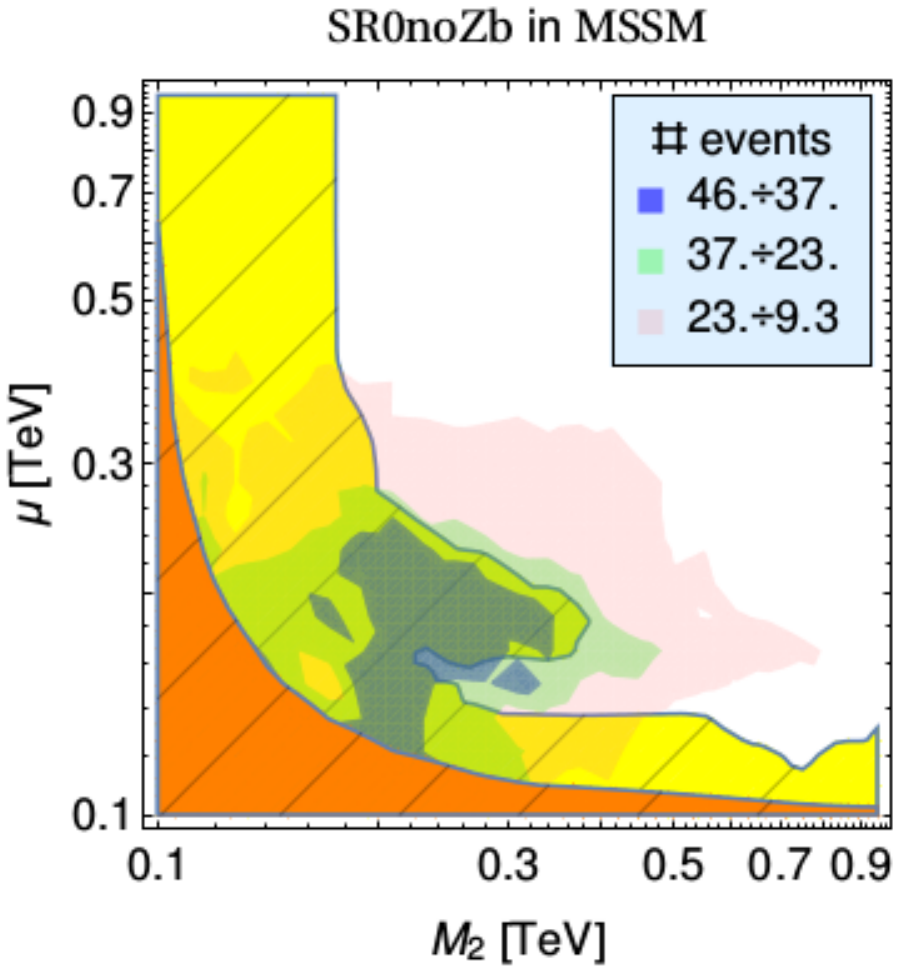}
\includegraphics[width=0.325\columnwidth]{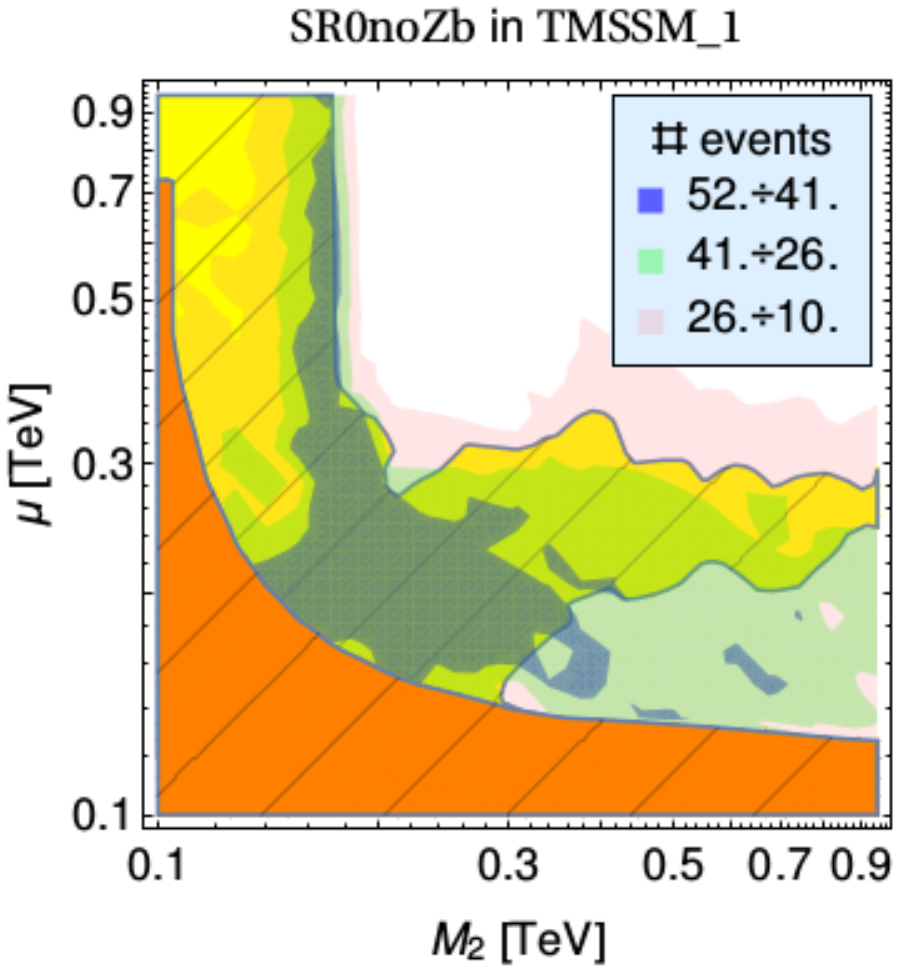}
\includegraphics[width=0.325\columnwidth]{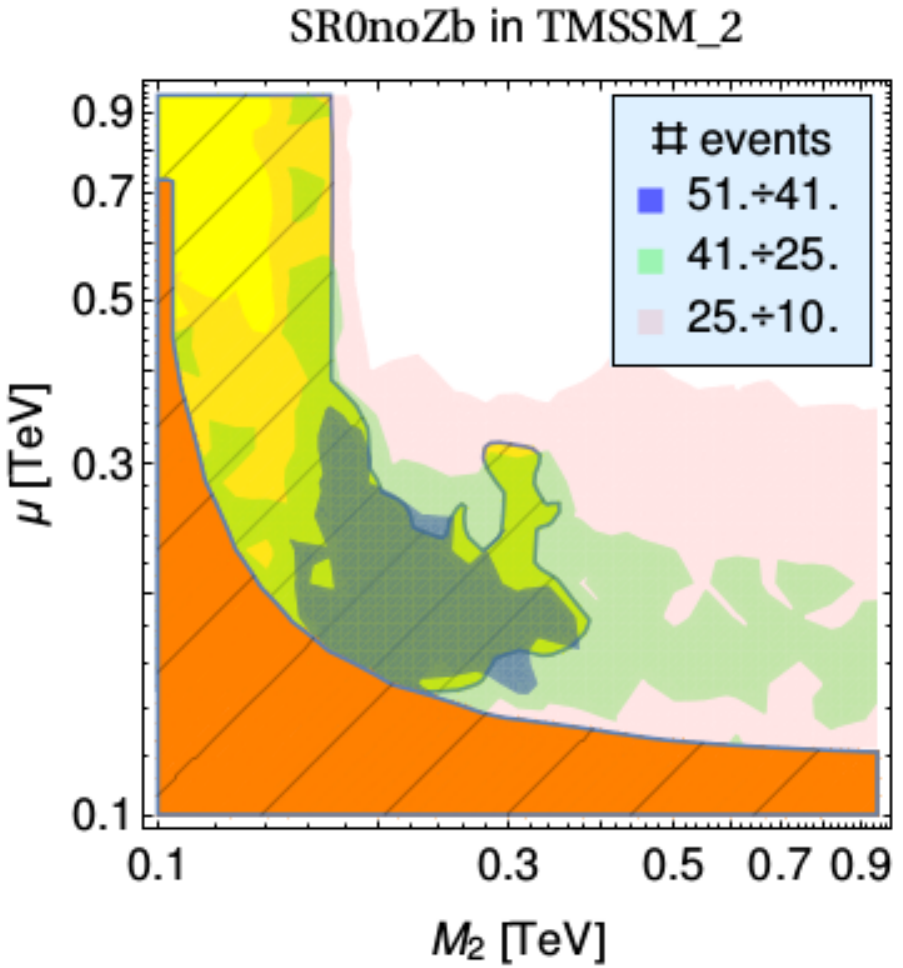}
\end{center}
\caption{Same as  Fig.~\ref{fig:excesses} but for the SRnoZb region of the four-lepton analysis.}\label{fig:excesses3}
\end{figure}
In several cases the forecasts allow to identify whether an anomaly in
the multi-lepton data can or cannot be ascribed to the EWinos (with a
Bino-like DM LSP). For illustrative purposes, we restrict the present
discussion to the SRWWa, SR0$\tau$a-20 and SRnoZb SRs and
we assume an excess of $\sim$150 events in SRWWa for a $100$
fb$^{-1}$ luminosity. As shown in the Appendix~\ref{sec:appA}, if
this excess is (exclusively) due the EWinos, no excess above 10 events
is possible in SR0$\tau$a-20 (here the background is expected to be
around 3 events). On the other hand, if this $\sim$150 excess is
accompanied by an anomaly of $\sim$50 events in SRnoZb (here the
background is about 13 events) and a few events in SR0$\tau$a-20, the
deviation is compatible with the MSSM, TMSSM\_1 and TMSSM\_2 EWino
production, at least for what concerns these three SRs. In
particular, the compatibility is possible in the part of
$\{\mu,M_2\}$-plane where, for a given model, the blue regions in
Figs.~\ref{fig:excesses} and~\ref{fig:excesses3} overlap. Of course, to
fully test the EWino hypothesis, the compatibility of the excesses in
all the SRs must be checked (see Appendix~\ref{sec:appA}
for more forecast plots). In addition, to quantify the compatibility,
the systematic uncertainties involved in the EWino production cannot
be disregarded~\footnote{We compute the EWino production cross sections at leading order.}. In this sense, all the numbers of expected
signal events in our forecast plots should be used up to a common
normalization factor. The statistical uncertainty must also be considered,
and this can be done as explained in the next section.

\subsection{Disentangling  the TMSSM from the MSSM}\label{sec:dis}

In light of the previous results, one might wonder whether the TMSSM
(in one of its scenarios) can be distinguished from the MSSM if an
excess is indeed observed in future data. The difficulties for
addressing this question are two-fold. On the one hand, the measurement of
such an excess is subject to statistical errors due to the potentially
large (depending on the SR) background fluctuations. In
order to estimate this uncertainty, let us call $D$ the total number
of observed events in a particular SR. Let $B$ stands for
the number of expected SM events. The number of measured signal events
is then given by $S = D-B$ and hence, under the assumption of gaussian
distributed events, the uncertainty in $S$ can be estimated to be
\begin{equation}\label{eq:error}
\Delta S = \sqrt{D + B} = \sqrt{S + 2B}.	
\end{equation}
In this way, we can link the uncertainty on $S$ with $S$ itself,
provided the number of background events is well known. A similar approach has been adopted in a different context in Ref.~\cite{delAguila:2013yaa}. Given that the
multi-lepton production in both the signal and the background is mainly
mediated by EW gauge bosons produced in quark-antiquark
collisions, we compute the latter simply rescaling the numbers quoted
in Sec.~\ref{sec:analysis} by the same amount found for the signal.

On the other hand, it is clear that even the precise measurement of
$S$ in a single SR cannot shed light on the nature of the
SUSY model, nor on the values of $M_2$ and $\mu$. As a matter of fact,
the whole border between the red and the white regions in
Fig.~\ref{fig:excesses} in the MSSM correspond to $S\sim 40$
events. A similar number is found in a wide region of the parameter
space of both TMSSM scenarios. Therefore, comparing several
(potentially all) SRs becomes necessary for disentangling
the TMSSM and the MSSM.

\begin{figure}[t]
\begin{center}
\includegraphics[width=0.49\columnwidth]{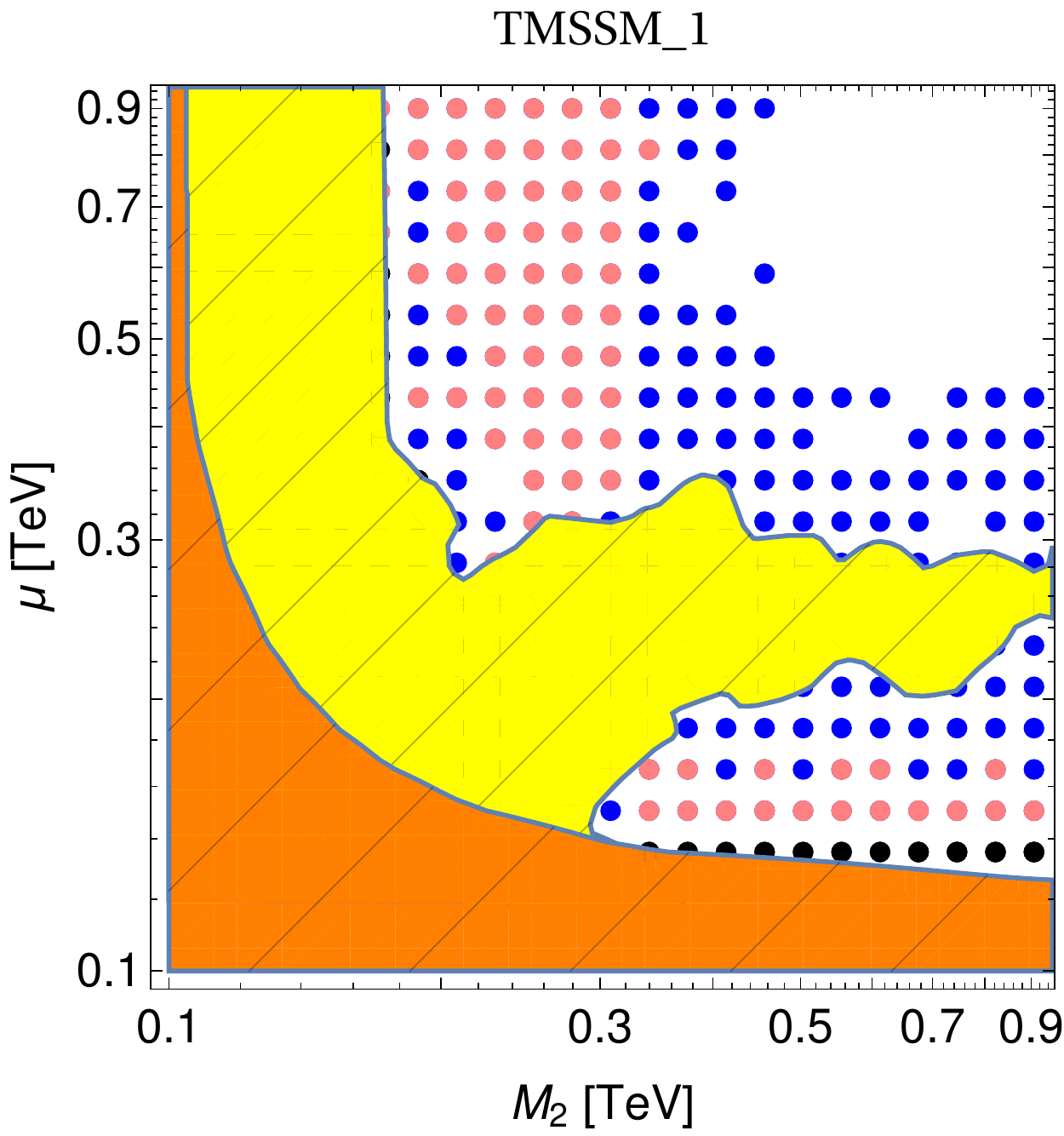}
\includegraphics[width=0.49\columnwidth]{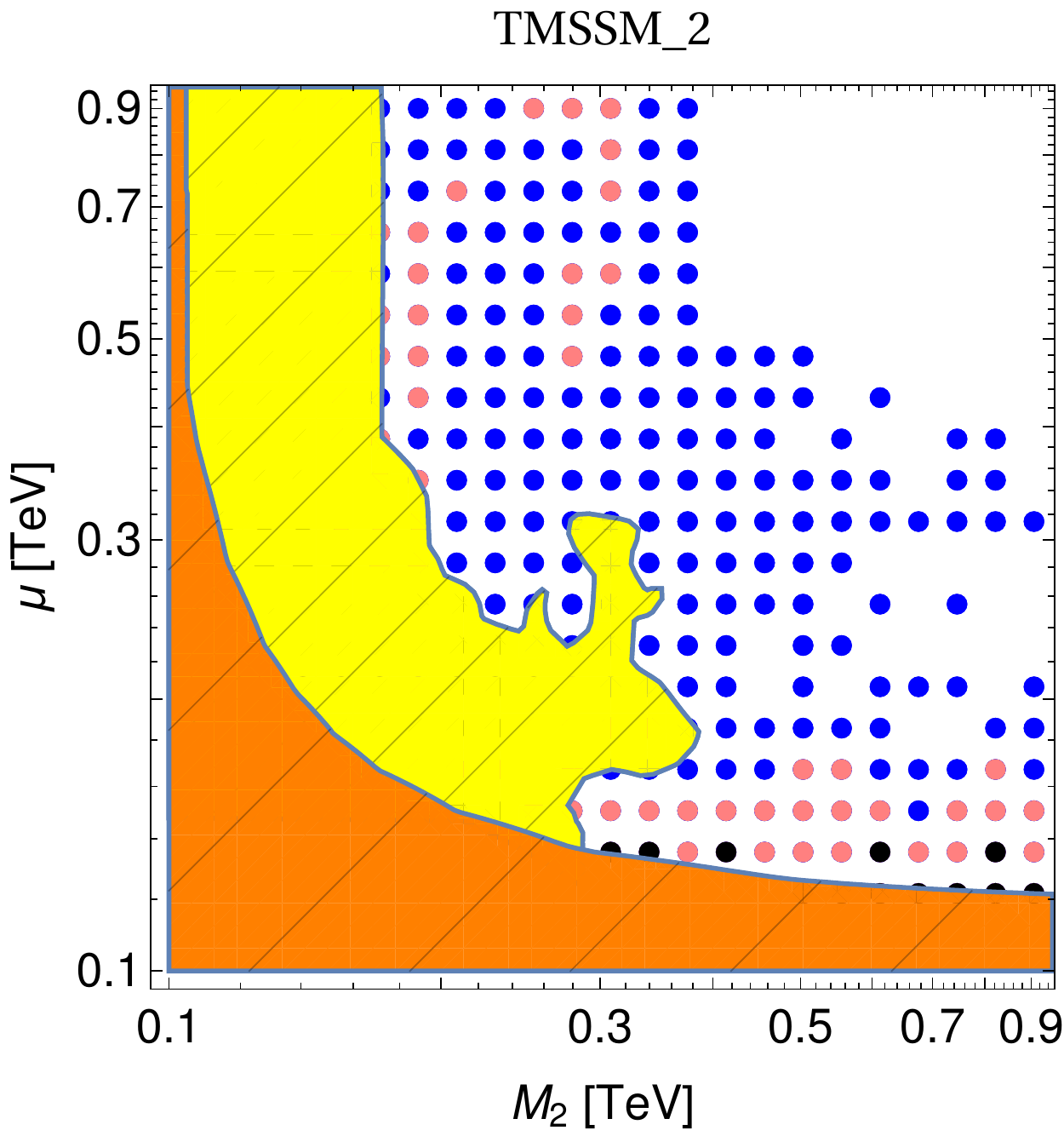}
\end{center}
\caption{Paramater space regions of the TMSSM\_1 (left) and the TMSSM\_2 (right) that can be disentangled from
the MSSM with 100 fb$^{-1}$ (black points), 300 fb$^{-1}$ (red points) and 3000 fb$^{-1}$ (blue points) by comparing bin by bin in all multi-lepton analyses.}\label{fig:final}
\end{figure}

Our suggested strategy is as follows. For each parameter space point
of the TMSSM, we check whether there exist \textit{at least one} point of the MSSM for which the expected numbers of events in \textit{all} SRs
\textit{separately} are compatible, within twice the standard
deviation given by Eq.~(\ref{eq:error}), with those predicted by the
selected value of the TMSSM parameters. If this is \textit{not} the
case, the latter can be discriminated from the MSSM. We adopt this
approach instead of comparing the whole SR distributions in order the
results to be conservative. Indeed, several (uncorrelated) SRs are
often involved when a TMSSM parameter point is discriminated from the
MSSM.

In the left (right) panel of Fig.~\ref{fig:final} we depict the
regions of the TMSSM\_1 (TMSSM\_2) that can be disentangled from the
MSSM with 100 fb$^{-1}$ (black points), 300 fb$^{-1}$ (red points) and
3000 fb$^{-1}$ (blue points). It is apparent that most of parameter
space region not yet excluded in both versions of the TMSSM would lead
to significantly different predictions from the MSSM ones at the LHC
in the long term.

\section{Conclusions}\label{sec:Concl}

What are the actual bounds on electroweakinos (EWinos) when the
simplified model spectra assumption adopted in the experimental analyses is
relaxed? Is there any pattern among the multi-lepton signal regions
that events produced by EWinos should follow?  In presence of an excess in the multi-lepton channels, is it possible to disentangle among supersymmetric models by means of the details of this pattern? 
These are the questions we have tackled in the
present paper. 

To answer to these questions, we have recast the present most
contraining analyses on EWinos and applied them to MSSM and
Triplet-extension-of-the-MSSM (TMSSM) scenarios with sfermions,
gluinos and non-SM Higgses well above the present bounds. Contrarily
to the experimental assumptions, no specific EWino mass hierarchy has
been assumed. To illustrate our procedure and findings we have chosen a scenario in which the lightest neutralino, Bino-like, is a good dark matter candidate and has a mass in the Higgs funnel region, where LHC is expected to have a better reach and sensitivity with respect to dark matter experiments.

By means of our study we have confirmed that searches for final states
with missing transverse energy and two, three or four
leptons~\cite{Aad:2014vma,Aad:2014nua,ATLAS-CONF-2013-036} are very
efficient to probe light-EWino scenarios~\cite{Martin:2015hra}.
In particular, in the considered MSSM case the present strongest
constraints on EWinos, which still come from Run 1 analyses, always
result more stringent than the LEP chargino mass bound, {\it i.e.}~$m_{\widetilde{\chi}^\pm} \gtrsim  104\,$GeV. 
On the contrary, in the TMSSM, there
exists a parameter region (with the Wino and Higgsino mass parameters
at $M_2\gtrsim 300\,$GeV and $\mu\simeq 120\,$GeV,
respectively) that evades the LHC multi-lepton
constraints and is limited only by the LEP chargino bound.

We have also provided forecasts for the multi-lepton signals produced
by the EWino sectors of the MSSM and the TMSSM. For both models we have
determined the number of signal events that are expected in each of
the signal regions of the multi-lepton analyses above.  Irrespectively
of the particular MSSM or TMSSM realization, these numbers exhibit
some qualitative correlations. These should allow to easily understand
whether future anomalies in multi-lepton data are or are not
ascribable to EWinos and, in case, what typical values of $\mu$ and
$M_2$ can explain the signal.

We have moreover proven that with large enough luminosity
the above correlations become precise and sensitive to details of the
EWino sector. In some cases, given an EWino signal, it is possible to
understand whether the underling theory is the MSSM or some other
supersymmetric model with an extended EWino sector. For instance,
already at 100\,fb$^{-1}$, there exist a few TMSSM configurations
whose EWino signals have correlations that cannot be produced by the
MSSM EWinos. This drastically improves at 3000\,fb$^{-1}$: in the
considered TMSSM scenario, only the region with $\mu\gtrsim 400\,$GeV
and $M_2\gtrsim 400\,$GeV leads to signals that can be ascribed also
to EWinos of the MSSM.

Interestingly, some searches that we have not investigated in this
paper should be sensitive to the large $\mu$ and large $M_2$ parameter
region where our procedure fails to disentangle the MSSM from the
TMSSM. For instance, the CMS analysis on final states with one lepton
and $b$-jets~\cite{Aad:2015jqa} partially covers this parameter space, at
least in the MSSM~\cite{Martin:2015hra}. Also the kinematic
observables discussed in Ref.~\cite{Cabrera:2012gf} are
efficient in probing cases where $\mu$ is large. Therefore, including
these extra observables and the one-lepton plus $b$-jet search
seems promising and worth investigating in the future.

\acknowledgments CA and GN would like to thank Jes\'us Mar\'ia Moreno for useful
  discussions and hospitality at the beginning of this work. We
acknowledge T. Martin that kindly provided us output files to validate
our analysis and J. Soo Kim for discussions. CA would like to thank
L. Quertenmont and P. Demin for support for running the simulations on
the CP3 cluster. MC would like to thank the Depto. de F\'isica Te\'orica y del Cosmos in the University of Granada for the hospitality during the completion of this work. VML would like to thank the CP3 in Louvain-la-Neuve,
where a part of this project was completed, and Fabio Maltoni for kind
hospitality. The research of CA is supported by the ATTRACT - Brains
back to Brussels 2015 grant at UCL (Innoviris / 2015 BB2B 4). GN is supported by the Swiss
National Science Foundation (SNF) under grant 200020-15593. VML
acknowledges the support of the Consolider-Ingenio 2010 programme
under grant MULTIDARK CSD2009-00064, the Spanish MICINN under Grant
No. FPA2015- 65929-P, the Spanish MINECO ``Centro de excelencia Severo
Ochoa'' programme under Grant No.  SEV-2012-0249, and the European
Union under the ERC Advanced Grant SPLE under contract
ERC-2012-ADG-20120216-320421 and the BMBF under project
05H15PDCAA.

\appendix
\section{Relevance of the signal regions in the multi-lepton searches at 13 TeV}\label{sec:appA}

In this section we provide a detailed look at the sensitivity of the SRs of each multi-lepton search considered in the analysis for the 13 TeV configuration of the LHC. Before starting to describe the figures, let us define the information contained in each panel. Let us consider at first the top left panel of Fig.~\ref{fig:2lepton_mssm13tev_bins} as an example. This plot shows the number of signal events produced by the MSSM model in the SRm$_{T2,90}$ of the two-lepton search, in the $\{\mu, M_2\}$-plane. In particular, the  blue region shows the number of signal events starting from the 80\% of the maximum number of MSSM events up to its maximum, which is in this case 133. The green and pink regions display the number of signal events in between the 50\% and 80\% of the maximum and in between the 20\% and the 50\% of the maximum respectively. The panel also shows the exclusion contour (yellow) coming from Run~1, computed as described in Sec.~\ref{sec:numdet} as well as the exclusion limit (orange) from LEP on the chargino mass~\cite{PDG} ($m_{\widetilde{\chi}_1^+} \gtrsim 104$ GeV). All panels for the two-lepton, three-lepton and four-lepton search are produced with these details. 

\subsection*{MSSM}

\begin{figure}[t!]
\begin{minipage}[t]{0.3\textwidth}
\centering
\includegraphics[width=1.\columnwidth]{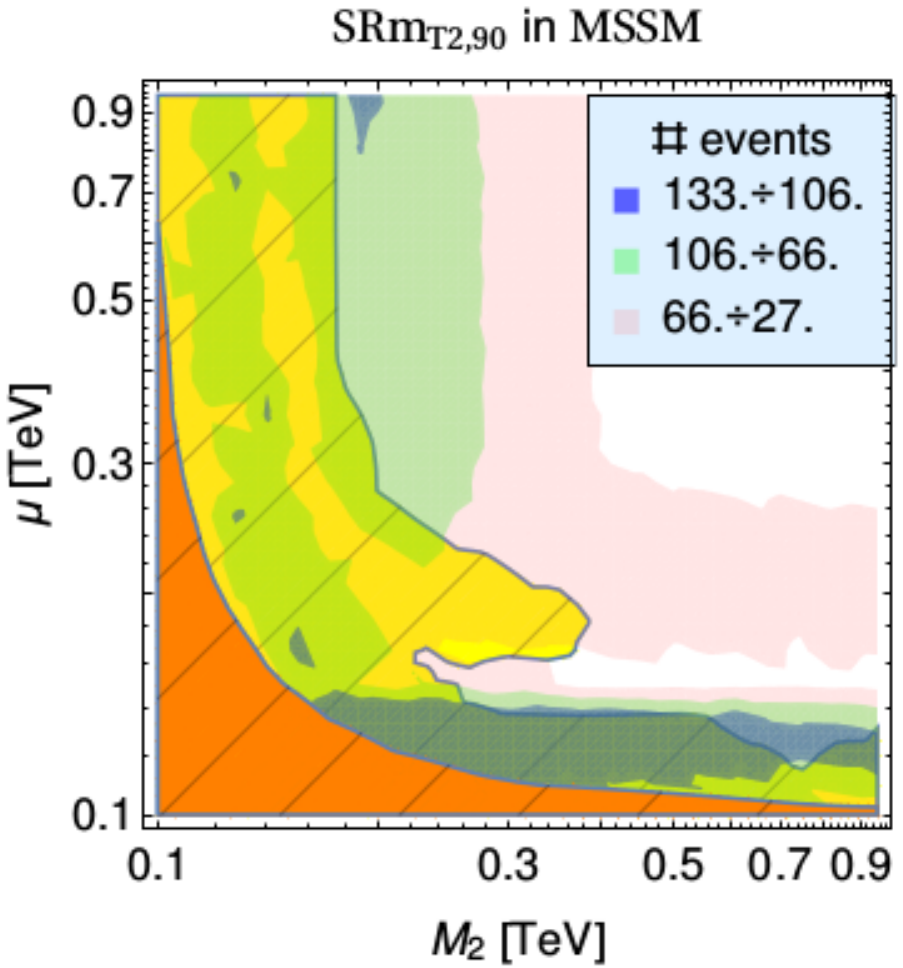}
\end{minipage}
\begin{minipage}[t]{0.3\textwidth}
\centering
\includegraphics[width=1.\columnwidth]{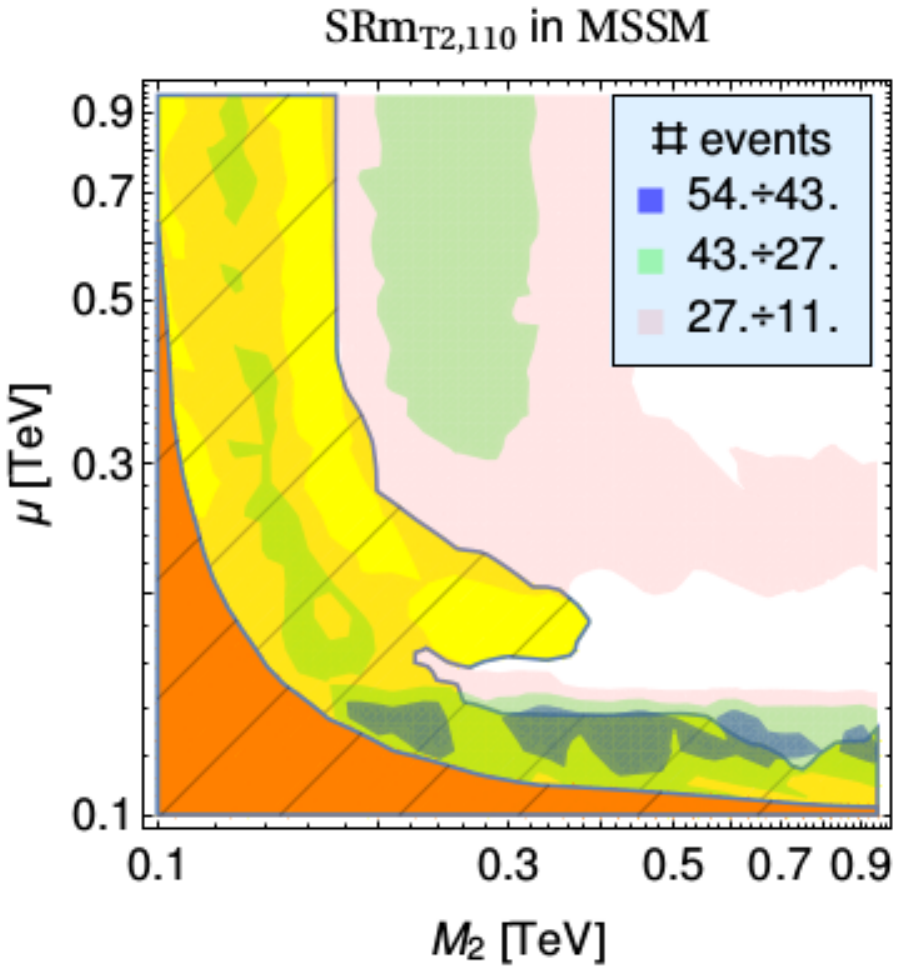}
\end{minipage}
\begin{minipage}[t]{0.3\textwidth}
\centering
\includegraphics[width=1.\columnwidth]{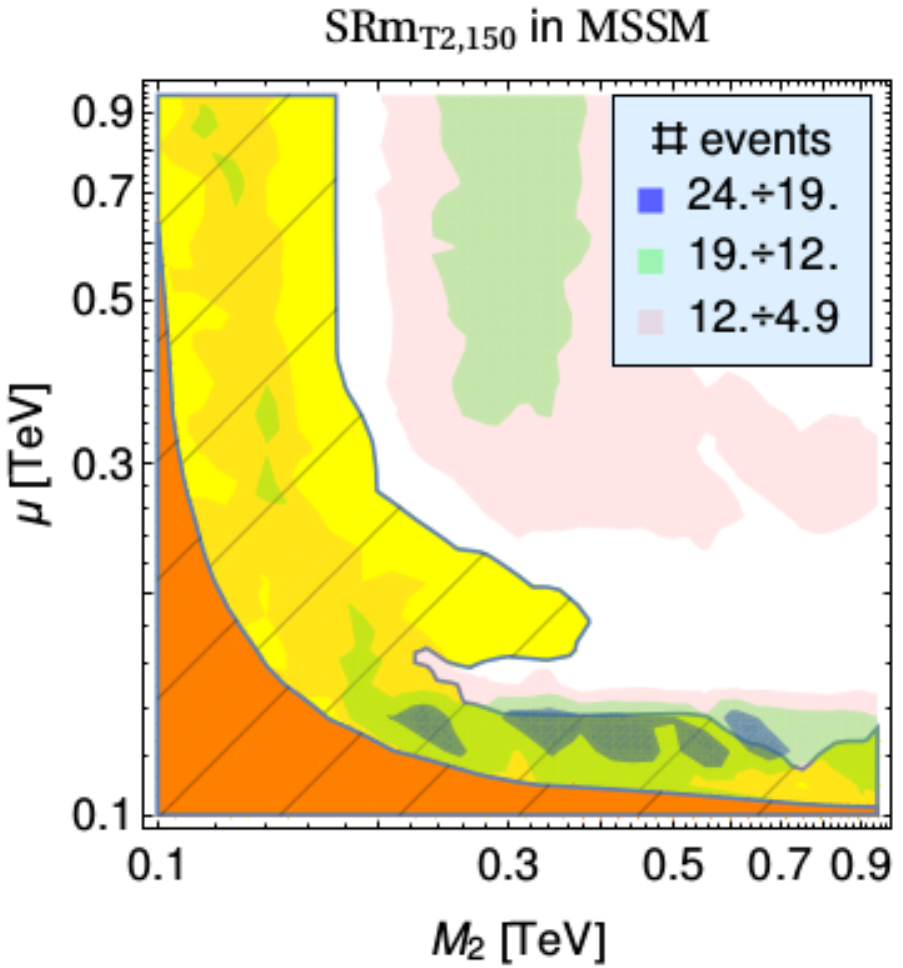}
\end{minipage}
\\
\begin{minipage}[t]{0.3\textwidth}
\centering
\includegraphics[width=1.\columnwidth]{./figs/MSSM_63_13TeV-2lept-bin4.pdf}
\end{minipage}
\begin{minipage}[t]{0.3\textwidth}
\centering
\includegraphics[width=1.\columnwidth]{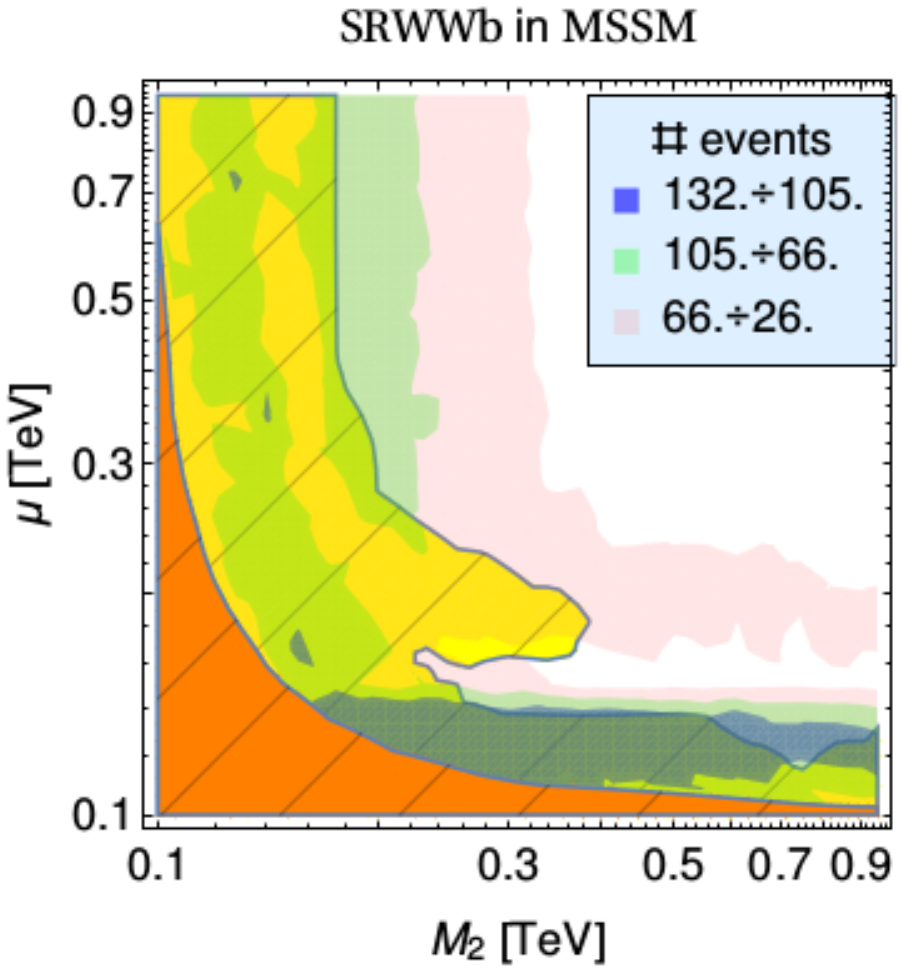}
\end{minipage}
\begin{minipage}[t]{0.3\textwidth}
\centering
\includegraphics[width=1.\columnwidth]{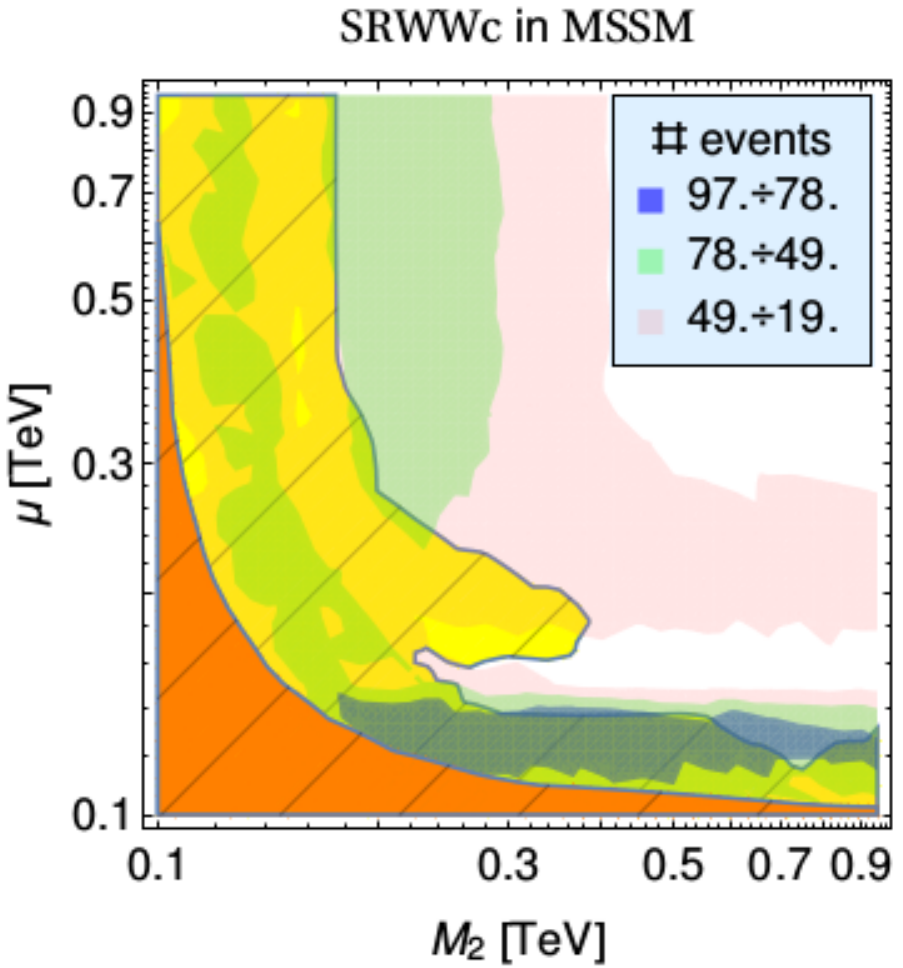}
\end{minipage}
\\
\begin{minipage}[t]{0.3\textwidth}
\centering
\includegraphics[width=1.\columnwidth]{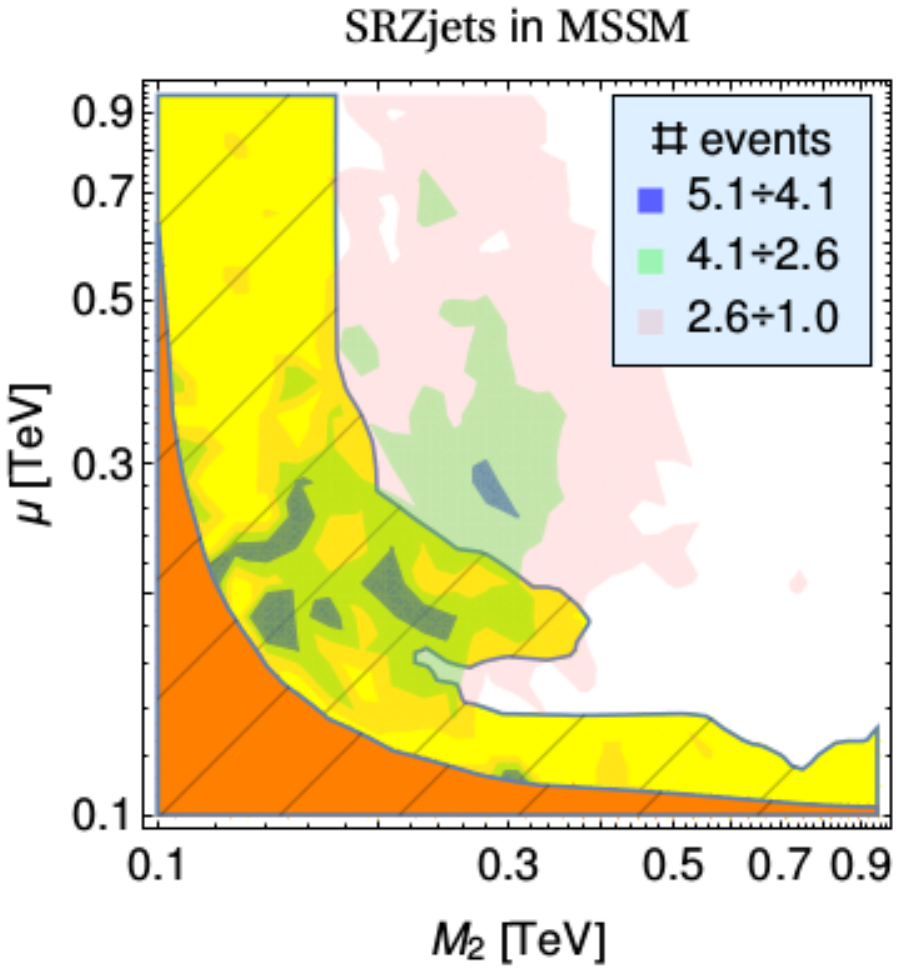}
\end{minipage}
\begin{minipage}[t]{0.3\textwidth}
\centering
\includegraphics[width=1.\columnwidth]{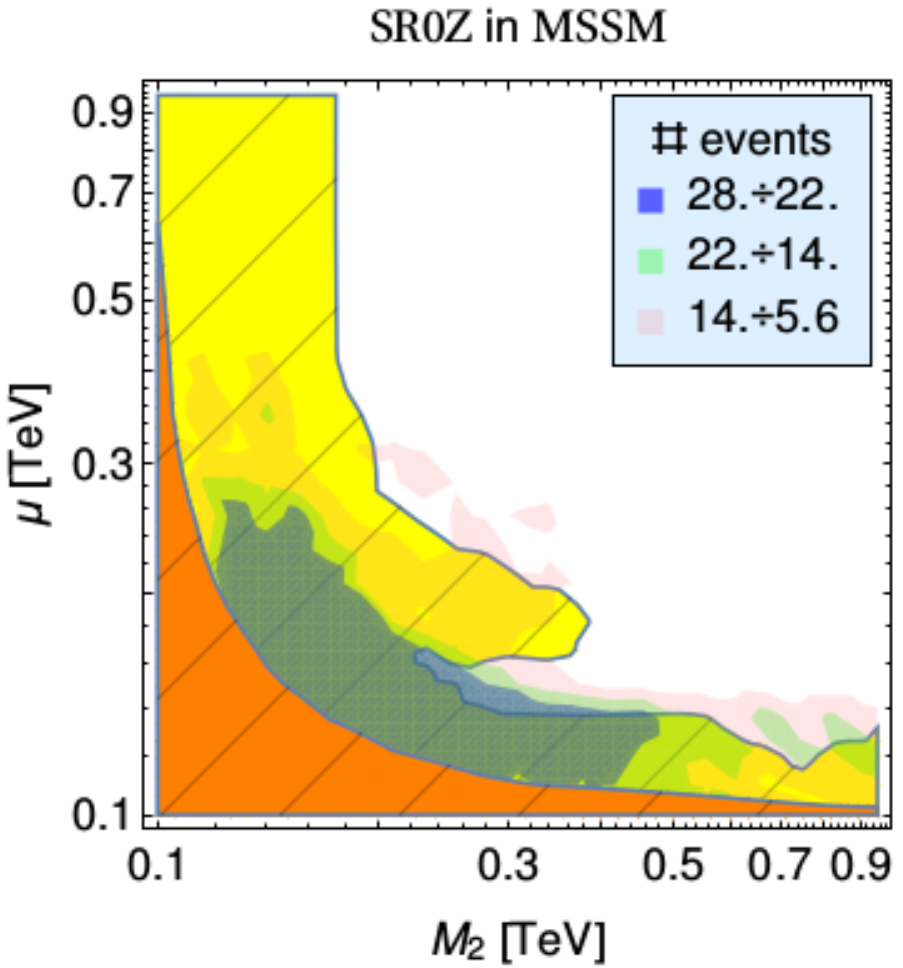}
\end{minipage}
\begin{minipage}[t]{0.3\textwidth}
\centering
\includegraphics[width=1.\columnwidth]{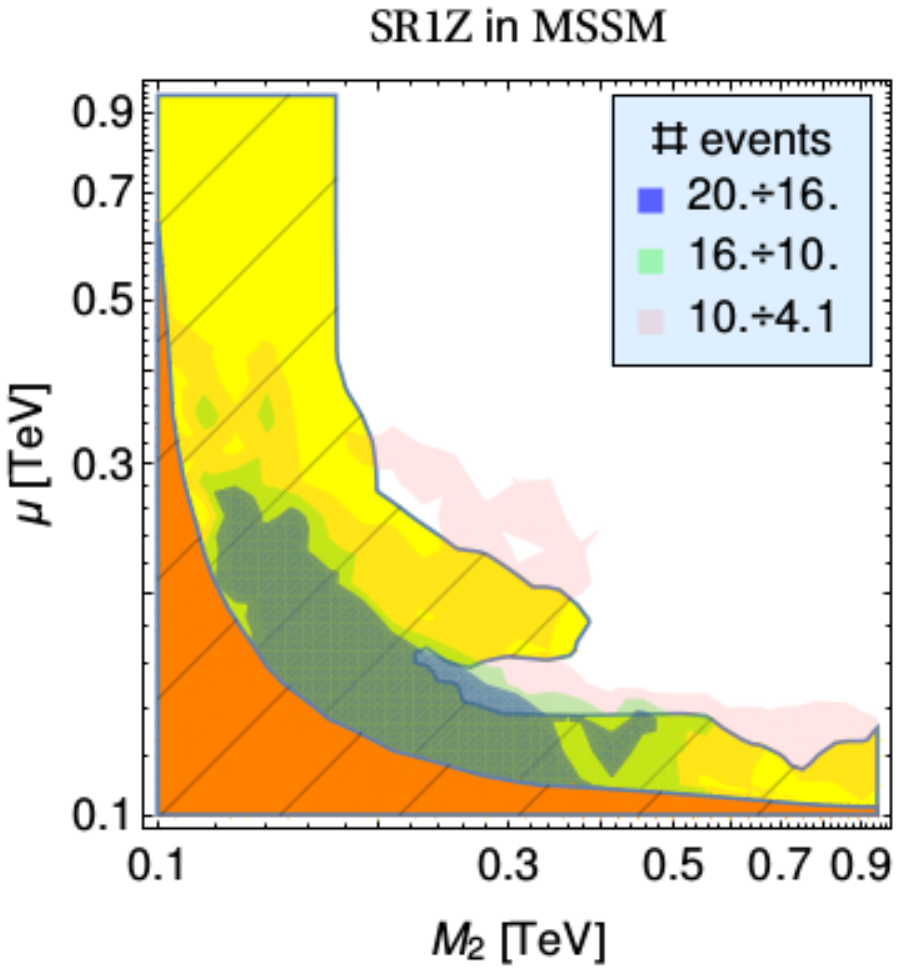}
\end{minipage}
\\
\begin{minipage}[t]{0.3\textwidth}
\centering
\includegraphics[width=1.\columnwidth]{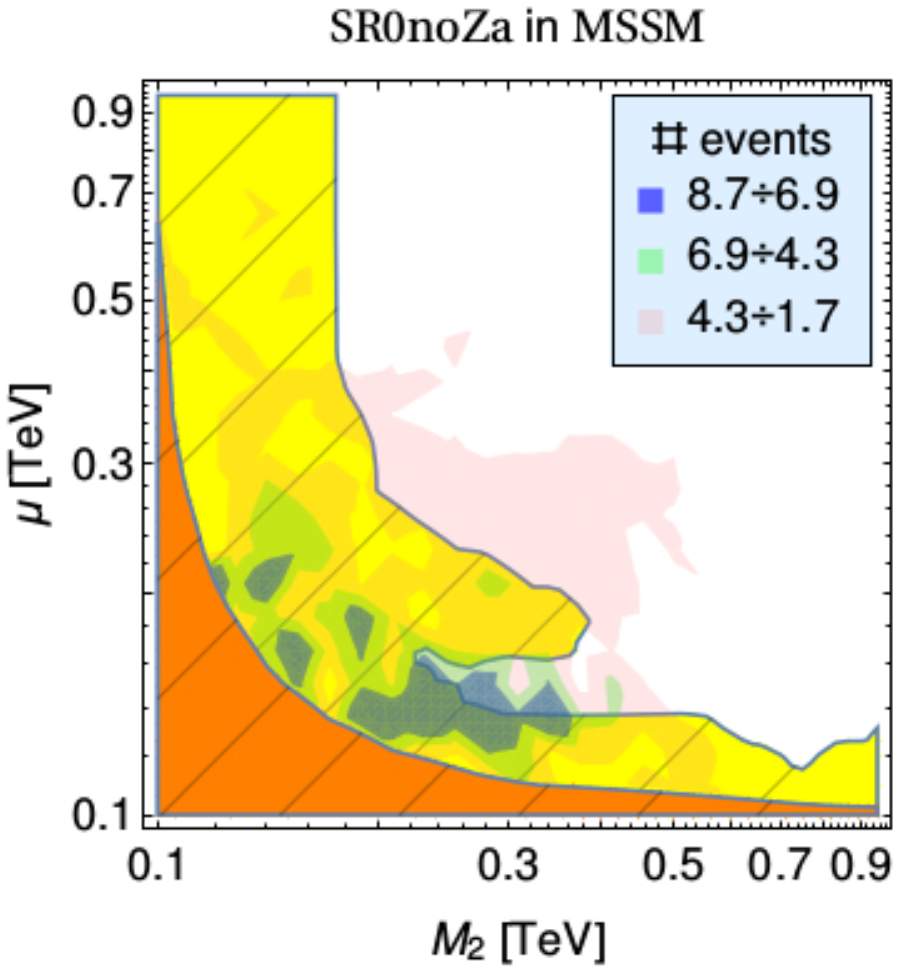}
\end{minipage}
\begin{minipage}[t]{0.3\textwidth}
\centering
\includegraphics[width=1.\columnwidth]{./figs/MSSM_63_13TeV-4lept-bin4.pdf}
\end{minipage}
\begin{minipage}[t]{0.3\textwidth}
\centering
\includegraphics[width=1.\columnwidth]{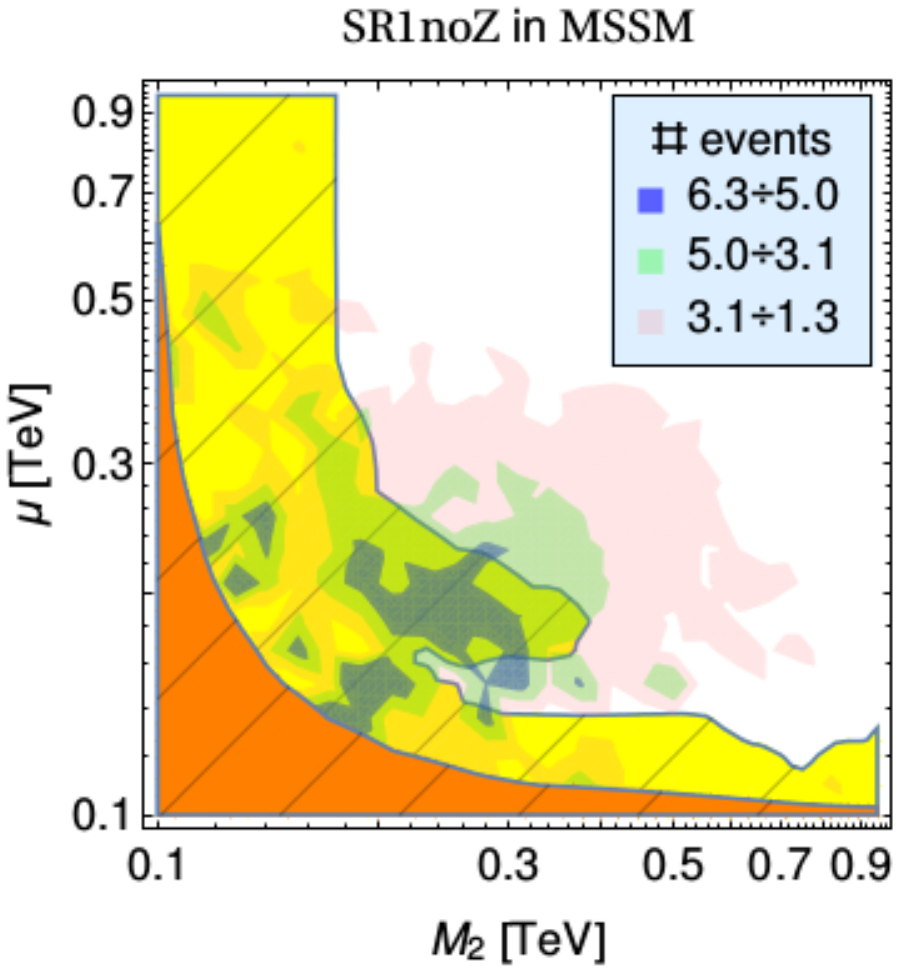}
\end{minipage}
\caption{{\bfseries Two-lepton \& four-lepton searches - MSSM - 13 TeV:} Number of events in the two-lepton and four-lepton search SRs in the $\{\mu,M_2\}$-plane for the MSSM. From left to right and top to bottom for the two-lepton + MET:  SRm$_{T2,90}$, SRm$_{T2,110}$, SRm$_{T2,150}$, SRWWa, SRWWb, SRWWc and SRZjets, as labelled. For the four-lepton case from left to right and top to bottom following the two-lepton SRs: SR0Z, SR1Z, SR0noZa, SR0noZb and SR1noZ. The blue region denotes the number of signal events in the SR in between the maximum and its 80\%. The green (pink) regions indicate the number of signal events in between the 50\% (20\%) and 80\% (50\%) of its maximum. In yellow we show the region excluded by Run~1, while in orange we denote the excluded region for charginos from LEP.}
\label{fig:2lepton_mssm13tev_bins}
\end{figure}

Figure~\ref{fig:2lepton_mssm13tev_bins} shows the sensitivity of the two-lepton and four-lepton searches of the MSSM scenario. For the two-lepton search we have seven SRs (SRm$_{T2,90}$, SRm$_{T2,110}$, SRm$_{T2,150}$, SRWWa, SRWWb, SRWWc, SRZjets) that corresponds to the first seven panels of Fig.~\ref{fig:2lepton_mssm13tev_bins}. The most relevant SRs are SRWWa, SRm$_{T2,90}$ and SRWWb, in terms of expecting the largest number of signal events. In this scenario, SRWWa is by far the SR with the largest number of expected events, however its sensitivity covers a parameter space in the $\{\mu , M_2\}$-plane which is basically almost completely excluded by LHC Run 1 (if considering as guideline the blue and green shaded regions). Concerning the four-lepton searches the SRs are represented in the last five panels of Fig.~\ref{fig:2lepton_mssm13tev_bins} (SR0Z, SR1Z, SR0noZa, SR0noZb, SR1noZ). The two SRs with the largest number of predicted events are SR0noZb and SR0Z, however these are less sensitive with respect to the other multi-lepton searches described in our analysis.

\begin{figure}
\begin{minipage}[t]{0.24\textwidth}
\centering
\includegraphics[width=1.\columnwidth]{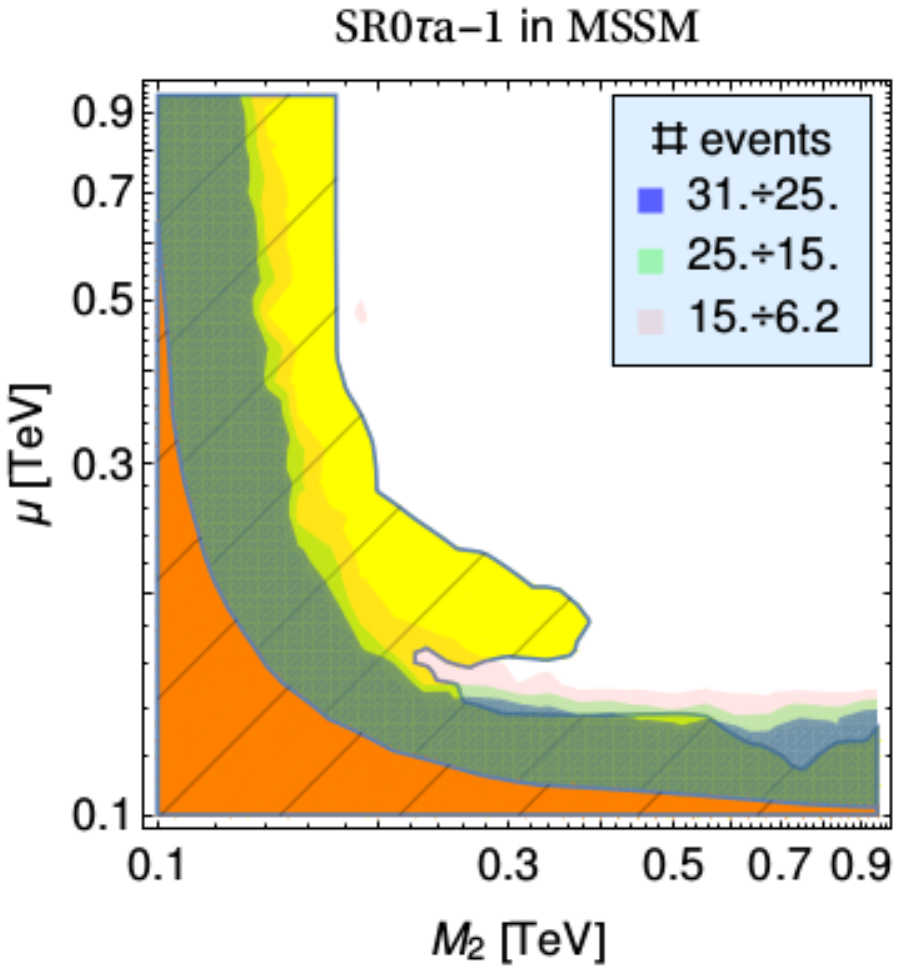}
\end{minipage}
\begin{minipage}[t]{0.24\textwidth}
\centering
\includegraphics[width=1.\columnwidth]{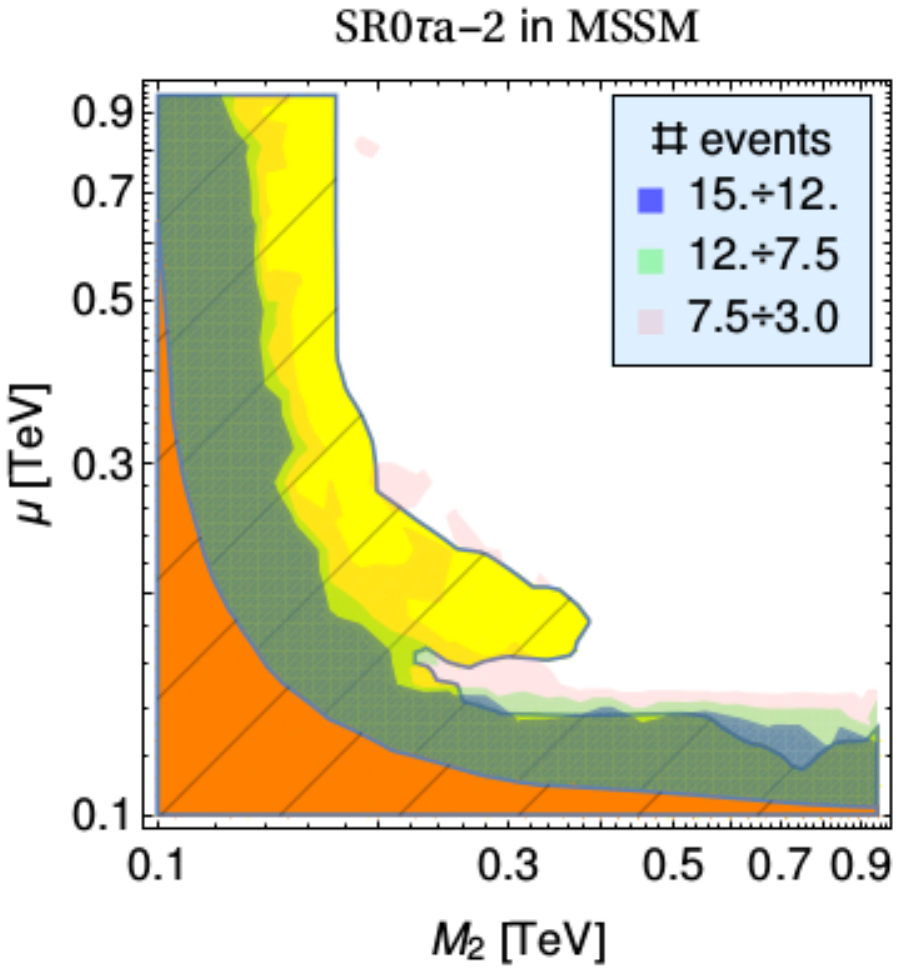}
\end{minipage}
\begin{minipage}[t]{0.24\textwidth}
\centering
\includegraphics[width=1.\columnwidth]{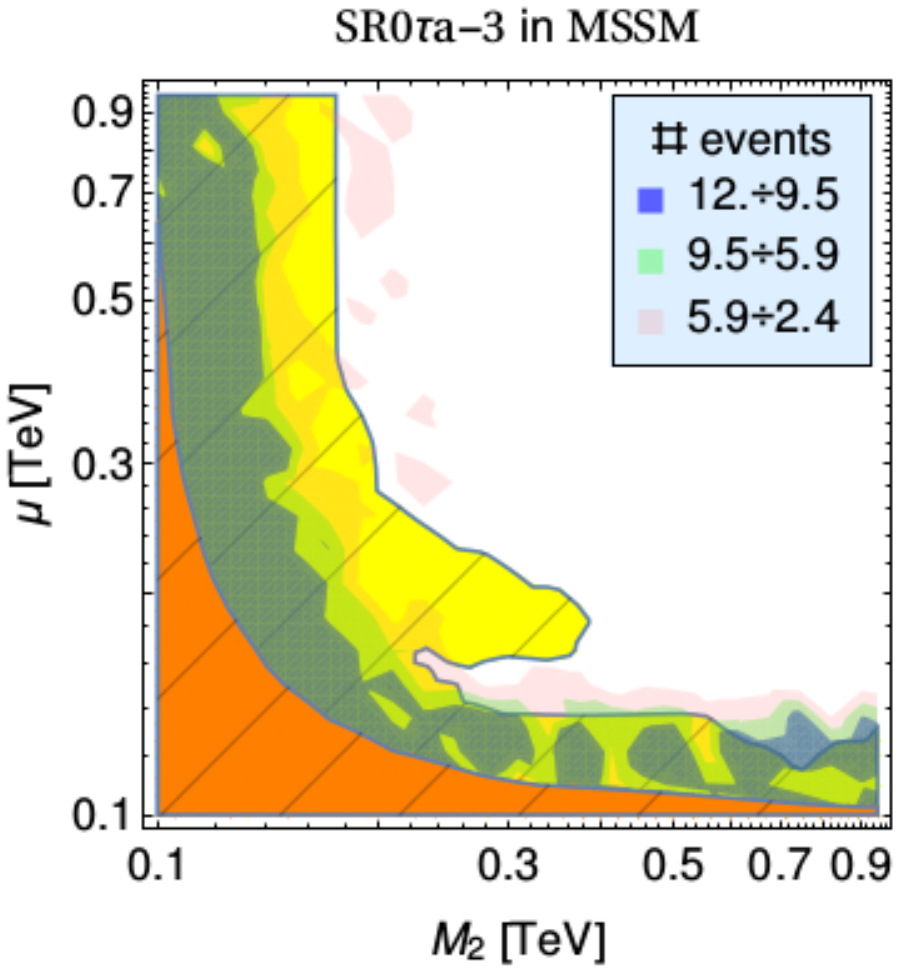}
\end{minipage}
\begin{minipage}[t]{0.24\textwidth}
\centering
\includegraphics[width=1.\columnwidth]{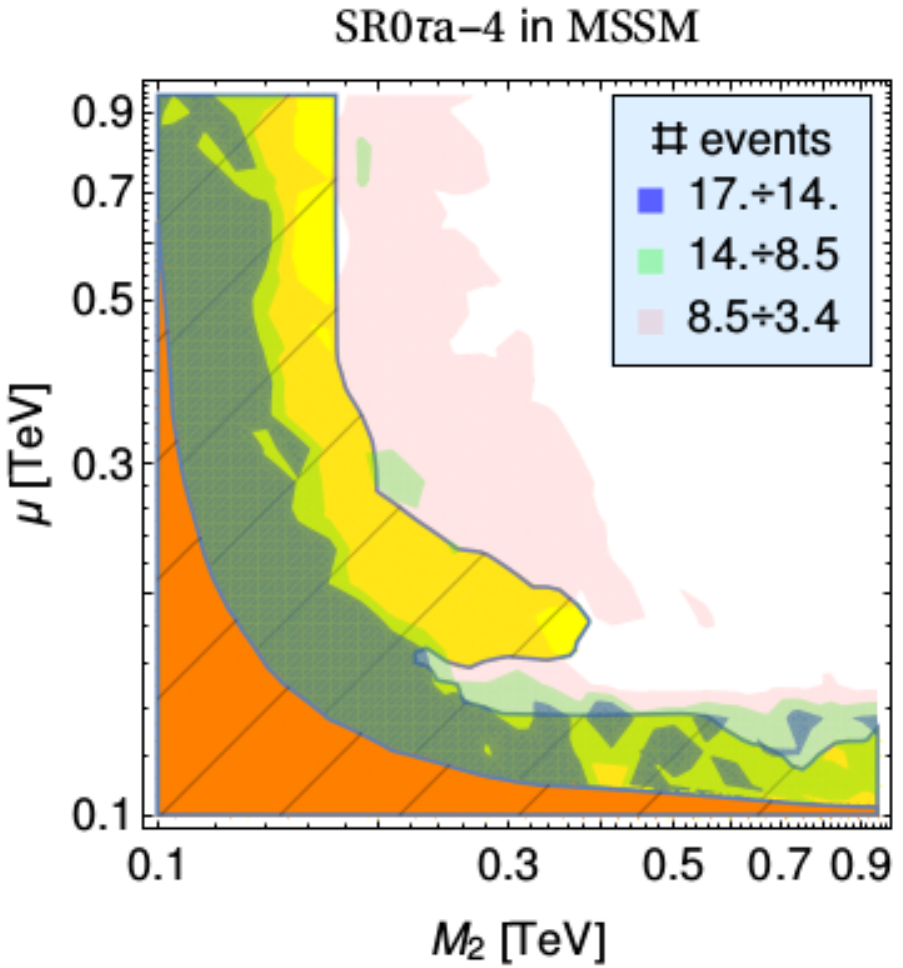}
\end{minipage}
\\
\begin{minipage}[t]{0.24\textwidth}
\centering
\includegraphics[width=1.\columnwidth]{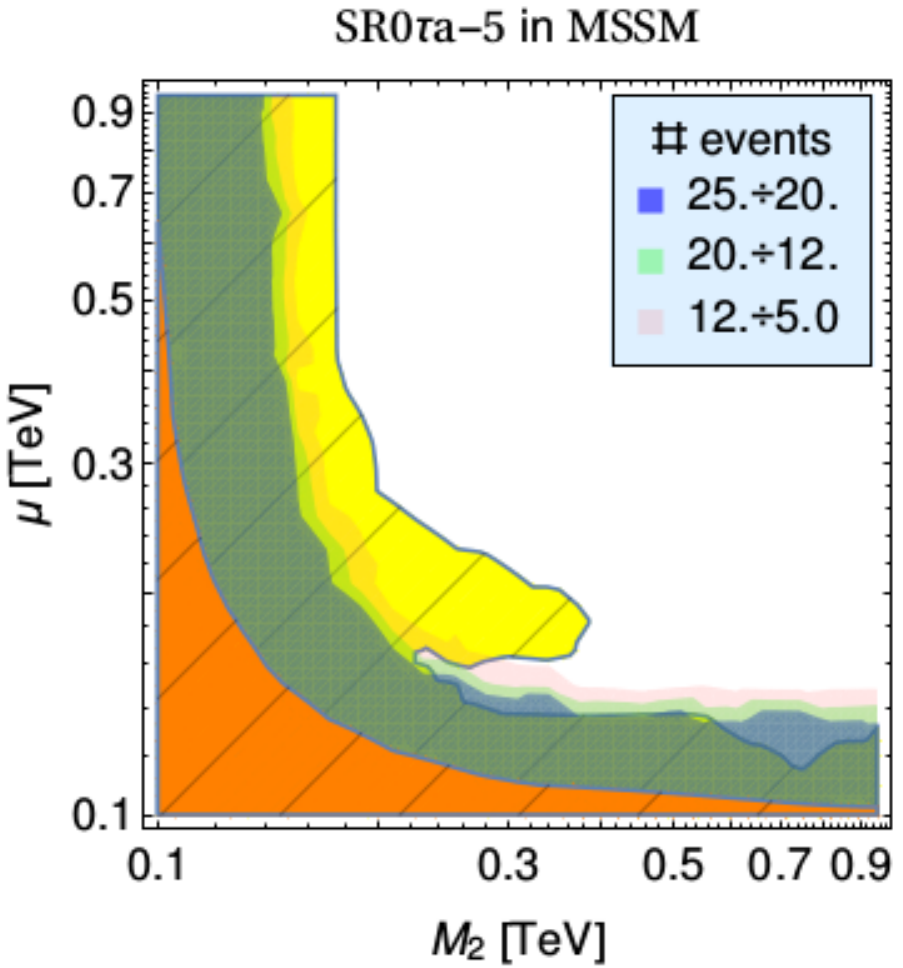}
\end{minipage}
\begin{minipage}[t]{0.24\textwidth}
\centering
\includegraphics[width=1.\columnwidth]{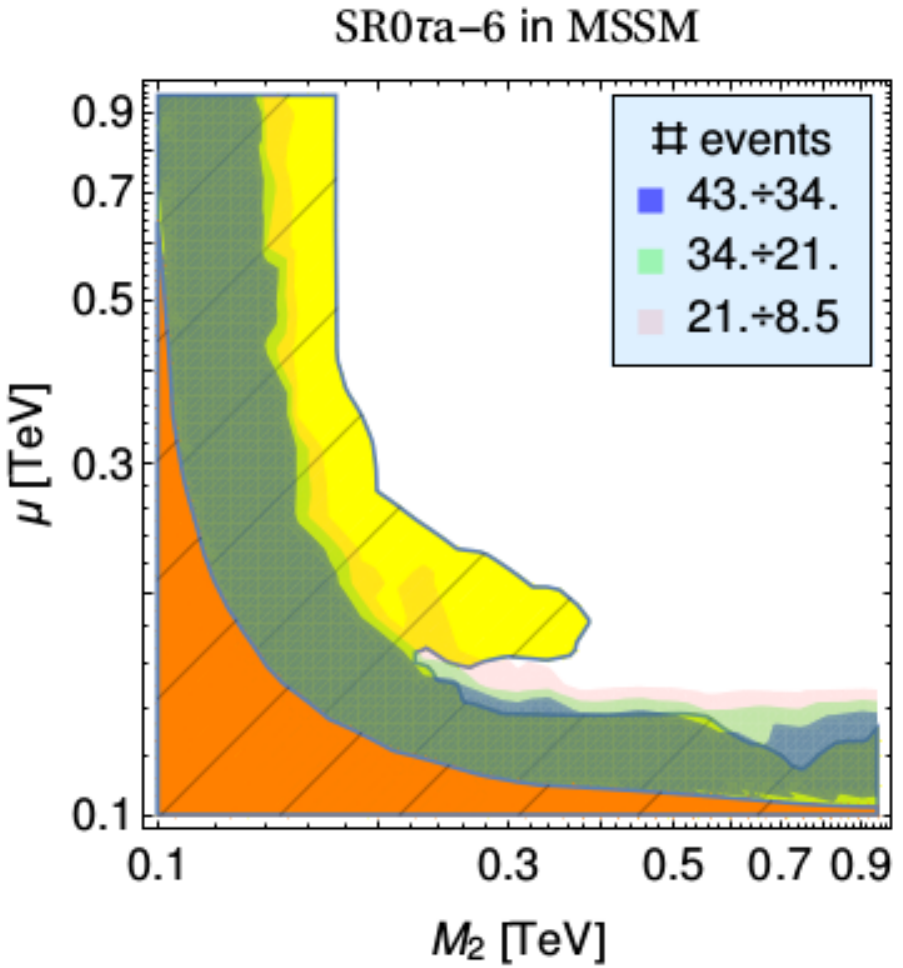}
\end{minipage}
\begin{minipage}[t]{0.24\textwidth}
\centering
\includegraphics[width=1.\columnwidth]{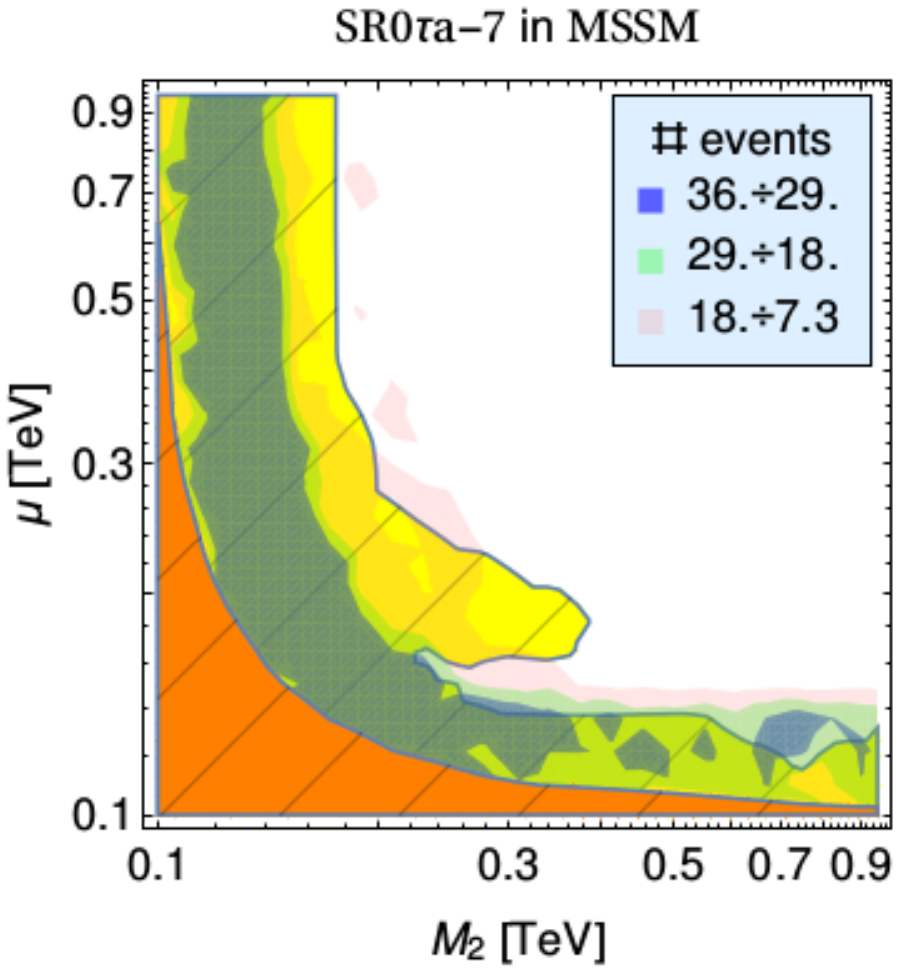}
\end{minipage}
\begin{minipage}[t]{0.24\textwidth}
\centering
\includegraphics[width=1.\columnwidth]{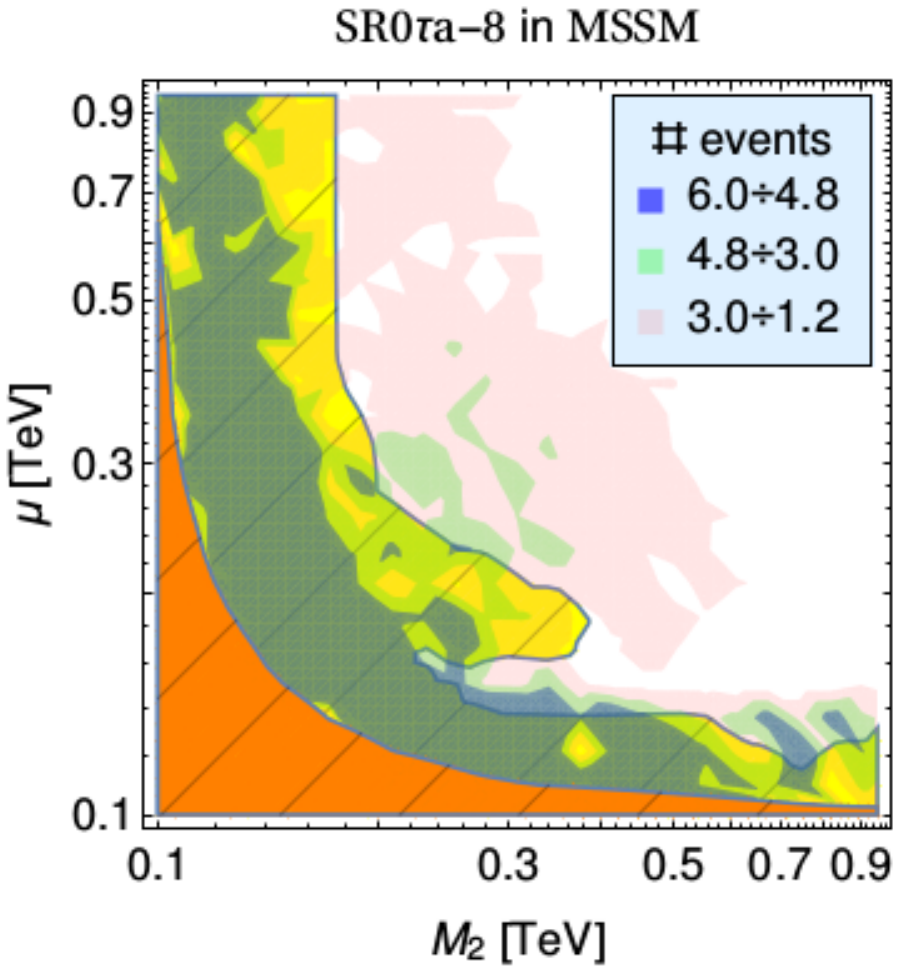}
\end{minipage}
\\
\begin{minipage}[t]{0.24\textwidth}
\centering
\includegraphics[width=1.\columnwidth]{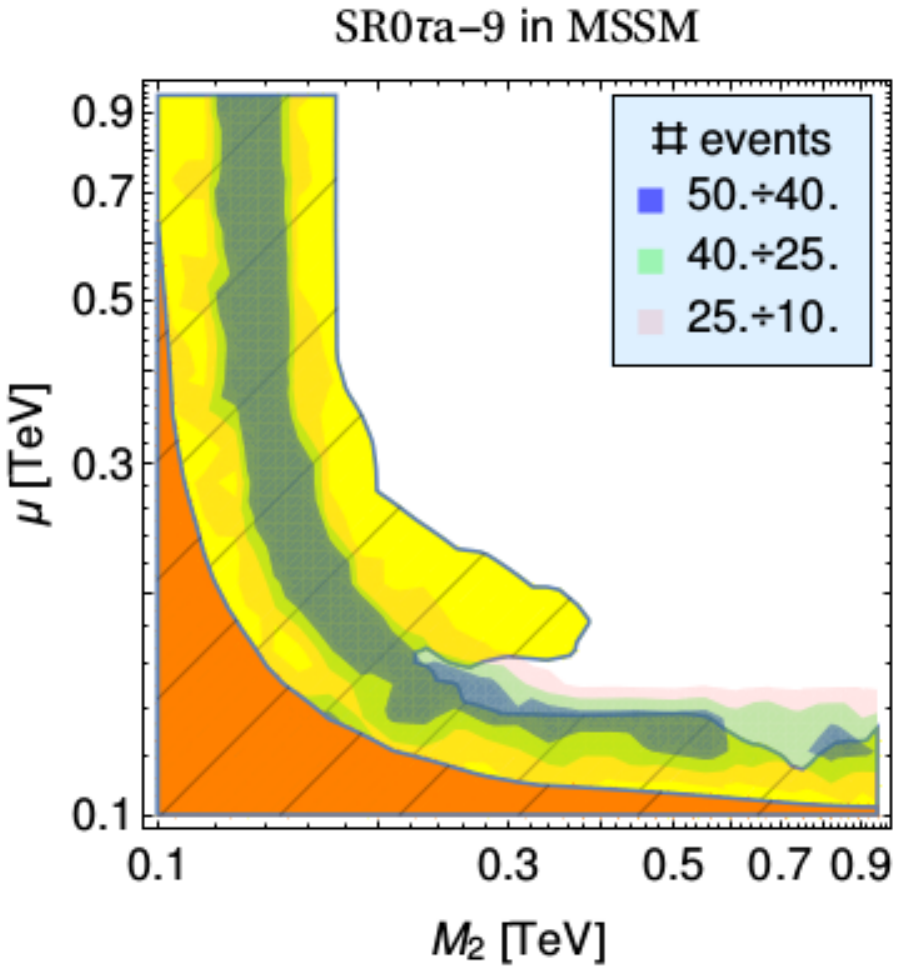}
\end{minipage}
\begin{minipage}[t]{0.24\textwidth}
\centering
\includegraphics[width=1.\columnwidth]{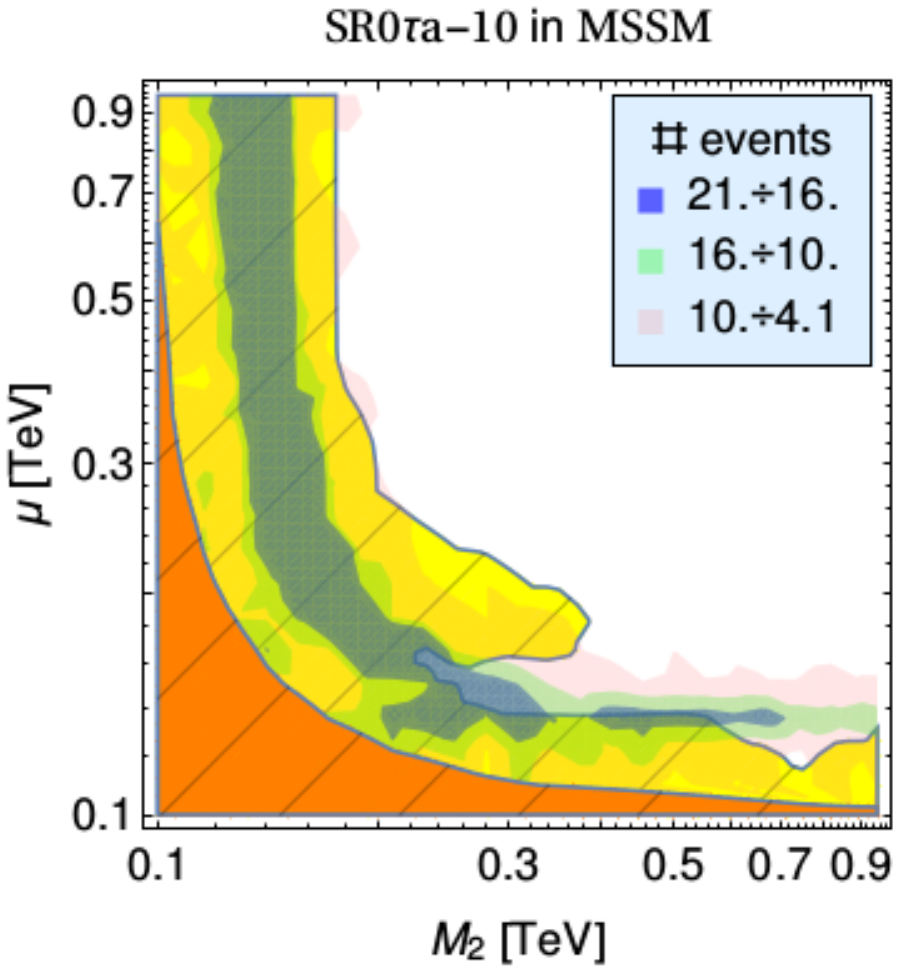}
\end{minipage}
\begin{minipage}[t]{0.24\textwidth}
\centering
\includegraphics[width=1.\columnwidth]{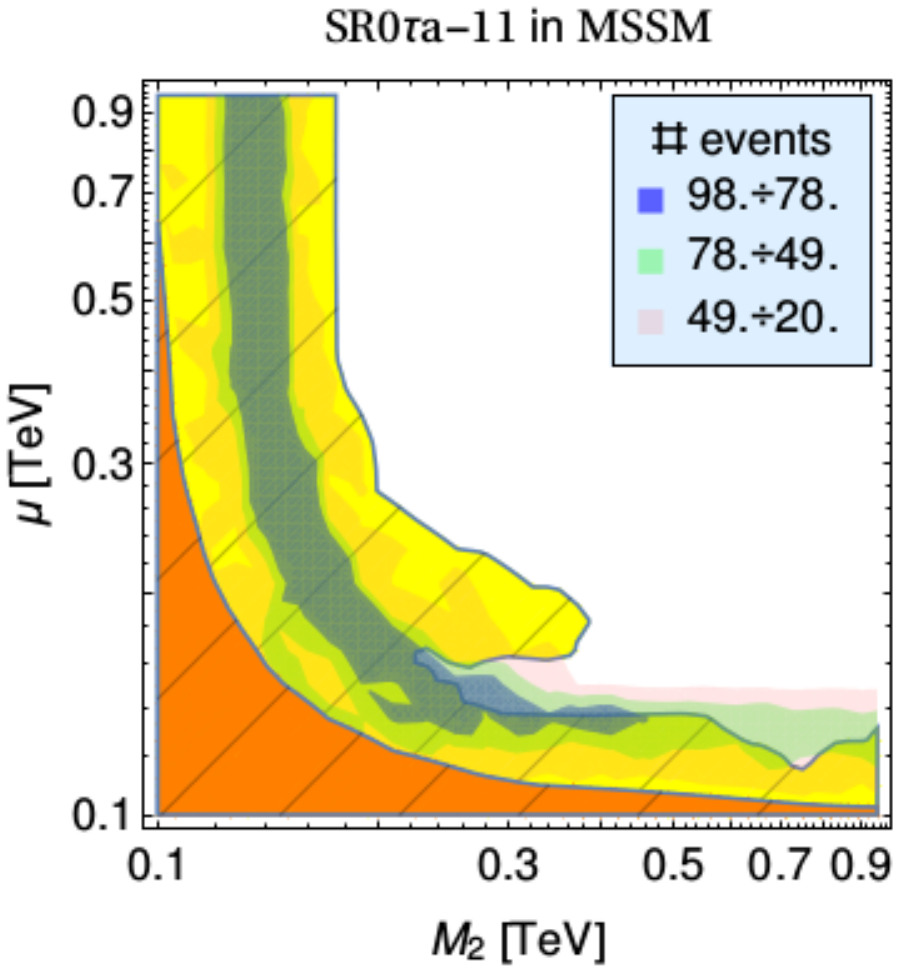}
\end{minipage}
\begin{minipage}[t]{0.24\textwidth}
\centering
\includegraphics[width=1.\columnwidth]{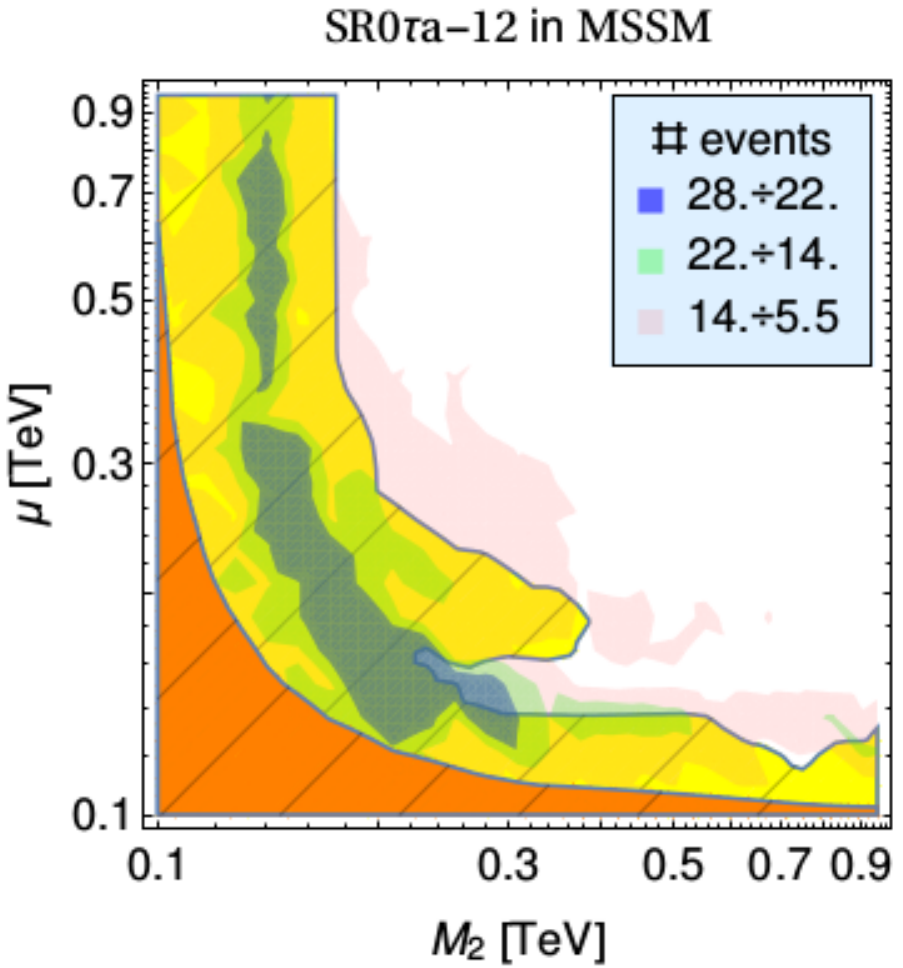}
\end{minipage}
\\
\begin{minipage}[t]{0.24\textwidth}
\centering
\includegraphics[width=1.\columnwidth]{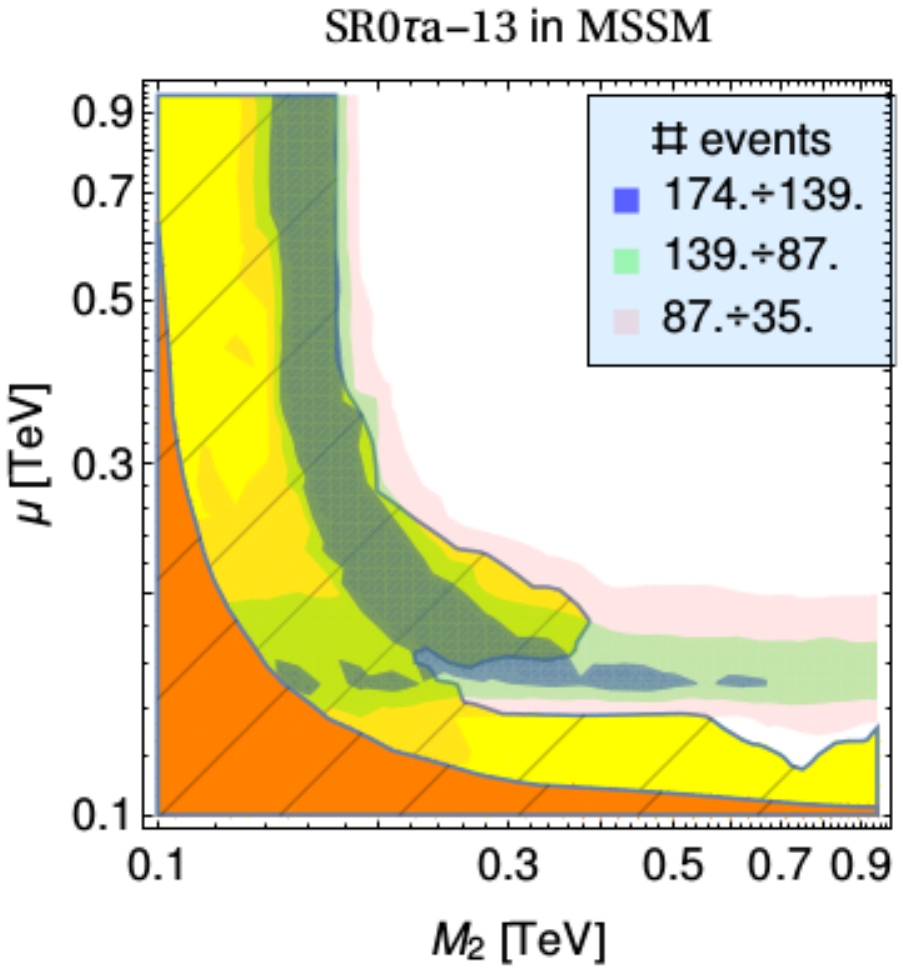}
\end{minipage}
\begin{minipage}[t]{0.24\textwidth}
\centering
\includegraphics[width=1.\columnwidth]{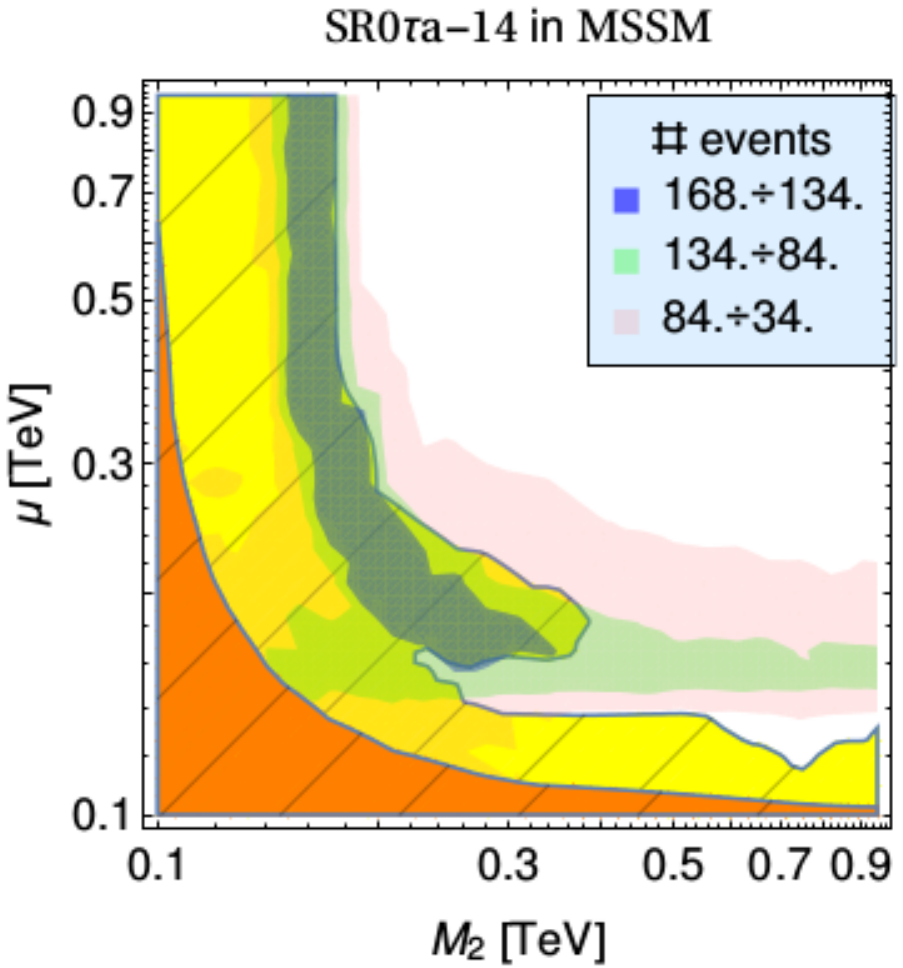}
\end{minipage}
\begin{minipage}[t]{0.24\textwidth}
\centering
\includegraphics[width=1.\columnwidth]{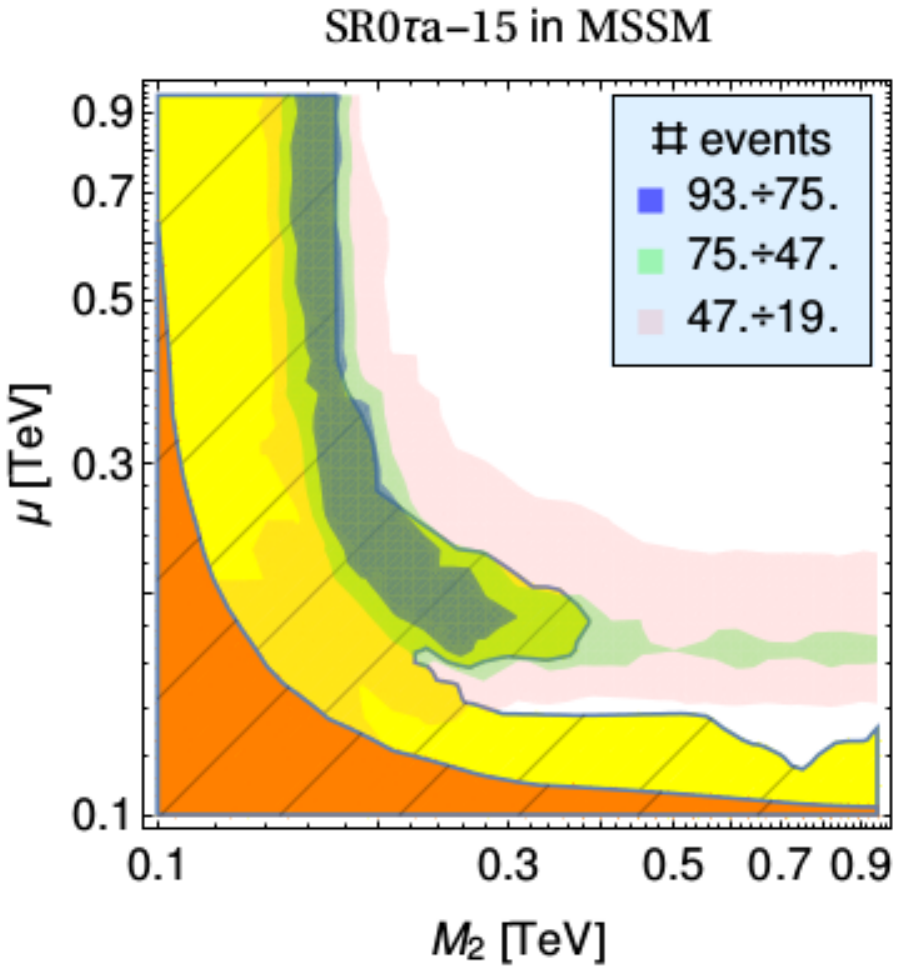}
\end{minipage}
\begin{minipage}[t]{0.24\textwidth}
\centering
\includegraphics[width=1.\columnwidth]{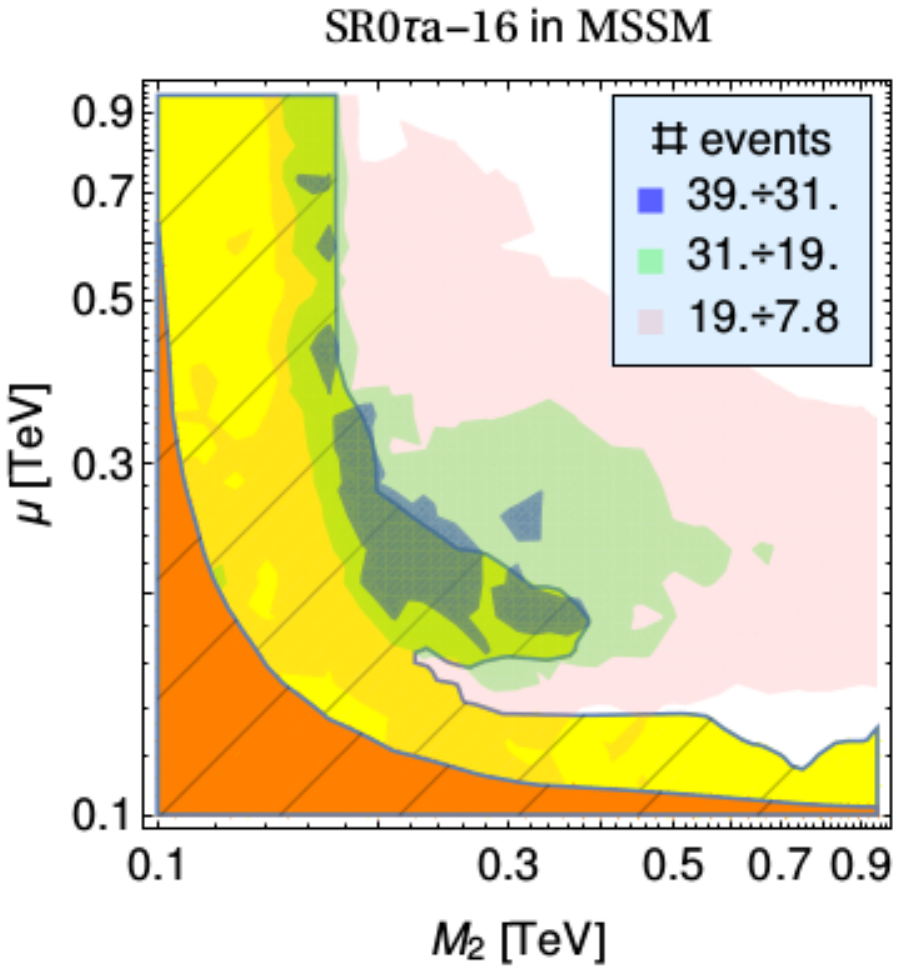}
\end{minipage}
\\
\begin{minipage}[t]{0.24\textwidth}
\centering
\includegraphics[width=1.\columnwidth]{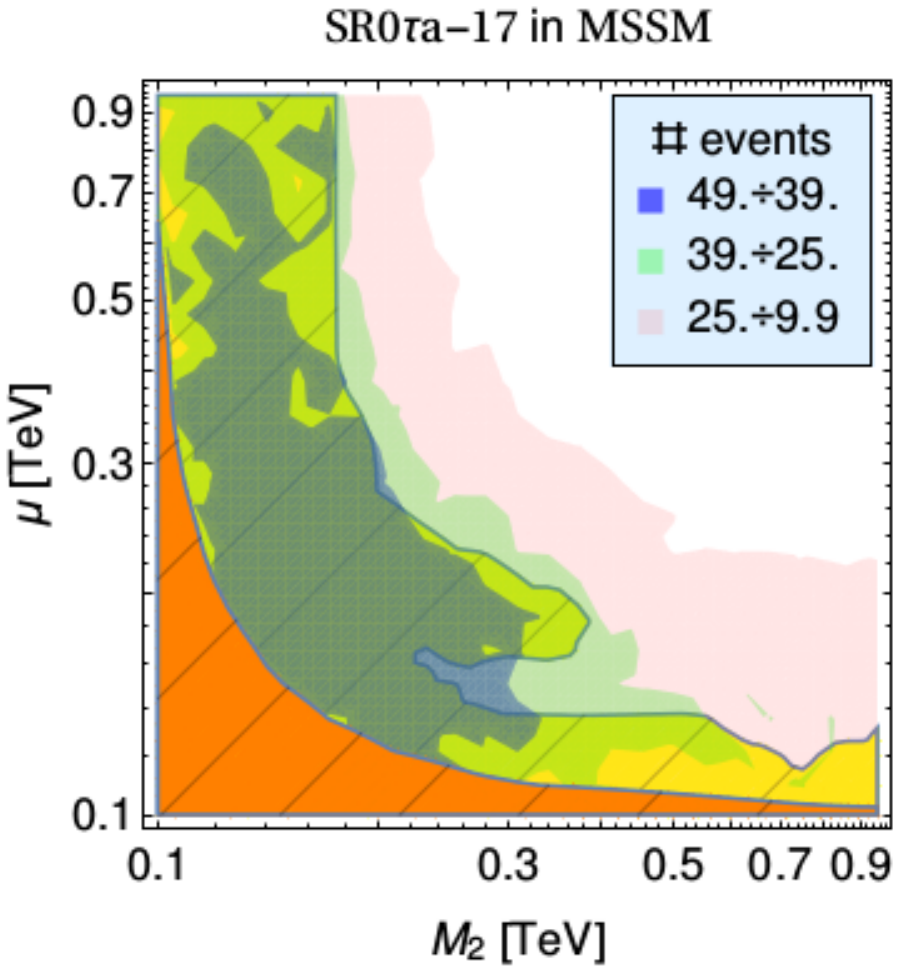}
\end{minipage}
\begin{minipage}[t]{0.24\textwidth}
\centering
\includegraphics[width=1.\columnwidth]{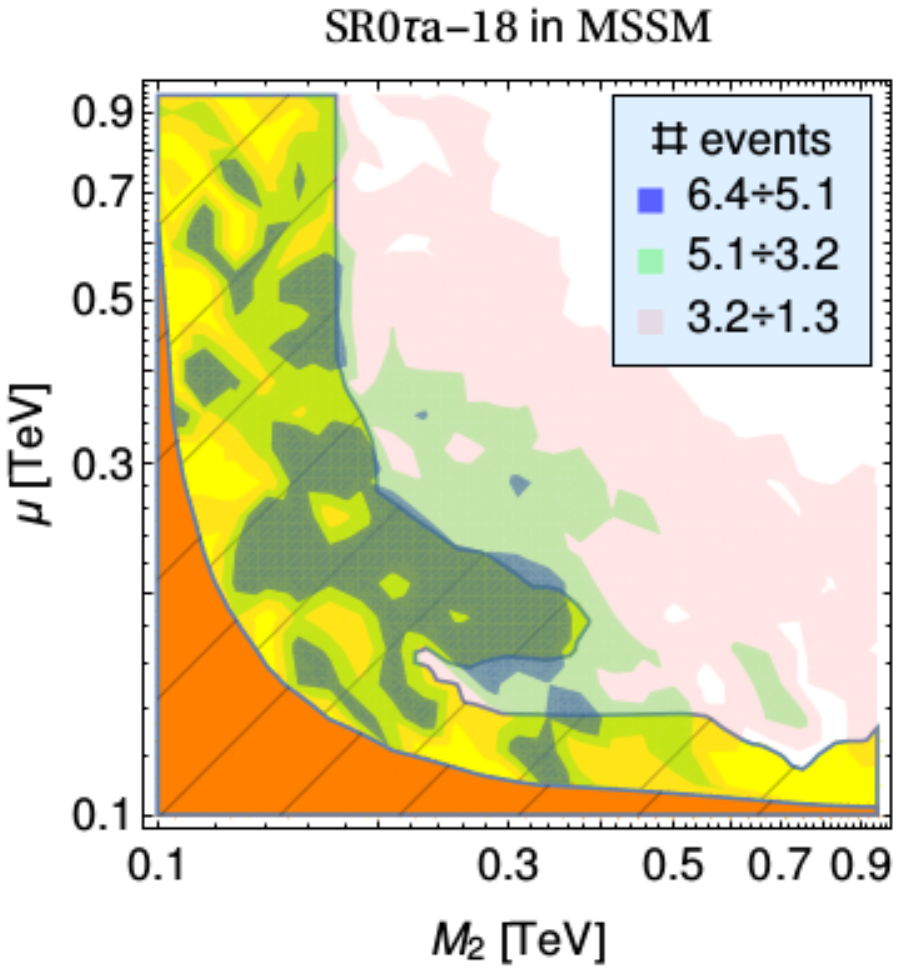}
\end{minipage}
\begin{minipage}[t]{0.24\textwidth}
\centering
\includegraphics[width=1.\columnwidth]{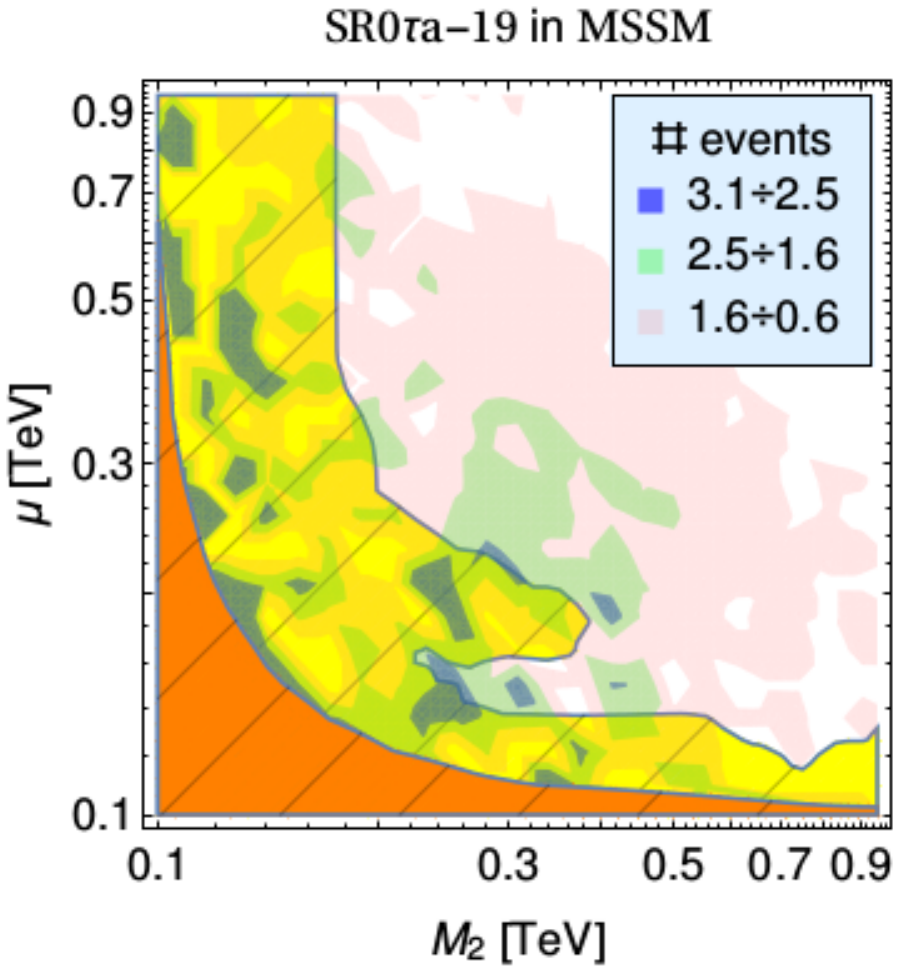}
\end{minipage}
\begin{minipage}[t]{0.24\textwidth}
\centering
\includegraphics[width=1.\columnwidth]{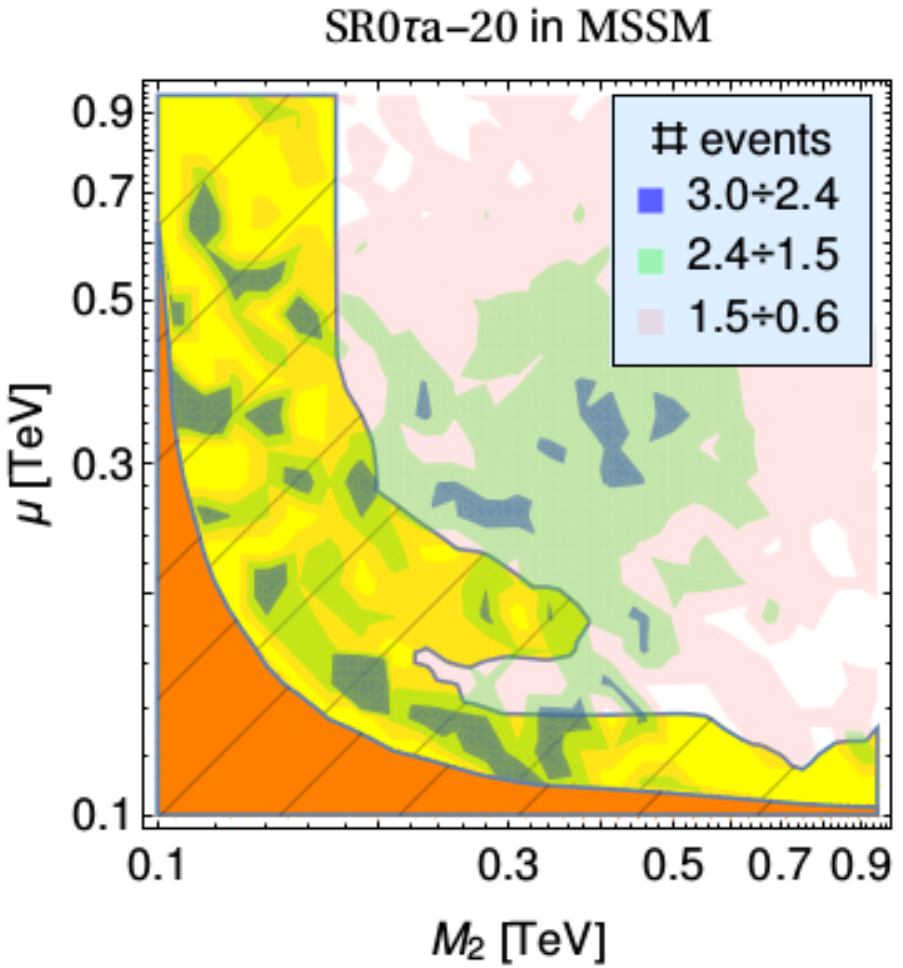}
\end{minipage}
\caption{{\bfseries Three-lepton search - MSSM - 13 TeV:} Number of events in the SR0$\tau$ of the three-lepton search in the $\{\mu,M_2\}$-plane for the MSSM. From left to right and top to bottom we show the 20 bins as labelled. The blue region denotes the number of signal events in the SR in between the maximum and its 80\%. The green (pink) regions indicate the number of signal events in between the 50\% (20\%) and 80\% (50\%) of its maximum. In yellow we show the region excluded by Run~1, while in orange we denote the excluded region for charginos from LEP.}
\label{fig:3lepton_mssm_13tev_bins}
\end{figure}

The bins of SR0$\tau$a of the three-lepton search for the MSSM scenario are represented in Fig.~\ref{fig:3lepton_mssm_13tev_bins}. The bins 13 and 14 have the largest predicted number of events and show as well a great capability in exploring the parameter space. It is not granted that the bins with the largest number of predicted events are also the most sensitive. For example, bin 11 has a very large number of signal events, $\mathcal{O}(10^2)$, however the exclusion/exploration potential is very limited and basically coincide with the region already excluded by Run 1. On the other hand we notice that bin 16 can perform well in exploring the model parameter space.

It should be noted that there are several bins with a negligible number of predicted events, which is an additional relevant information that can be used in EWinos searches. 

From Figs.~\ref{fig:2lepton_mssm13tev_bins} and~\ref{fig:3lepton_mssm_13tev_bins} we can infer that the two-lepton and the three-lepton searches are very important for the MSSM scenario in order to find EWino signals at the LHC, while the four-lepton search has a sensitivity which is reduced with respect to the other two searches.

\subsection*{TMSSM\_1 ($\lambda=0.65$, $\mu_\Sigma = 300$ GeV)}

\begin{figure}[t]
\begin{minipage}[t]{0.32\textwidth}
\centering
\includegraphics[width=1.\columnwidth]{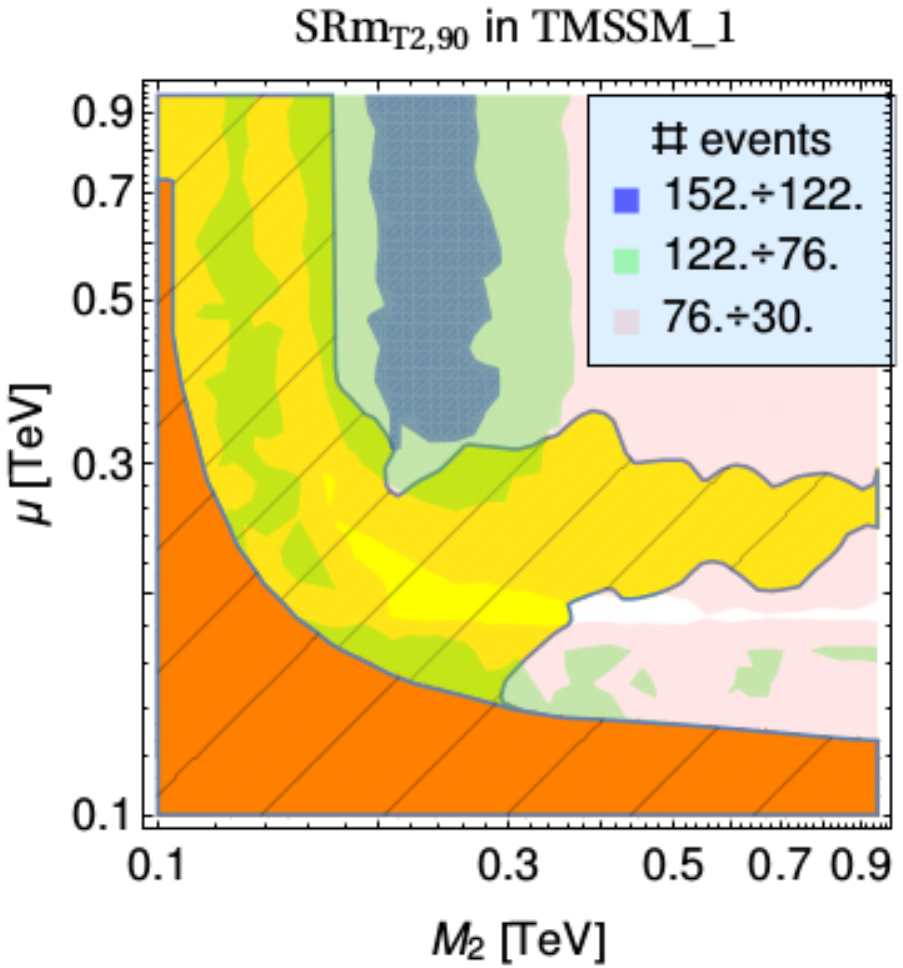}
\end{minipage}
\begin{minipage}[t]{0.32\textwidth}
\centering
\includegraphics[width=1.\columnwidth]{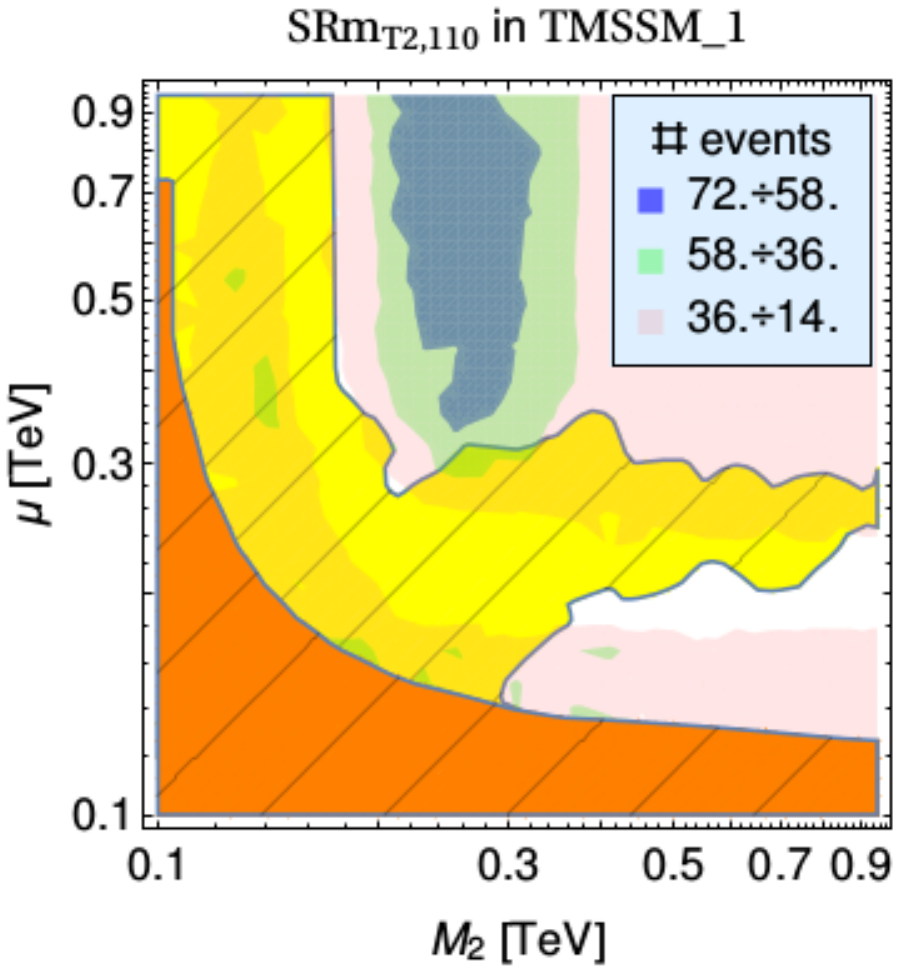}
\end{minipage}
\begin{minipage}[t]{0.32\textwidth}
\centering
\includegraphics[width=1.\columnwidth]{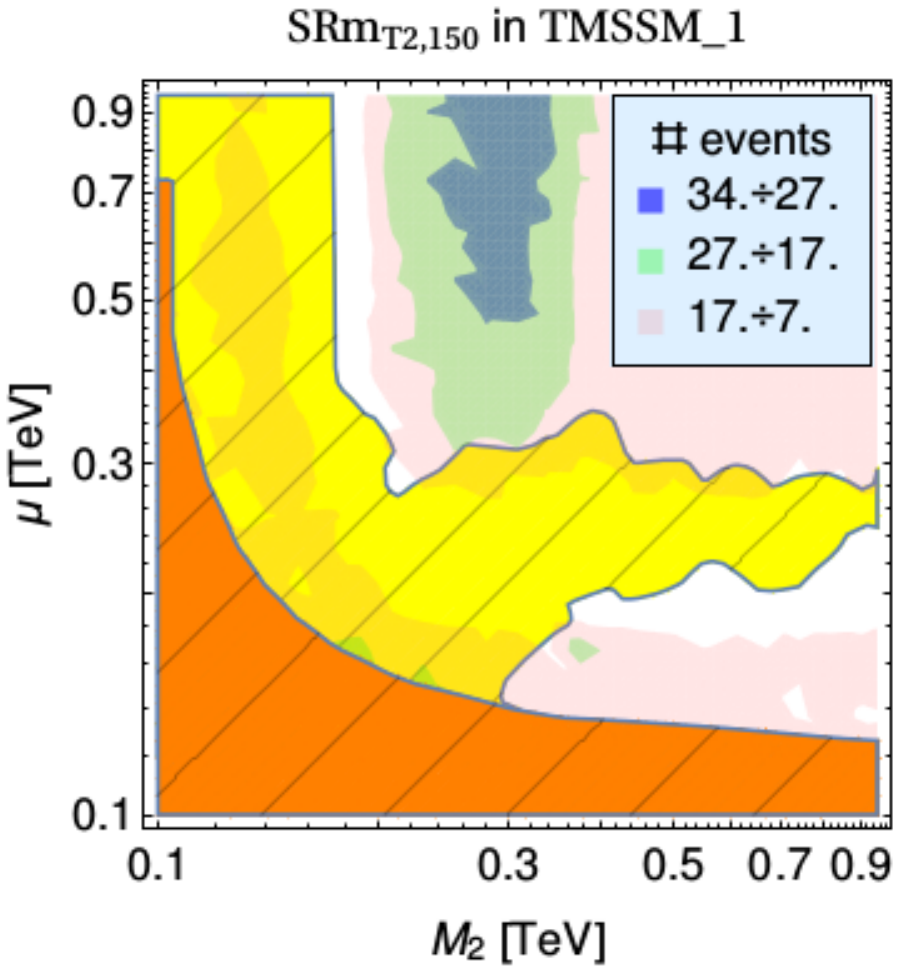}
\end{minipage}
\\
\begin{minipage}[t]{0.32\textwidth}
\centering
\includegraphics[width=1.\columnwidth]{./figs/TMSSM65_300_13TeV-2lept-bin4.pdf}
\end{minipage}
\begin{minipage}[t]{0.32\textwidth}
\centering
\includegraphics[width=1.\columnwidth]{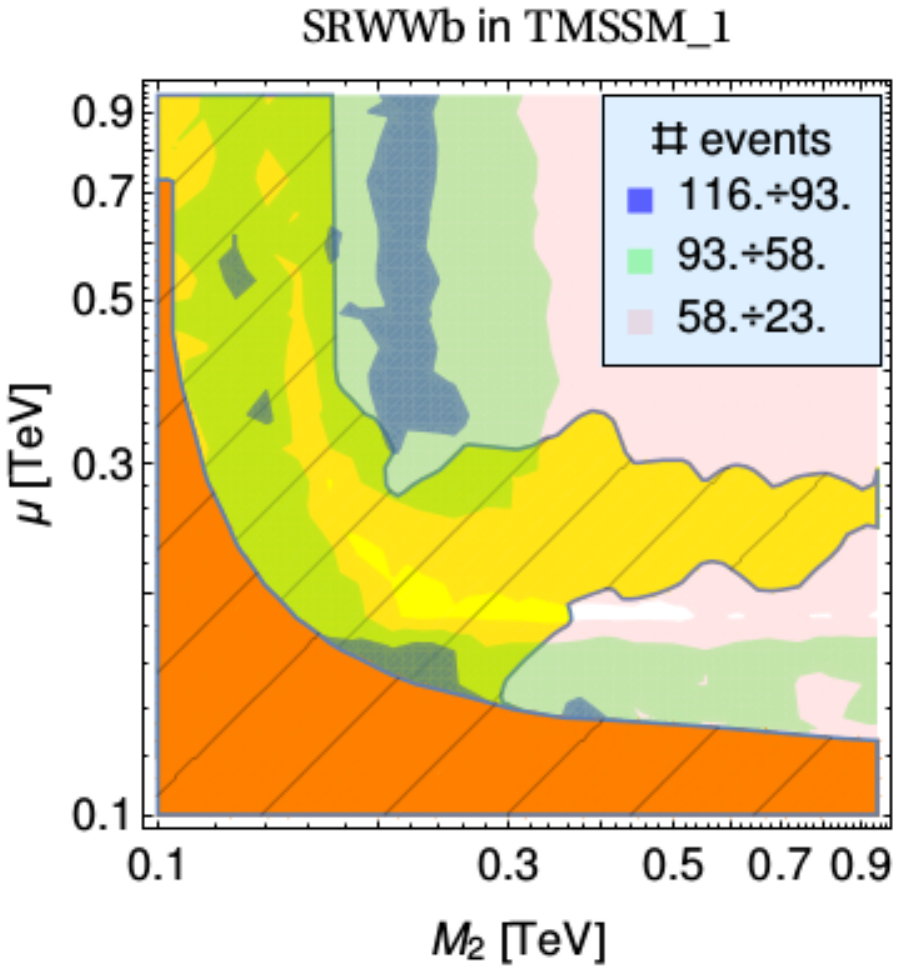}
\end{minipage}
\begin{minipage}[t]{0.32\textwidth}
\centering
\includegraphics[width=1.\columnwidth]{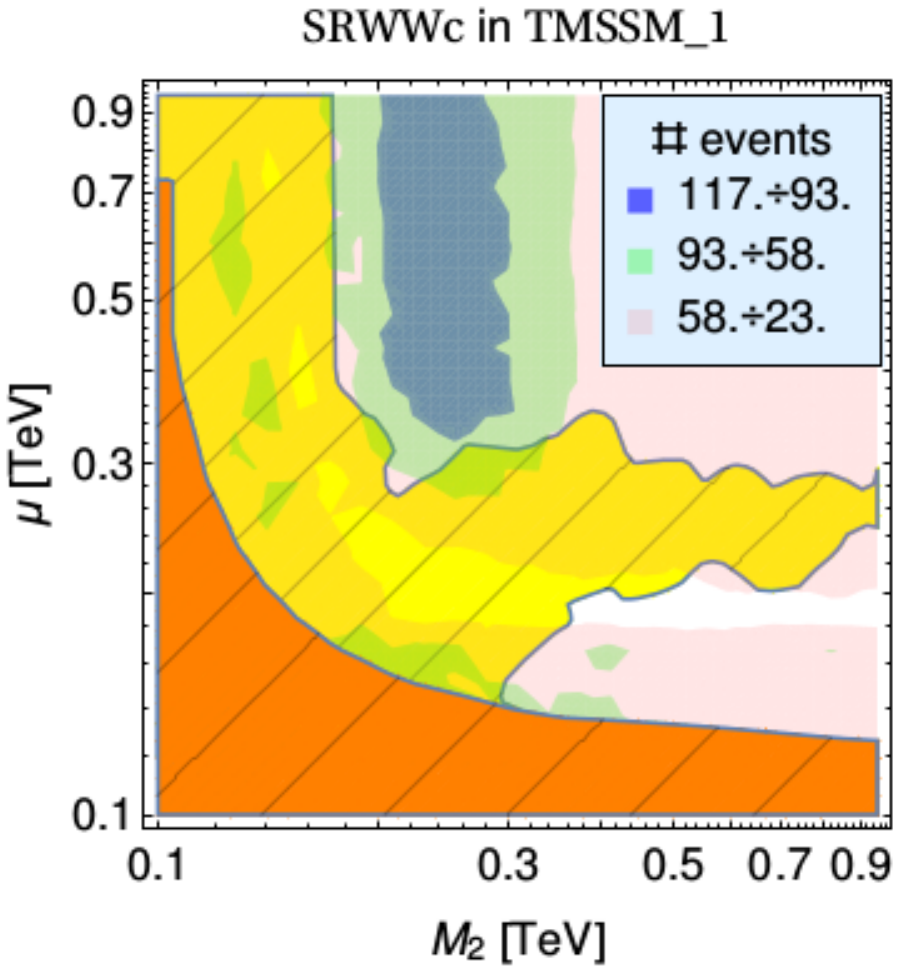}
\end{minipage}
\\
\begin{minipage}[t]{0.32\textwidth}
\centering
\includegraphics[width=1.\columnwidth]{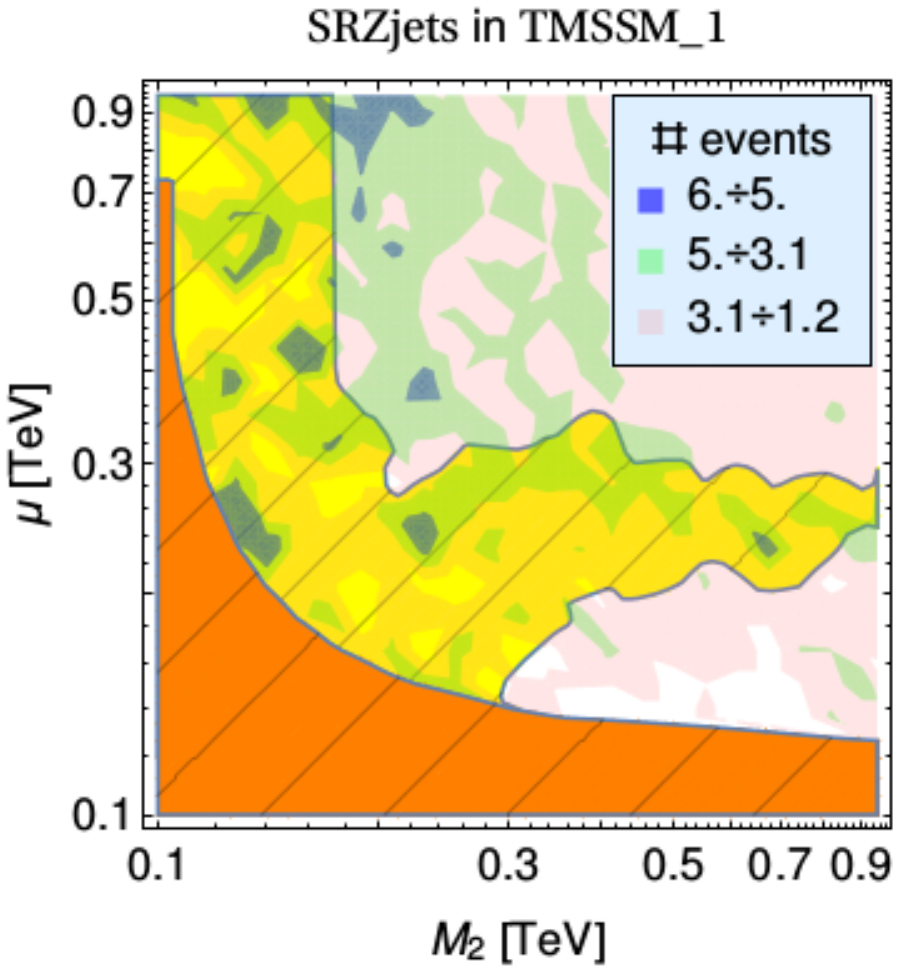}
\end{minipage}
\begin{minipage}[t]{0.32\textwidth}
\centering
\includegraphics[width=1.\columnwidth]{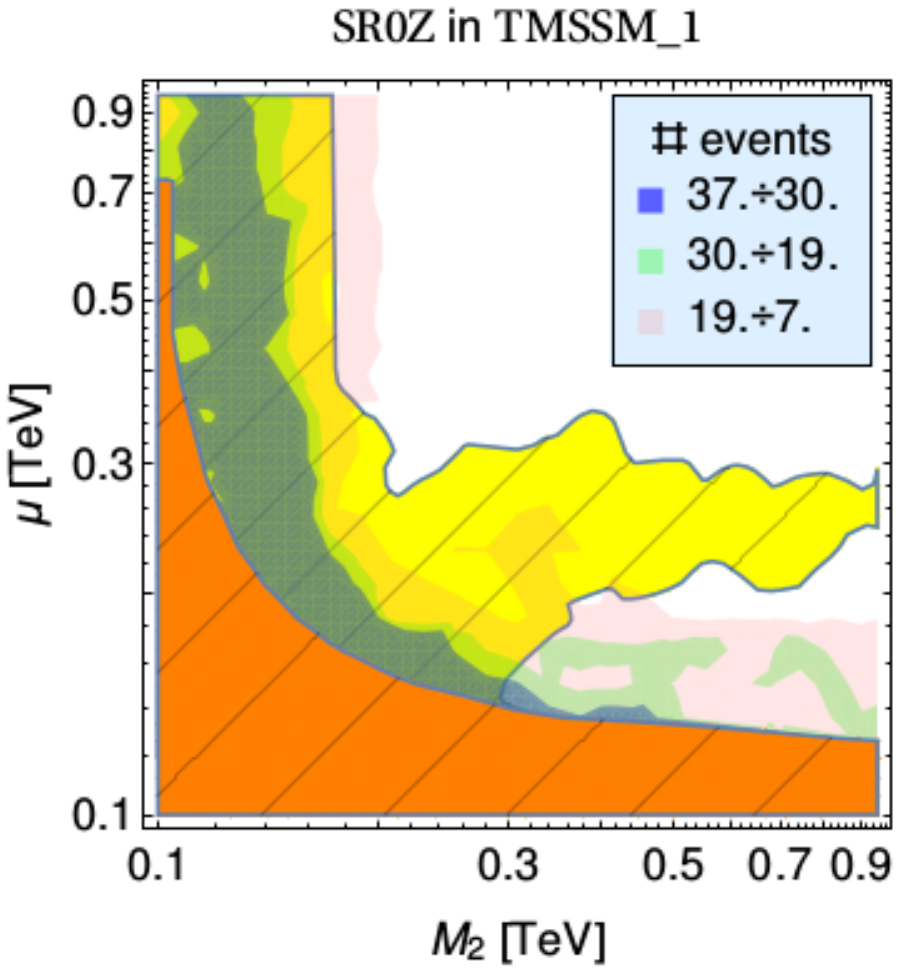}
\end{minipage}
\begin{minipage}[t]{0.32\textwidth}
\centering
\includegraphics[width=1.\columnwidth]{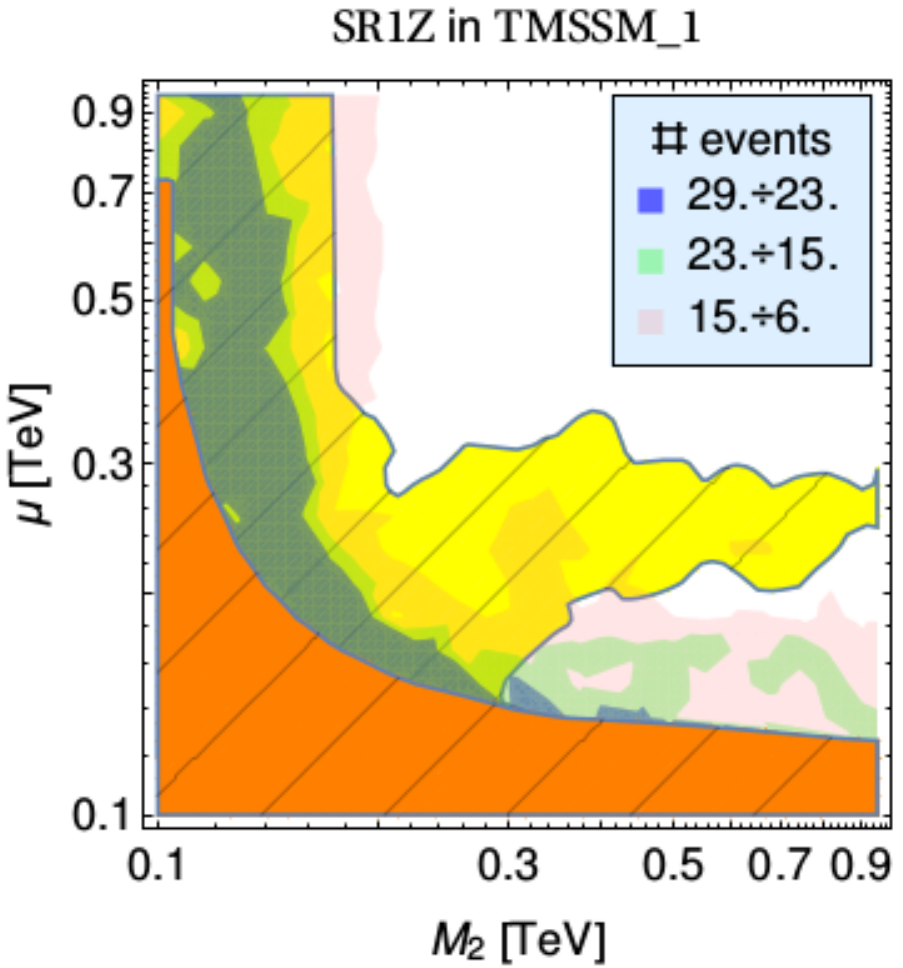}
\end{minipage}
\\
\begin{minipage}[t]{0.32\textwidth}
\centering
\includegraphics[width=1.\columnwidth]{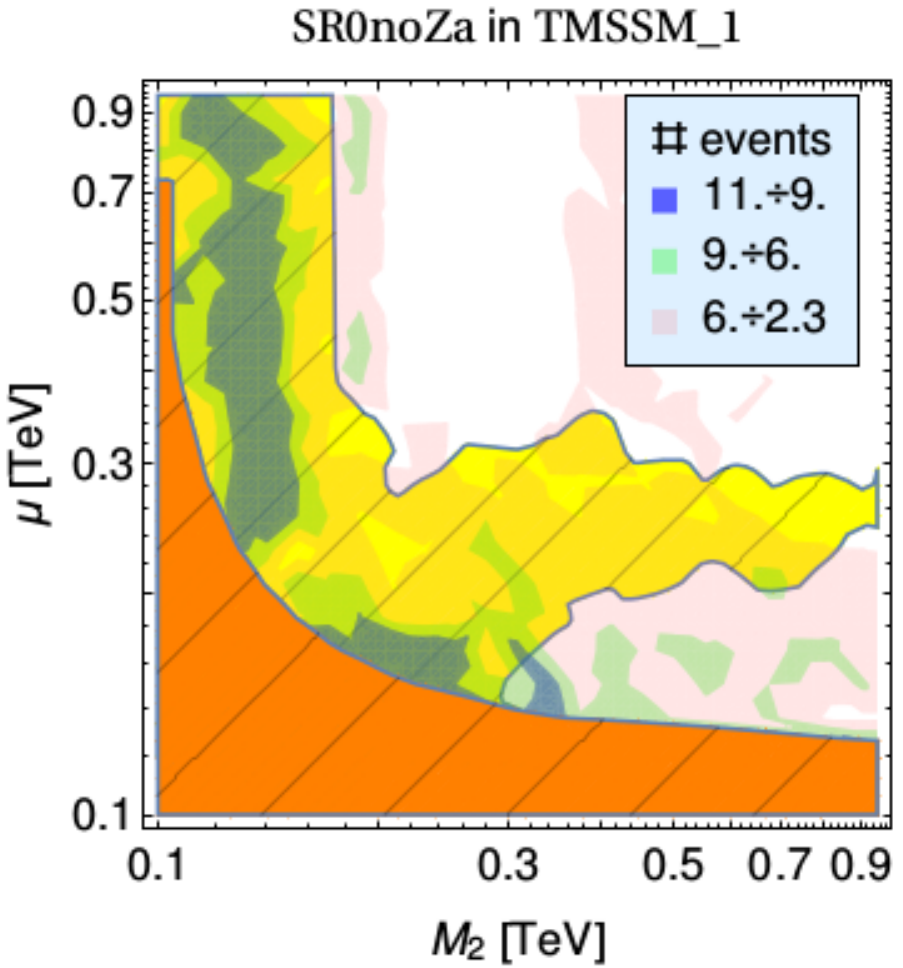}
\end{minipage}
\begin{minipage}[t]{0.32\textwidth}
\centering
\includegraphics[width=1.\columnwidth]{./figs/TMSSM65_300_13TeV-4lept-bin4.pdf}
\end{minipage}
\begin{minipage}[t]{0.32\textwidth}
\centering
\includegraphics[width=1.\columnwidth]{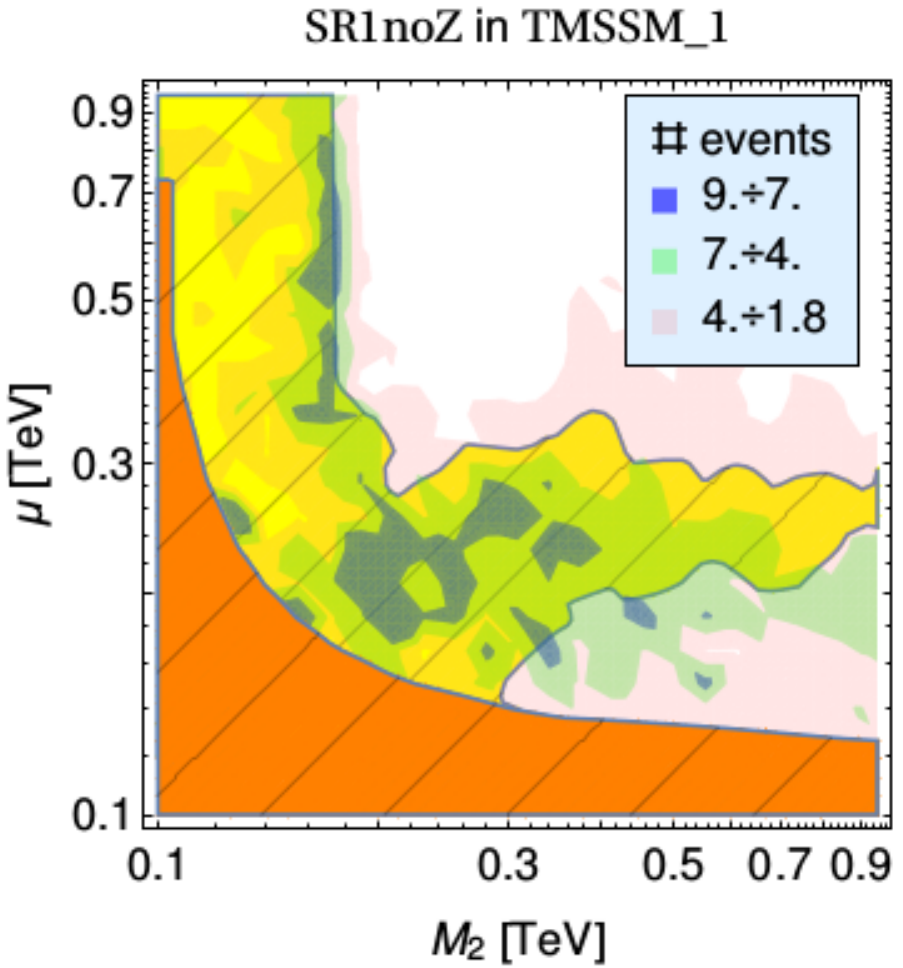}
\end{minipage}
\caption{{\bfseries Two-lepton \& four-lepton searches - TMSSM\_1 - 13 TeV:} Same as Fig.~\ref{fig:2lepton_mssm13tev_bins} for the TMSSM\_1 case ($\lambda = 0.65$ and $\mu_{\Sigma}=300$ GeV).}
\label{fig:2lepton_tmssm65300_13tev_bins}
\end{figure}

Figure~\ref{fig:2lepton_tmssm65300_13tev_bins} shows the sensitivity of the two-lepton and the four-lepton searches of the TMSSM\_1 scenario previously defined in Sec.~\ref{sec:analysis}. As it was the case in the MSSM scenario, the SRs expecting the largest number of signal events are SRWWa, SRm$_{T2,90}$ and SRWWb. However, in this scenario, SRWWc presents a comparable number of events to SRWWa and SRWWb (differently to what happens in the MSSM case).  In SRWWa, even though the blue shaded region mostly overlaps with the already existing exclusion limit of Run 1, the green region shows the potential to explore low values of $\mu$ up to 200 GeV irrespective of $M_2$. A complete new region of the parameter space shows up in SRm$_{T2,90}$ and SRWWc, indicating that these two SRs have the capabilities to explore low values of $M_2$ up to 300 GeV irrespective of the value of $\mu$ (blue region). In addition the number of predicted events is very different with respect to the MSSM. For the four-lepton case the bins with the largest number of predicted events are SR0noZb and SR0Z. Despite the fact that this search possesses a reduced sensitivity compared to the other multi-lepton searches, we notice that it seems to have an enhanced potential for discovery for this TMSSM\_1 scenario with respect to the MSSM case. This is certainly due to the enriched EWinos mass spectrum augmenting the number of combinations leading to four-lepton + MET signatures.

\begin{figure}
\centering
\includegraphics[scale=0.38]{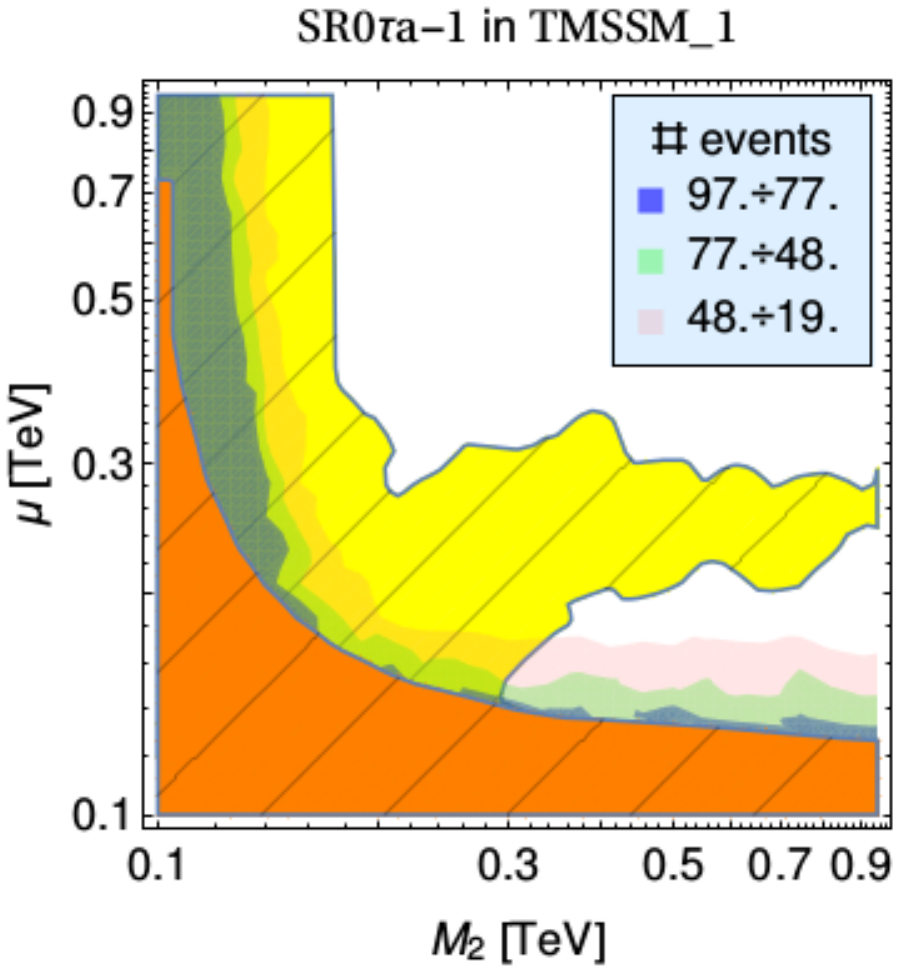}
\hspace*{0.1cm}
\includegraphics[scale=0.38]{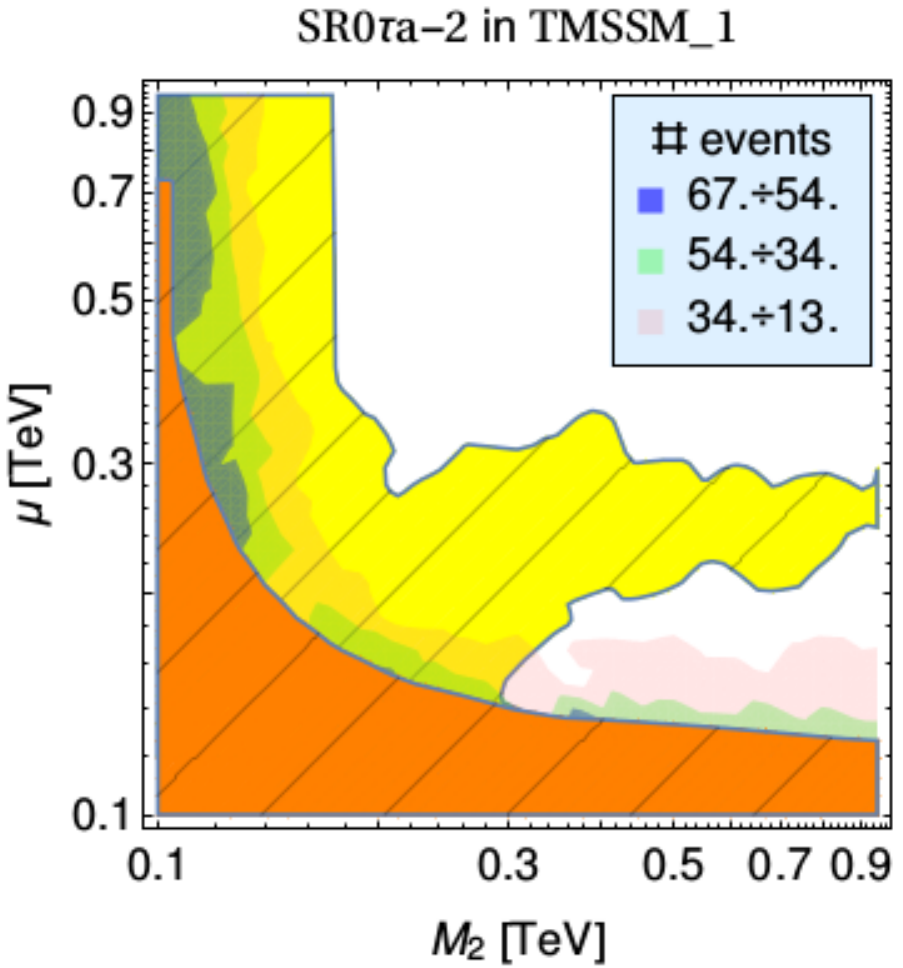}
\hspace*{0.1cm}
\includegraphics[scale=0.38]{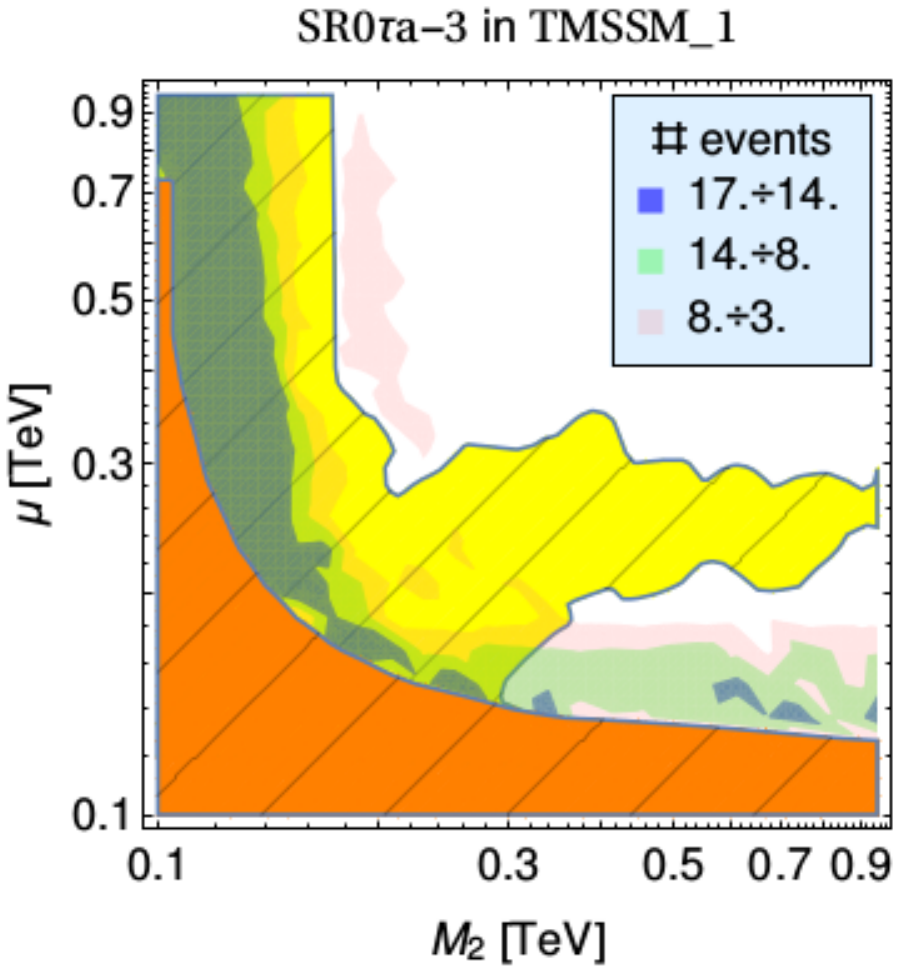}
\hspace*{0.1cm}
\includegraphics[scale=0.38]{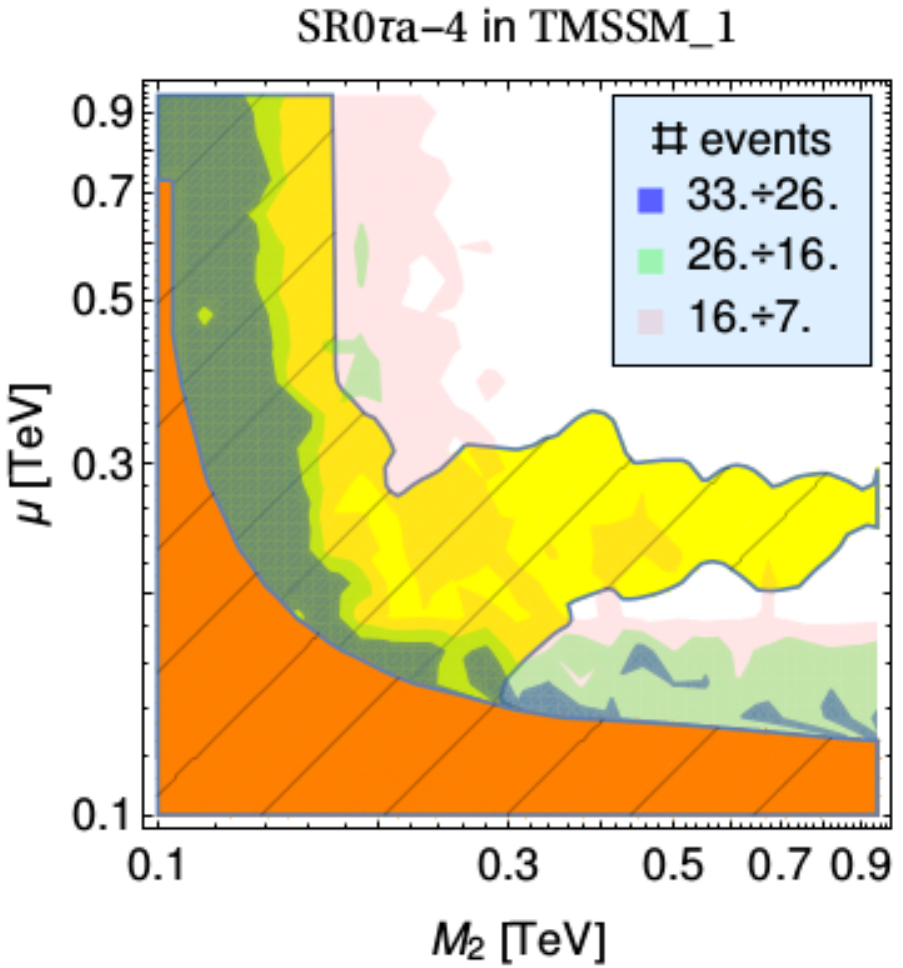}\\
\includegraphics[scale=0.38]{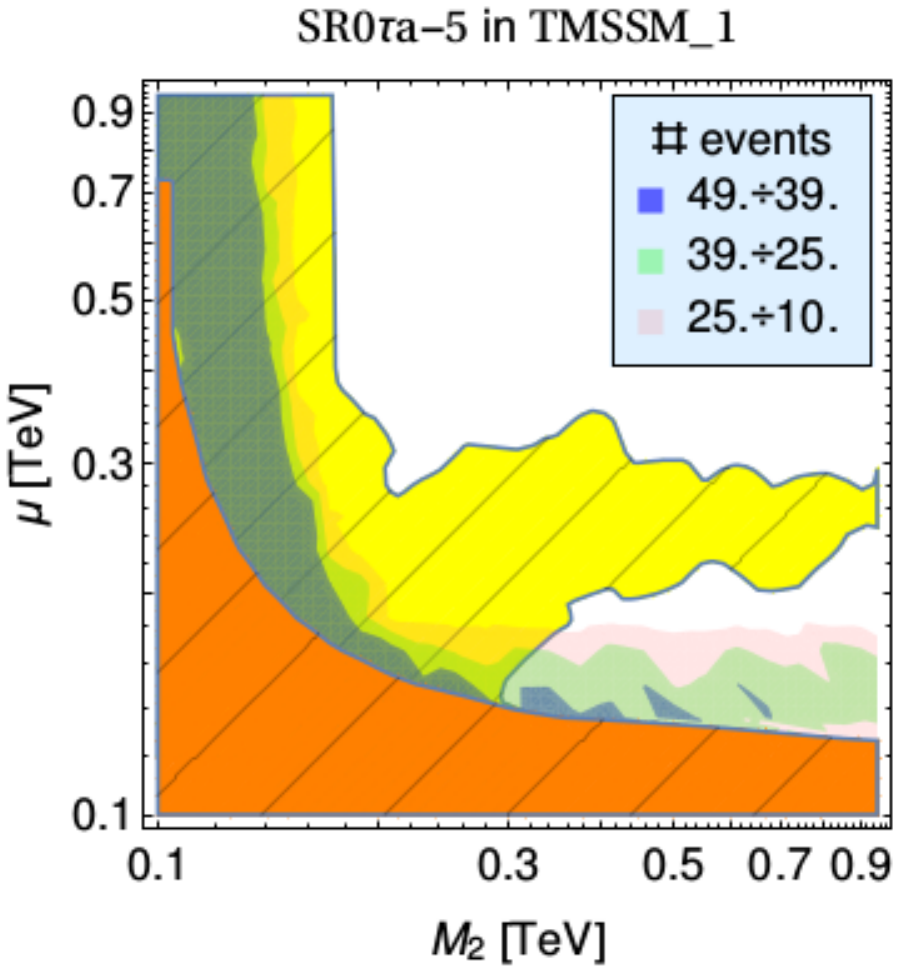}
\hspace*{0.1cm}
\includegraphics[scale=0.38]{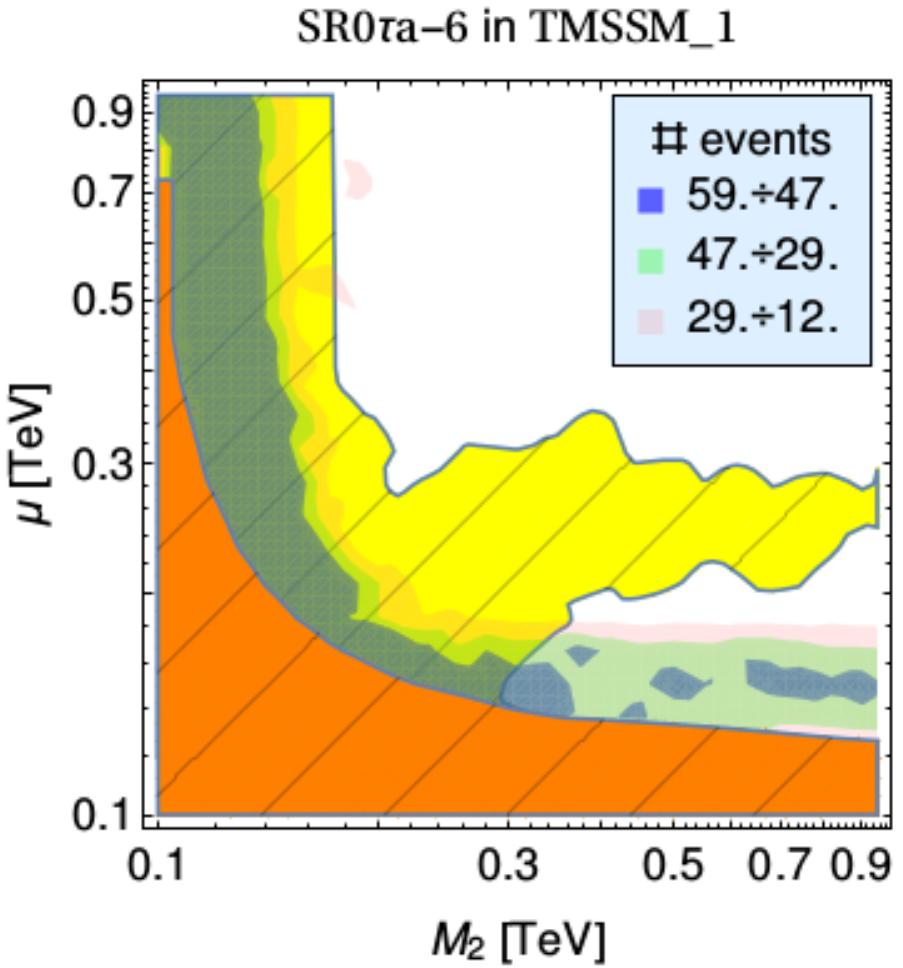}
\hspace*{0.1cm}
\includegraphics[scale=0.38]{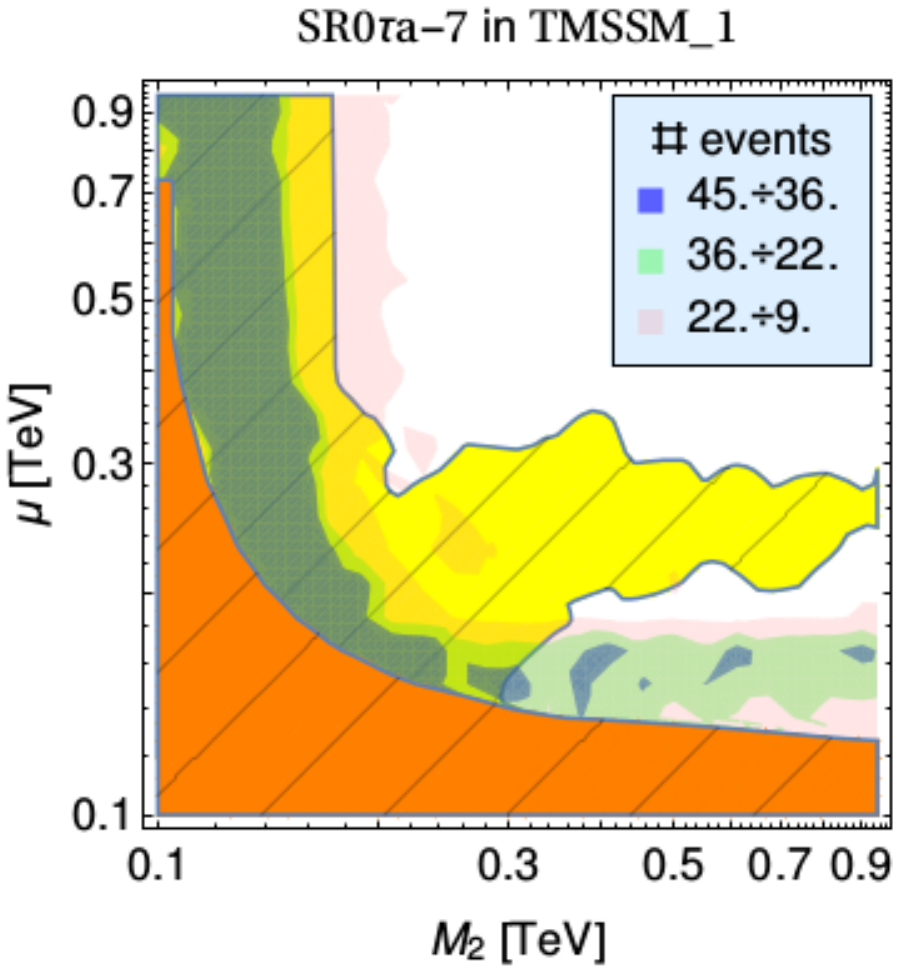}
\hspace*{0.1cm}
\includegraphics[scale=0.38]{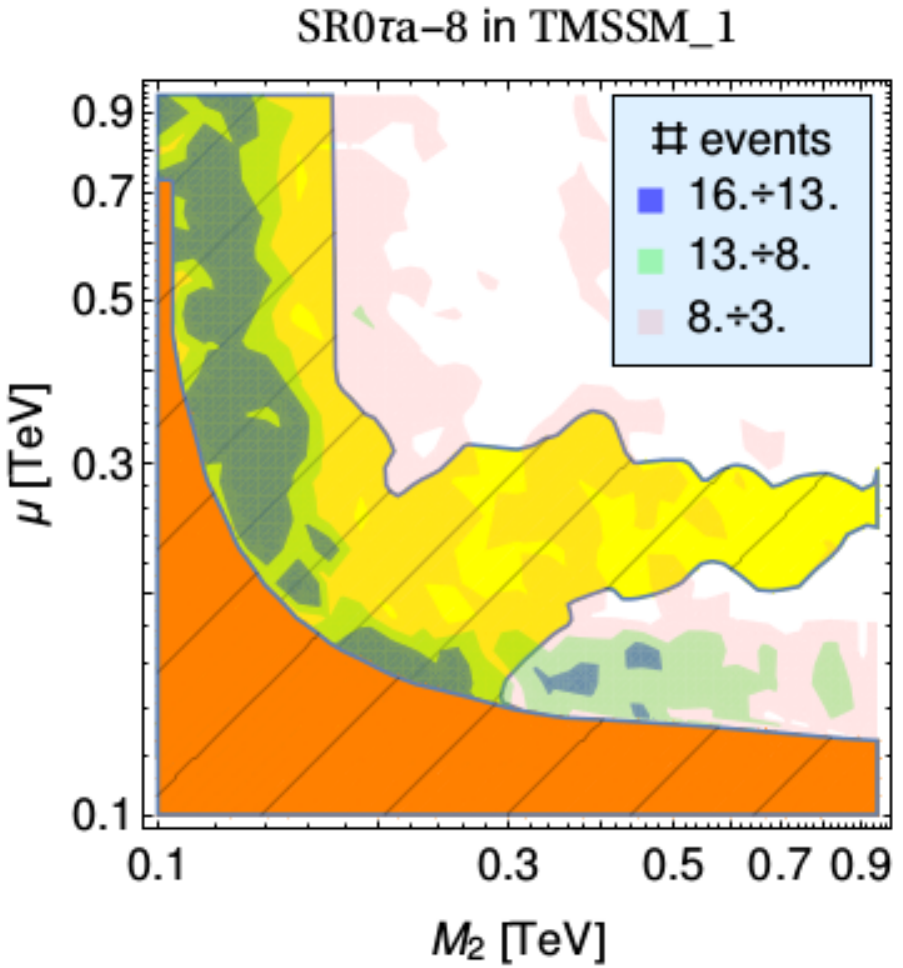}\\
\includegraphics[scale=0.38]{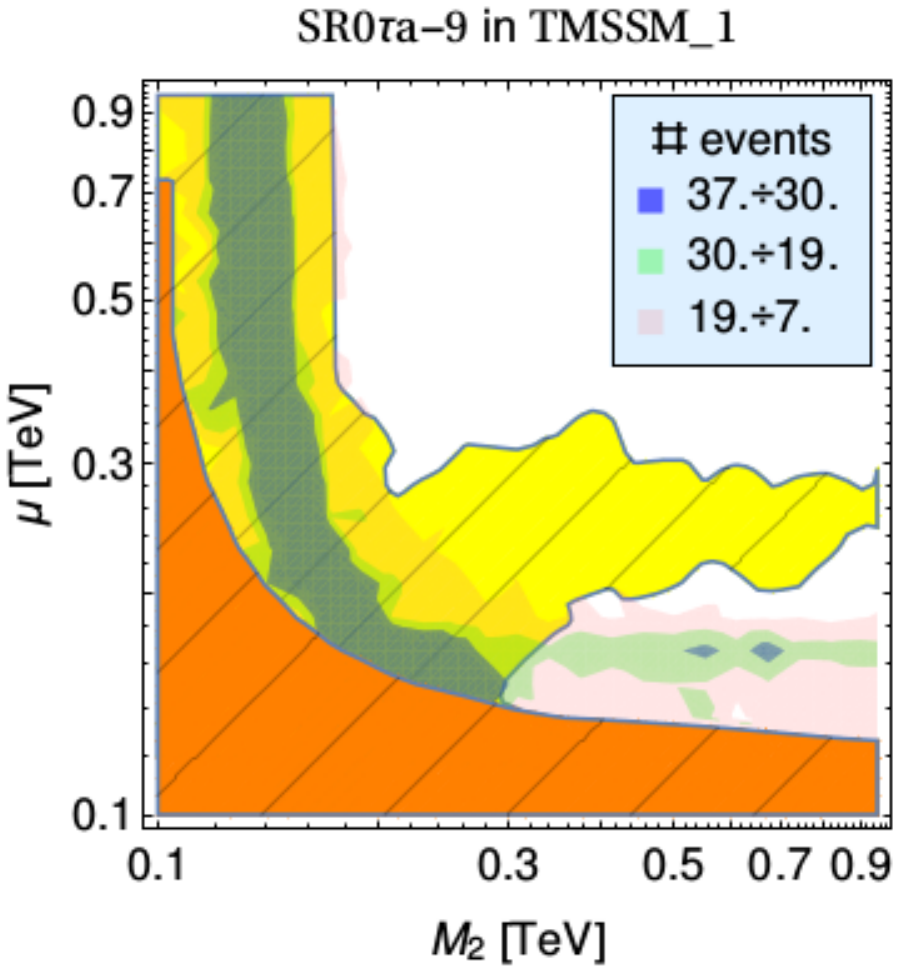}
\hspace*{0.1cm}
\includegraphics[scale=0.38]{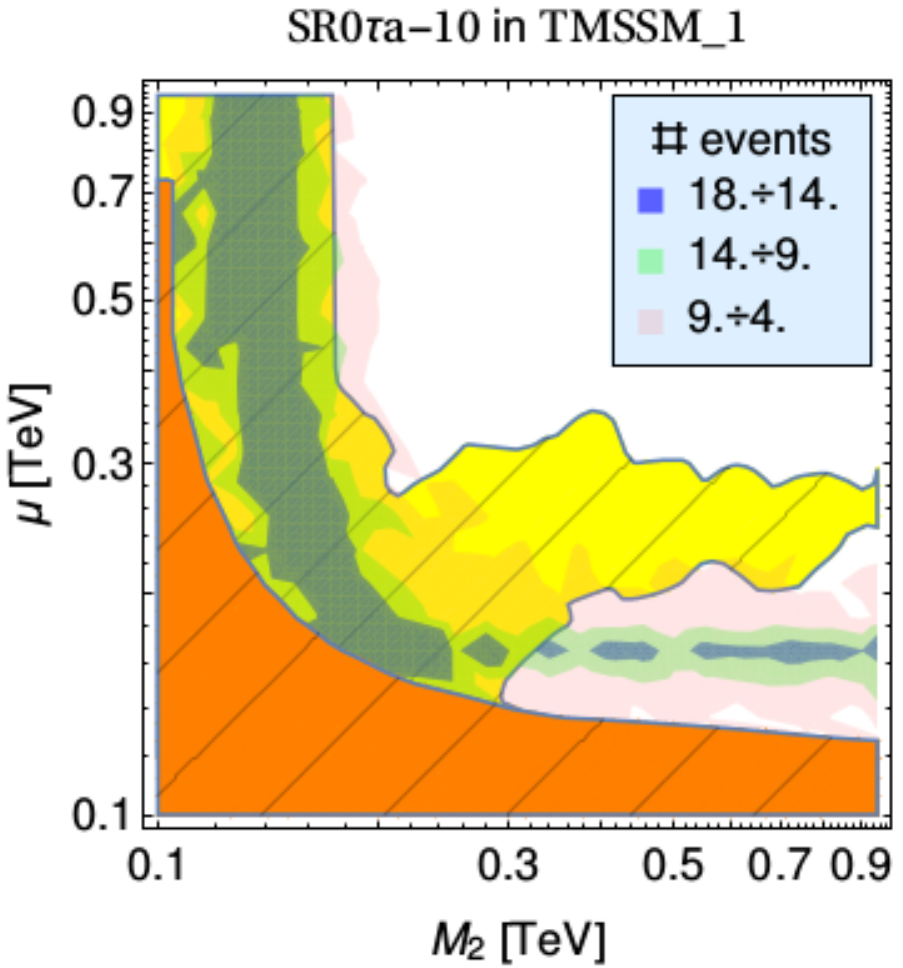}
\hspace*{0.1cm}
\includegraphics[scale=0.38]{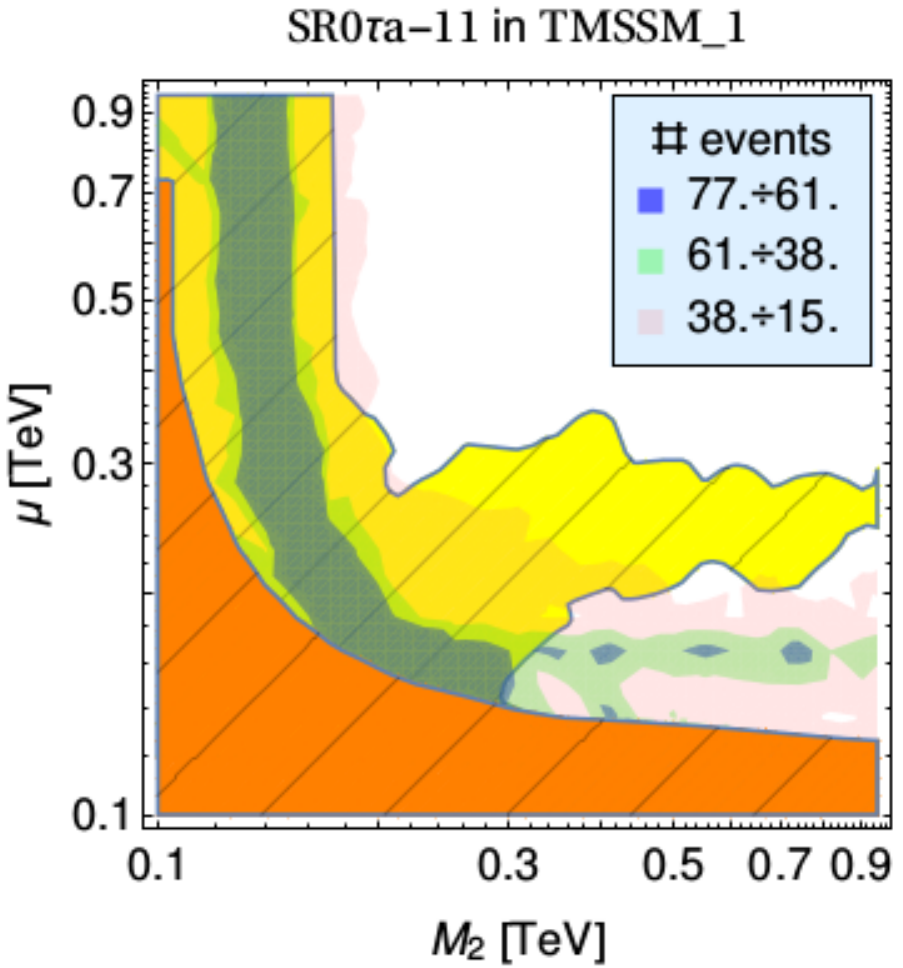}
\hspace*{0.1cm}
\includegraphics[scale=0.38]{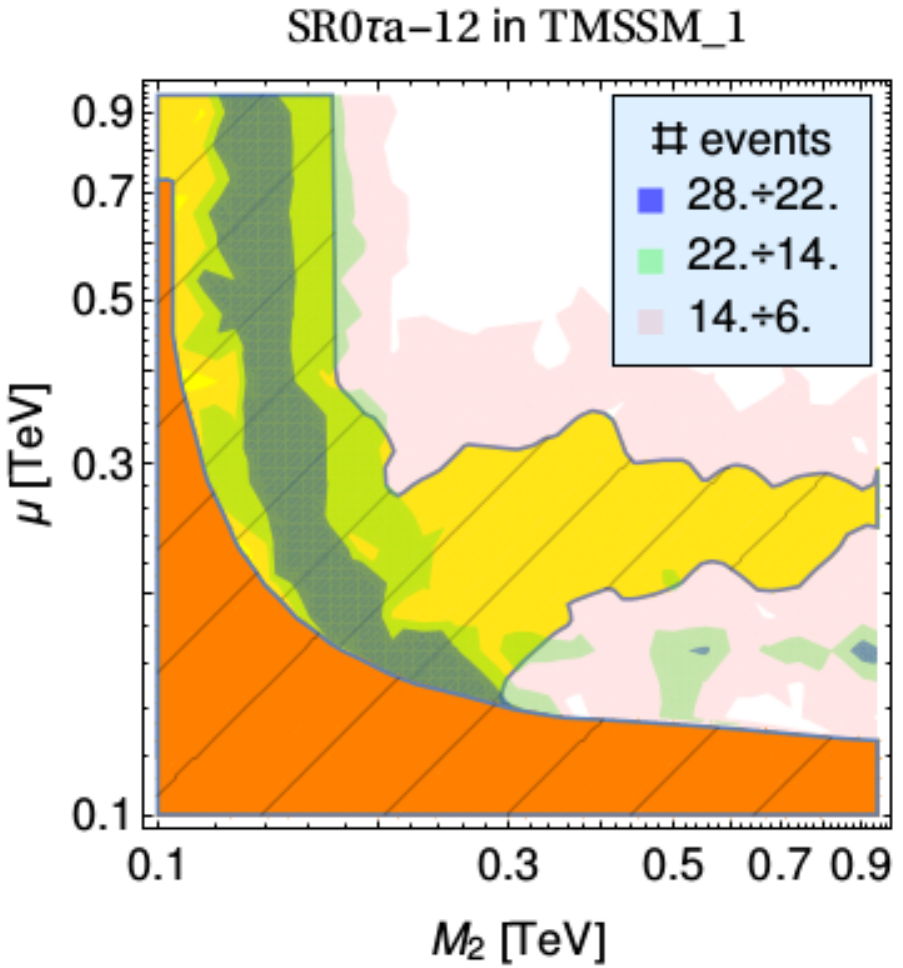}\\
\includegraphics[scale=0.38]{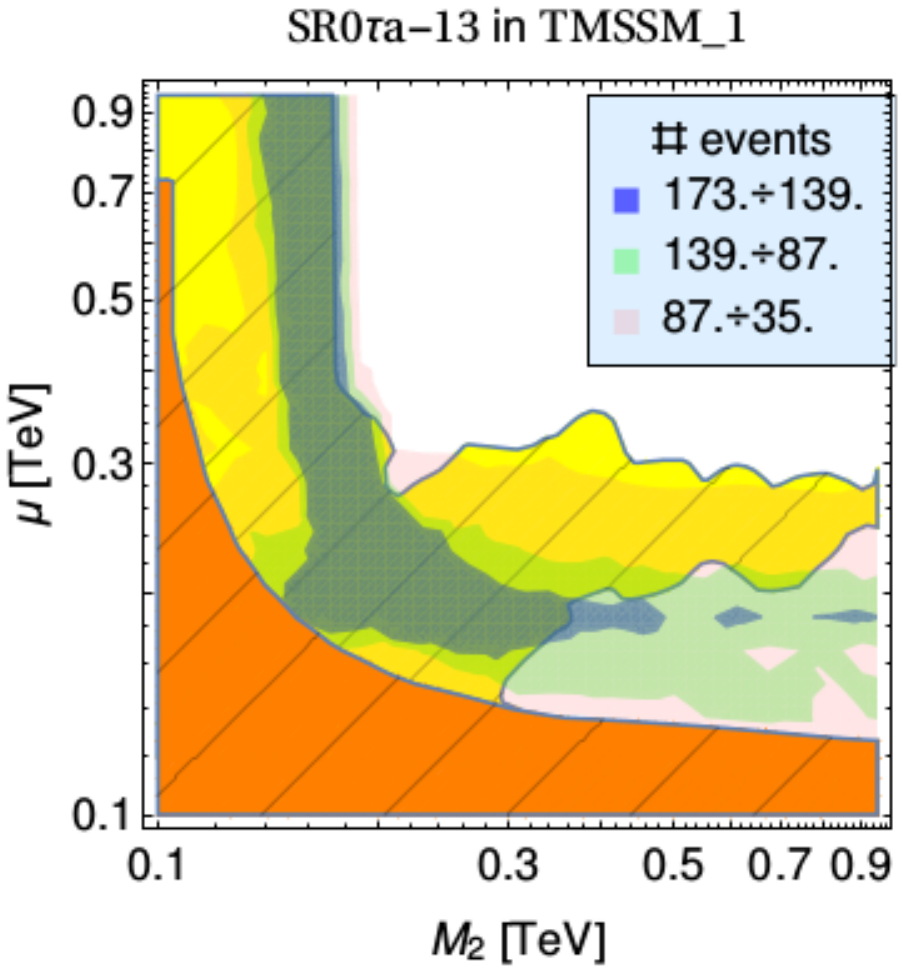}
\hspace*{0.1cm}
\includegraphics[scale=0.38]{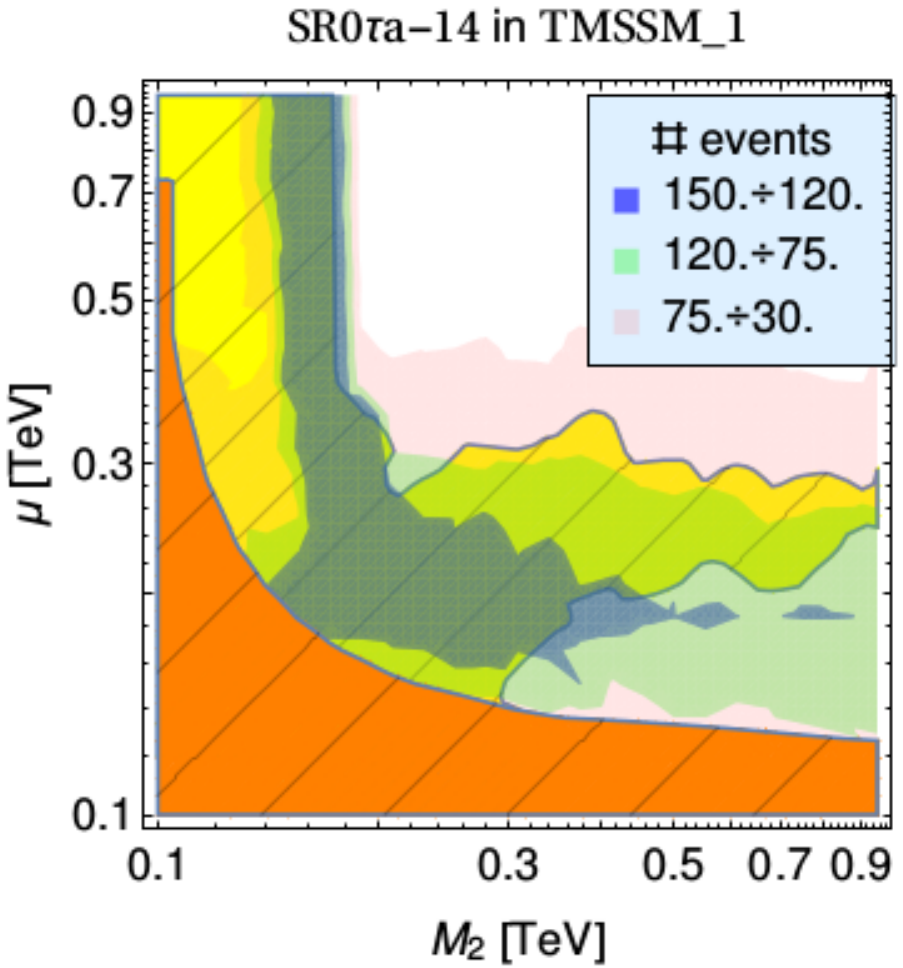}
\hspace*{0.1cm}
\includegraphics[scale=0.38]{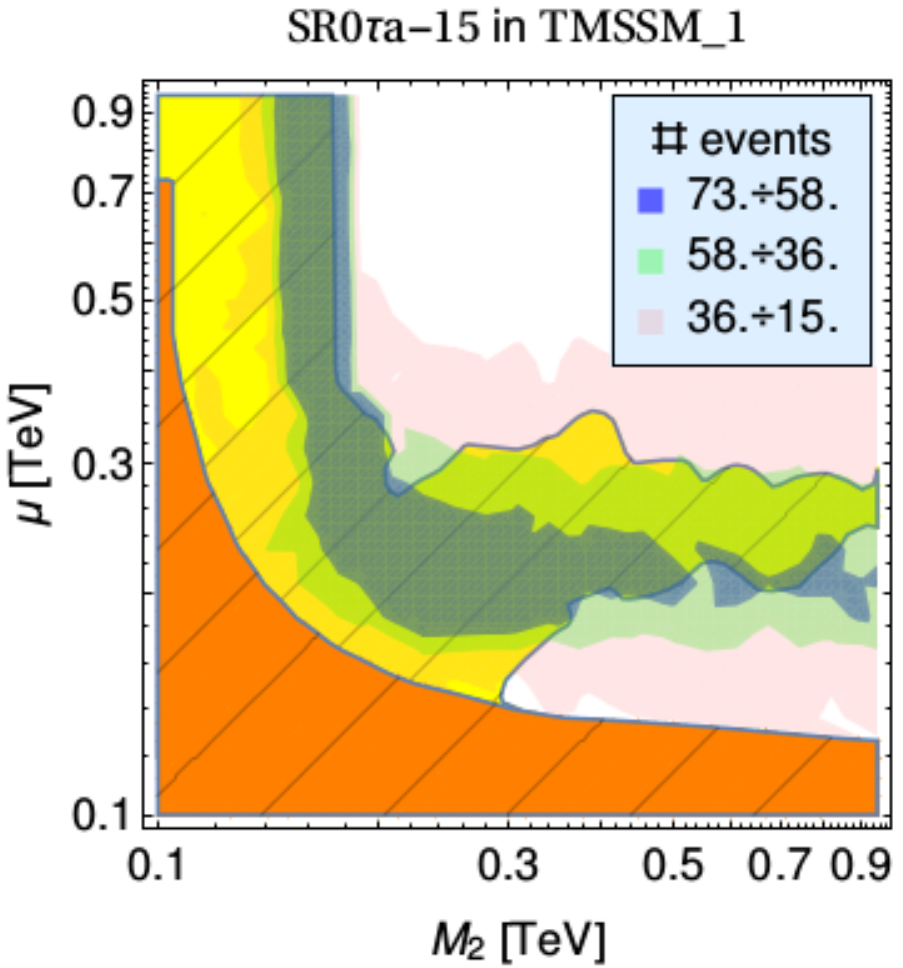}
\hspace*{0.1cm}
\includegraphics[scale=0.38]{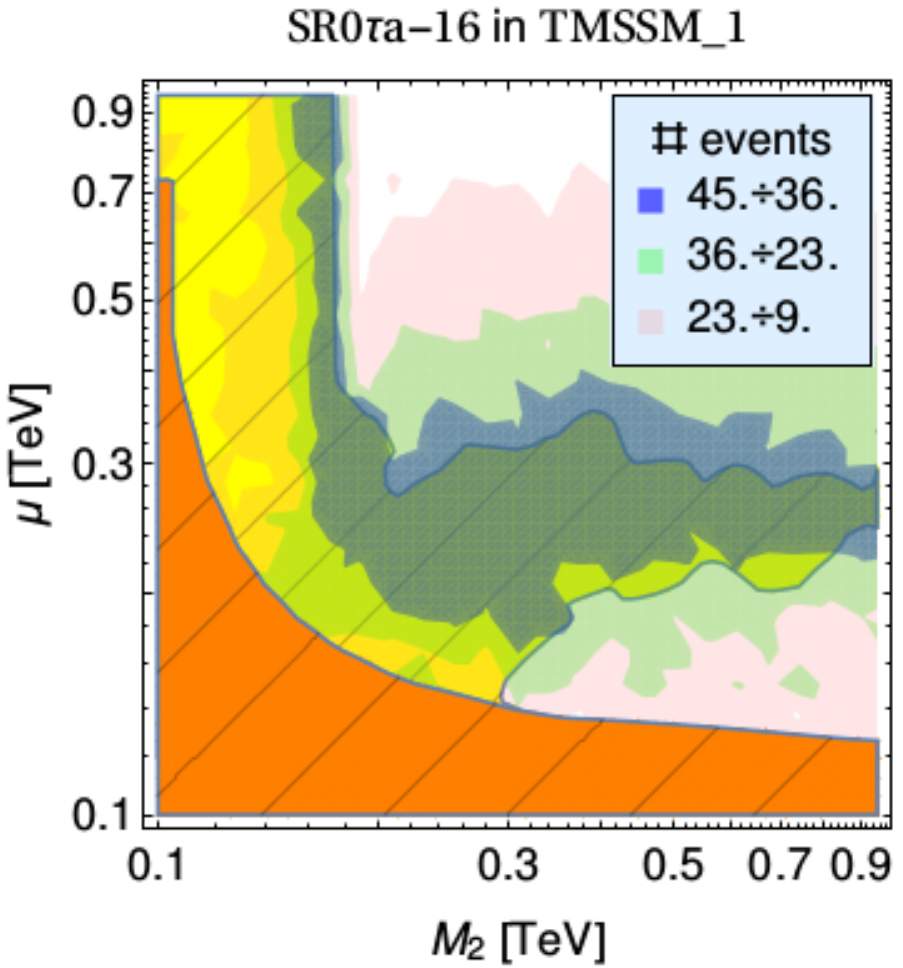}\\
\includegraphics[scale=0.38]{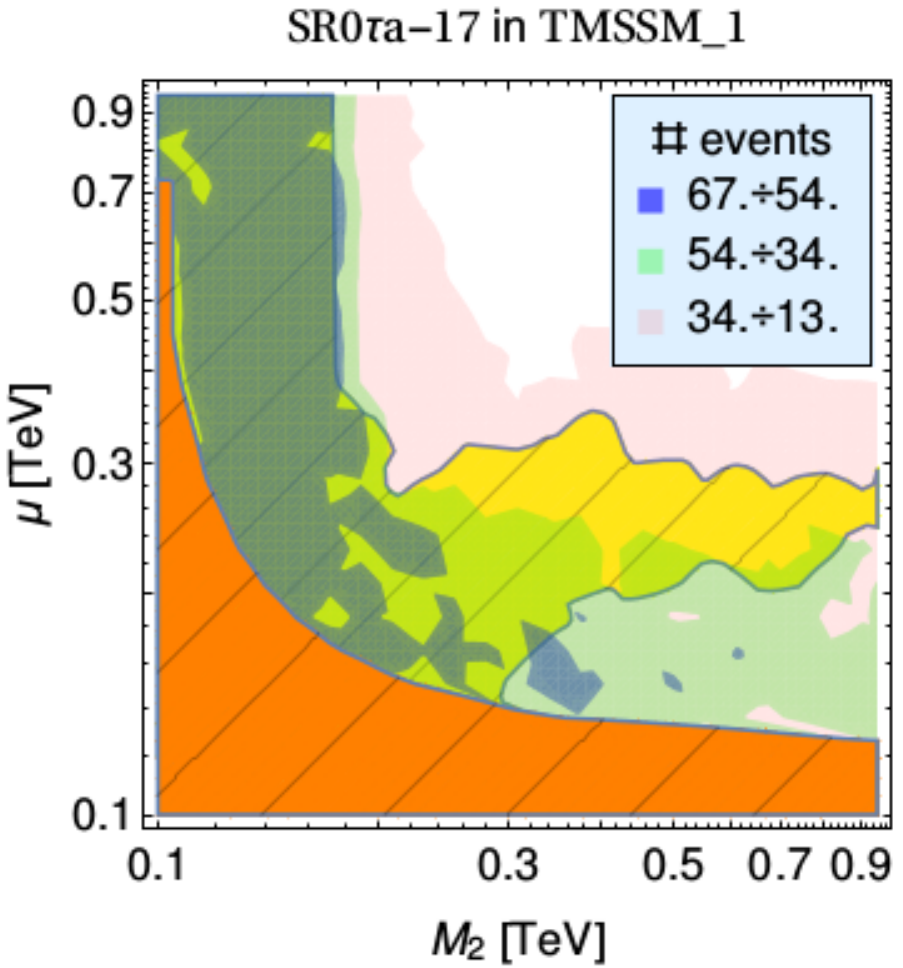}
\hspace*{0.1cm}
\includegraphics[scale=0.38]{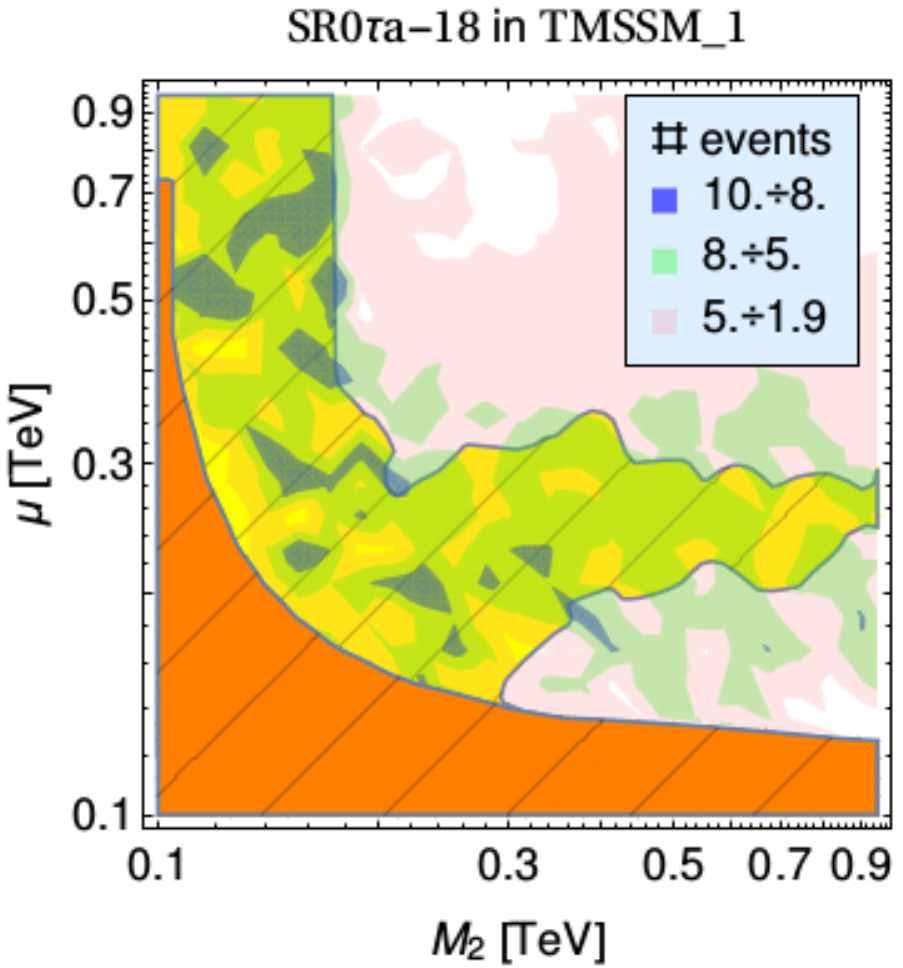}
\hspace*{0.1cm}
\includegraphics[scale=0.38]{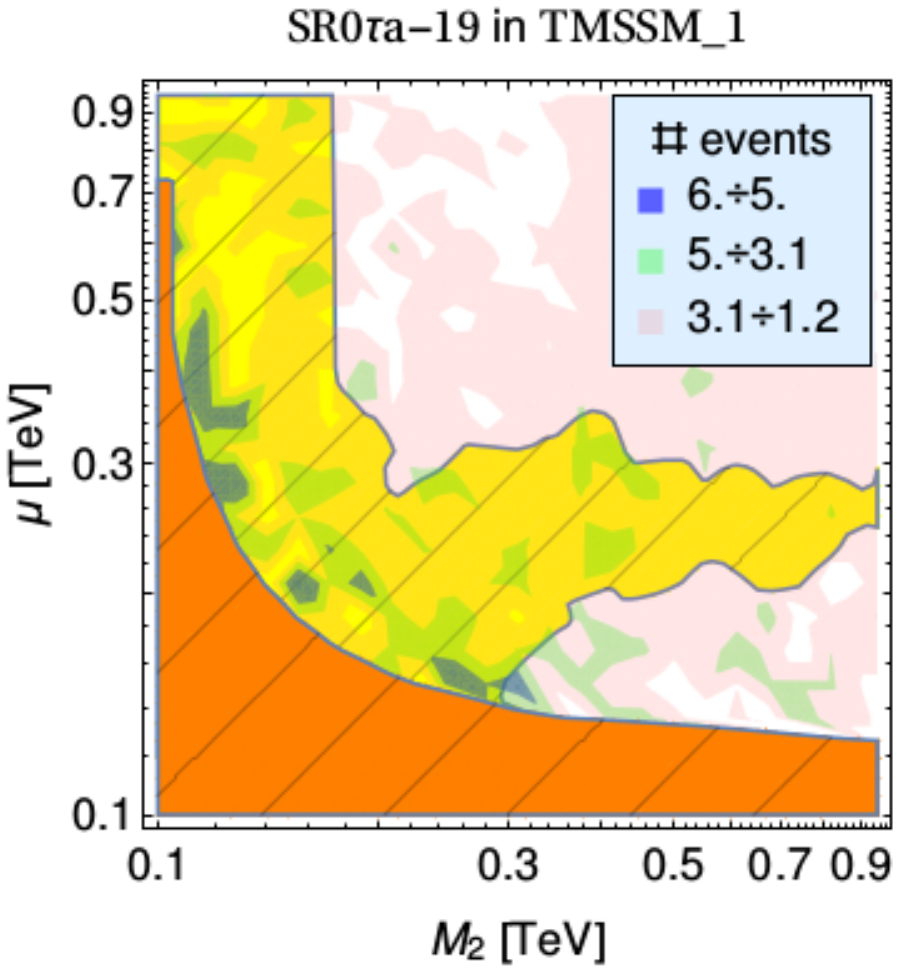}
\hspace*{0.1cm}
\includegraphics[scale=0.38]{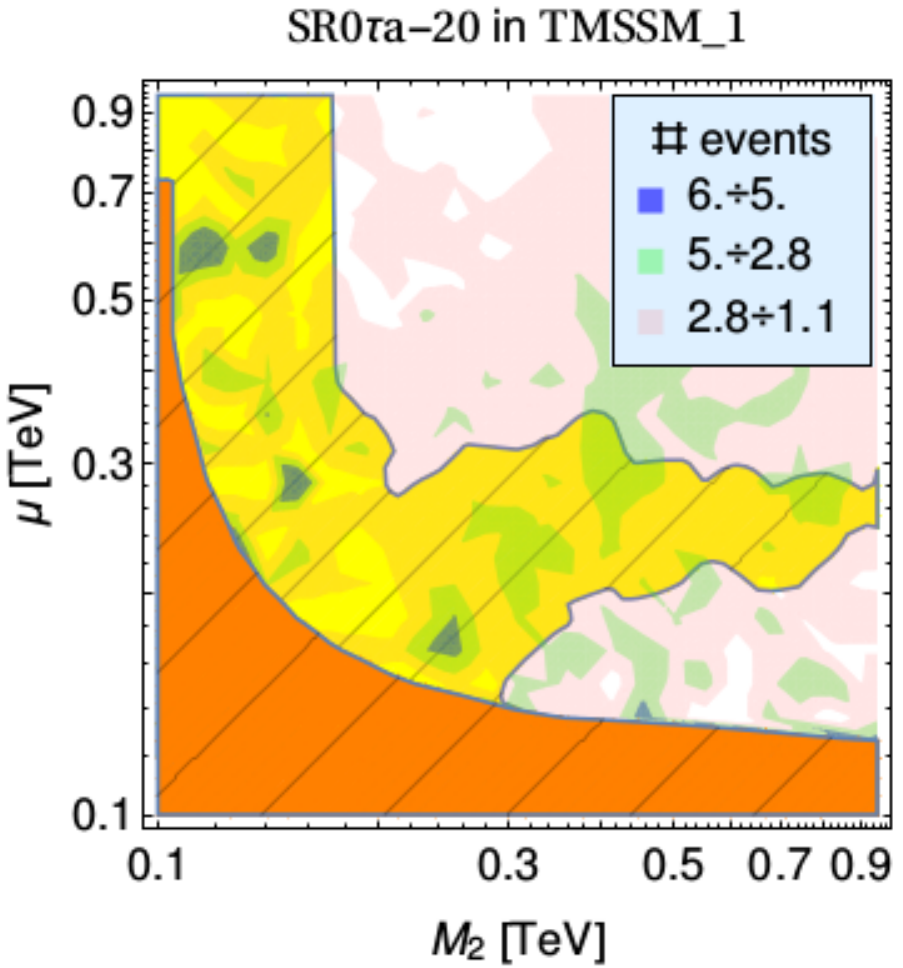}
\caption{{\bfseries Three-lepton search - TMSSM\_1 - 13 TeV:} Same as Fig.~\ref{fig:3lepton_mssm_13tev_bins} for the TMSSM\_1 case ($\lambda = 0.65$ and $\mu_{\Sigma}=300$ GeV).}
\label{fig:3lepton_tmssm65300_13tev_bins}
\end{figure}

Figure~\ref{fig:3lepton_tmssm65300_13tev_bins} shows the sensitivity of the bins of SR0$\tau$a to the three-lepton signatures of the TMSSM\_1 scenario. In this case the bins 13 and 14 have the largest predicted number of events, similarly to the MSSM case. For this scenario we notice that the number of predicted events is larger than in the MSSM case, as it was explained in Sec.~\ref{sec:excreg}. Complementary to those bins, notice that the bin 16 , even though having a much lower number of expected events, can perform better than bin 13 and bin 14 in the region with $\mu$ up to 500 GeV, irrespective on the value of $M_2$ (considering the green shaded region). 

\subsection*{TMSSM\_2 ($\lambda=0.65$, $\mu_\Sigma = 350$ GeV)}

\begin{figure}
\begin{minipage}[t]{0.32\textwidth}
\centering
\includegraphics[width=1.\columnwidth]{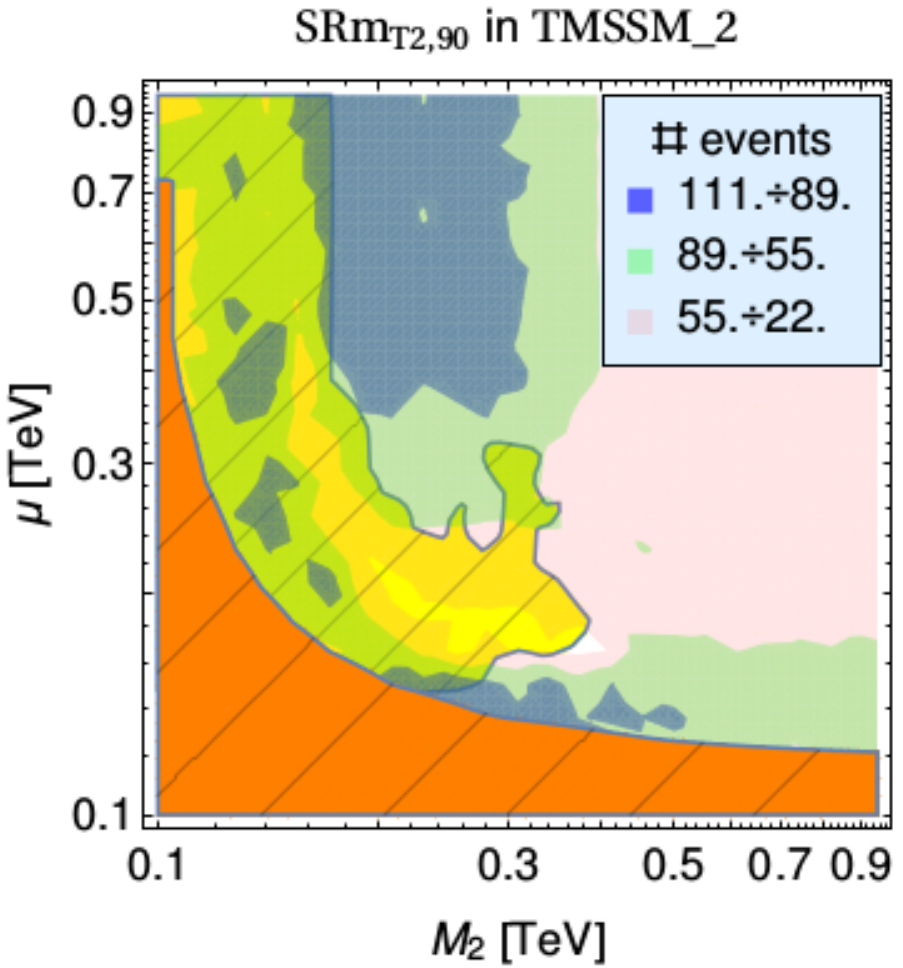}
\end{minipage}
\begin{minipage}[t]{0.32\textwidth}
\centering
\includegraphics[width=1.\columnwidth]{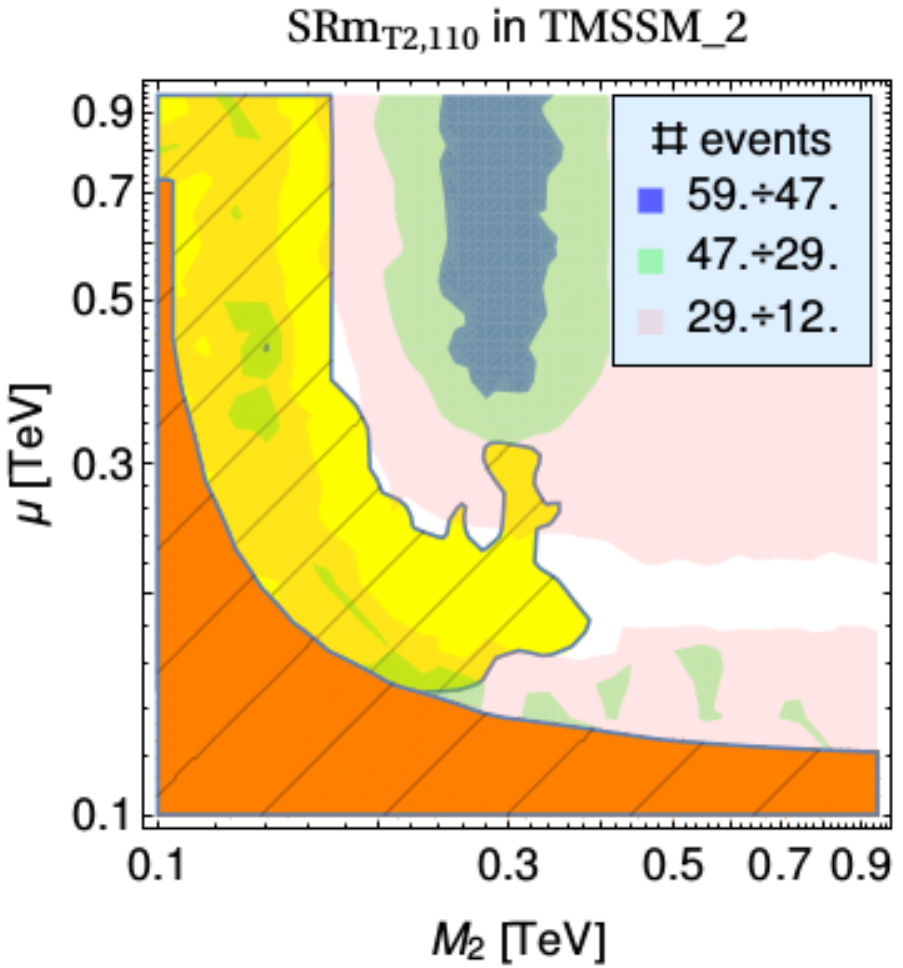}
\end{minipage}
\begin{minipage}[t]{0.32\textwidth}
\centering
\includegraphics[width=1.\columnwidth]{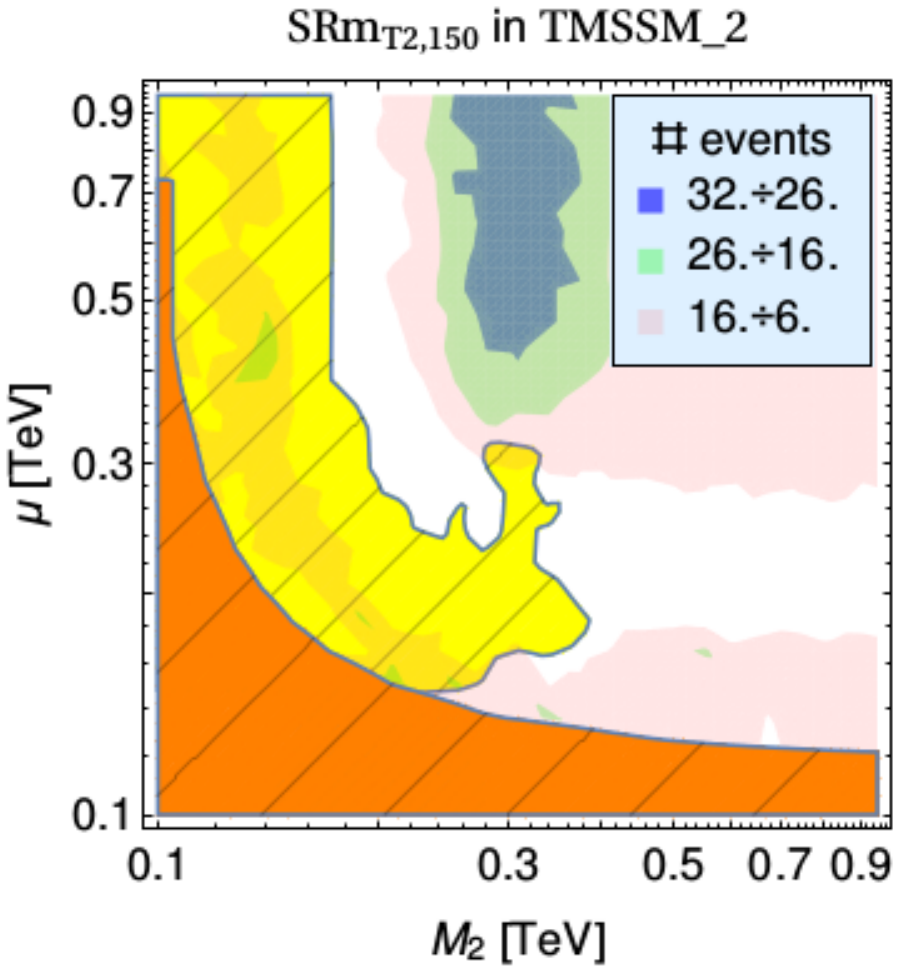}
\end{minipage}
\\
\begin{minipage}[t]{0.32\textwidth}
\centering
\includegraphics[width=1.\columnwidth]{./figs/TMSSM65_350_13TeV-2lept-bin4.pdf}
\end{minipage}
\begin{minipage}[t]{0.32\textwidth}
\centering
\includegraphics[width=1.\columnwidth]{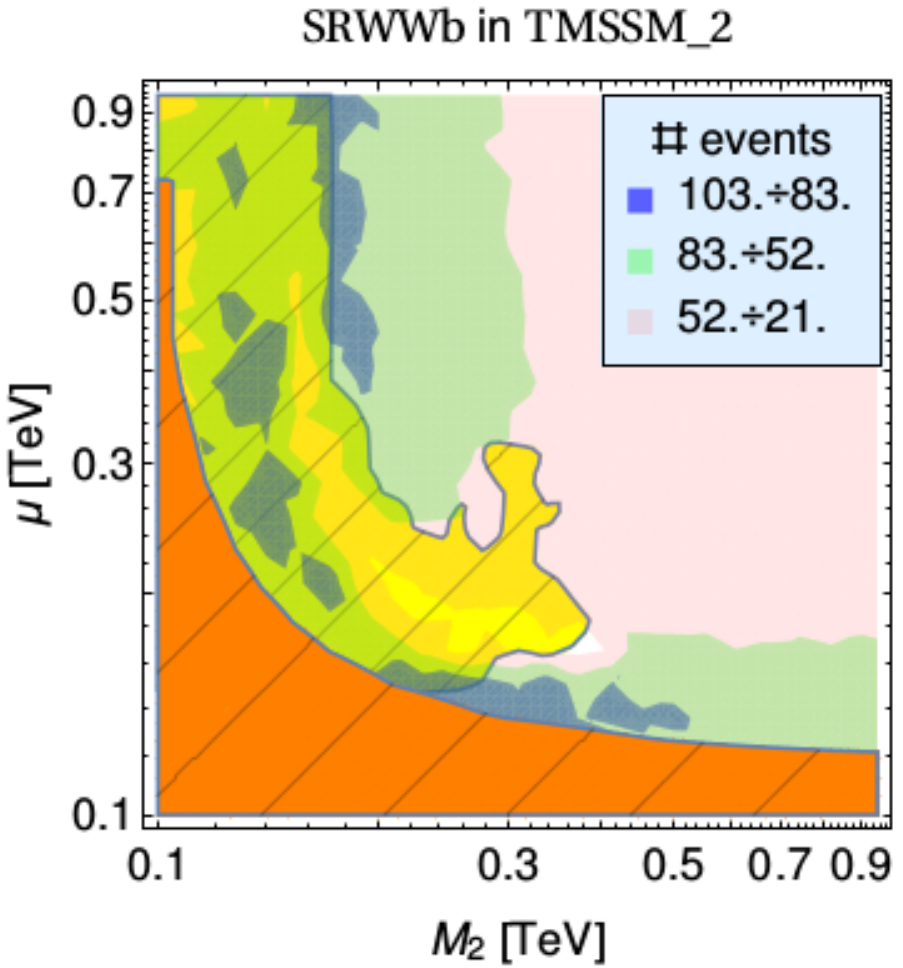}
\end{minipage}
\begin{minipage}[t]{0.32\textwidth}
\centering
\includegraphics[width=1.\columnwidth]{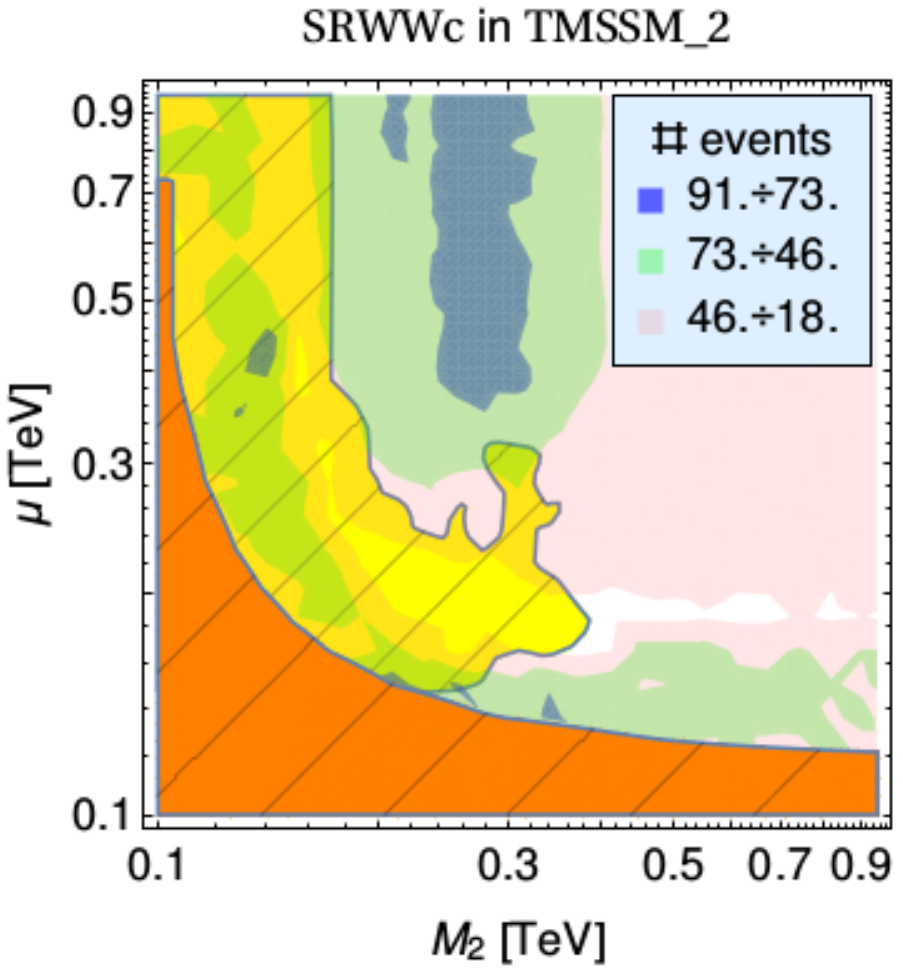}
\end{minipage}
\\
\begin{minipage}[t]{0.32\textwidth}
\centering
\includegraphics[width=1.\columnwidth]{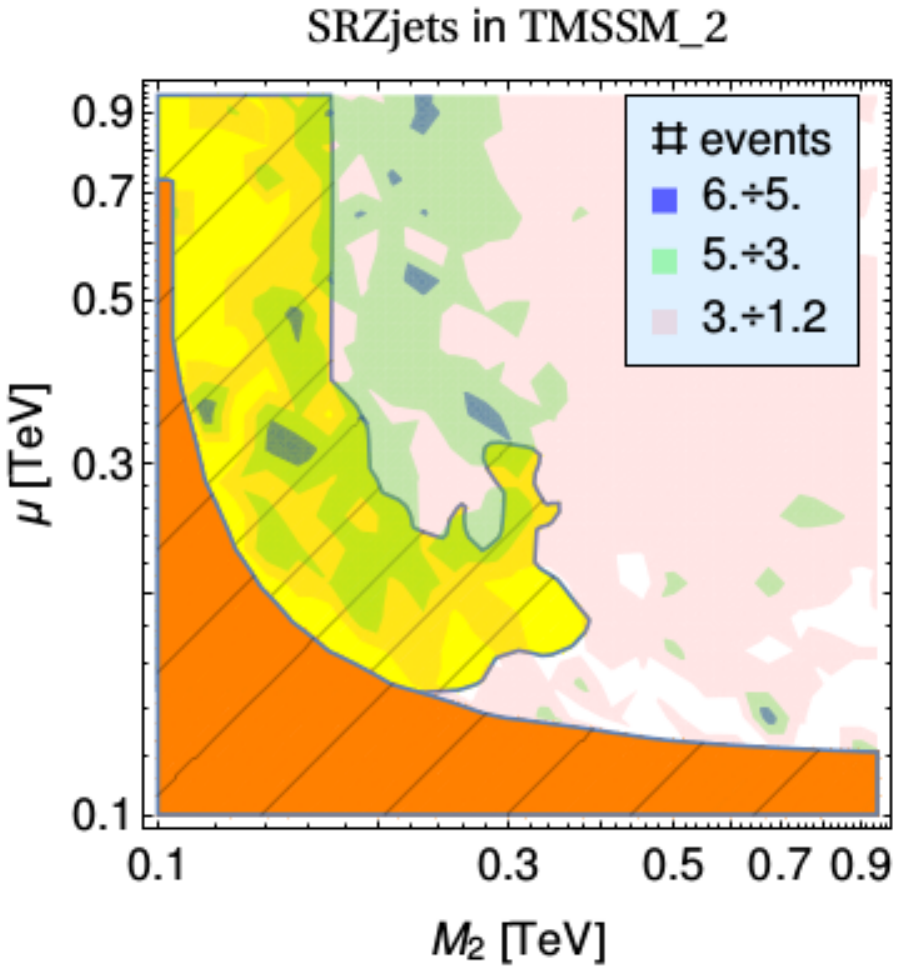}
\end{minipage}
\begin{minipage}[t]{0.32\textwidth}
\centering
\includegraphics[width=1.\columnwidth]{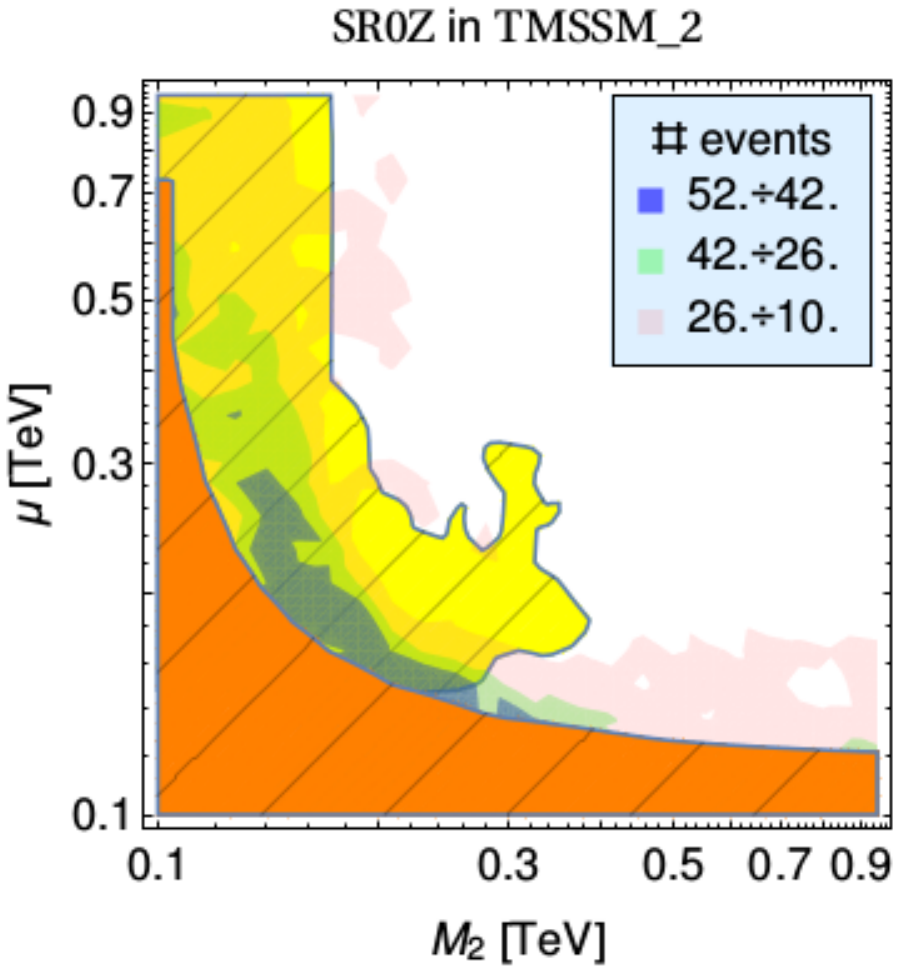}
\end{minipage}
\begin{minipage}[t]{0.32\textwidth}
\centering
\includegraphics[width=1.\columnwidth]{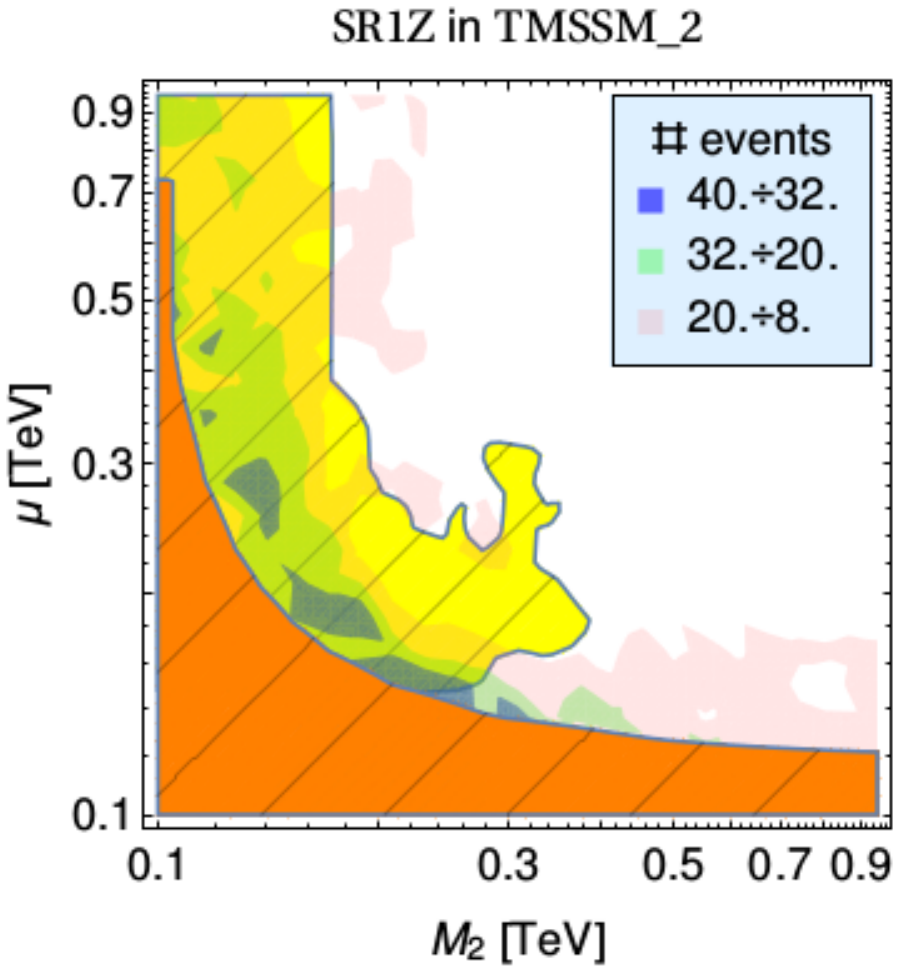}
\end{minipage}
\\
\begin{minipage}[t]{0.32\textwidth}
\centering
\includegraphics[width=1.\columnwidth]{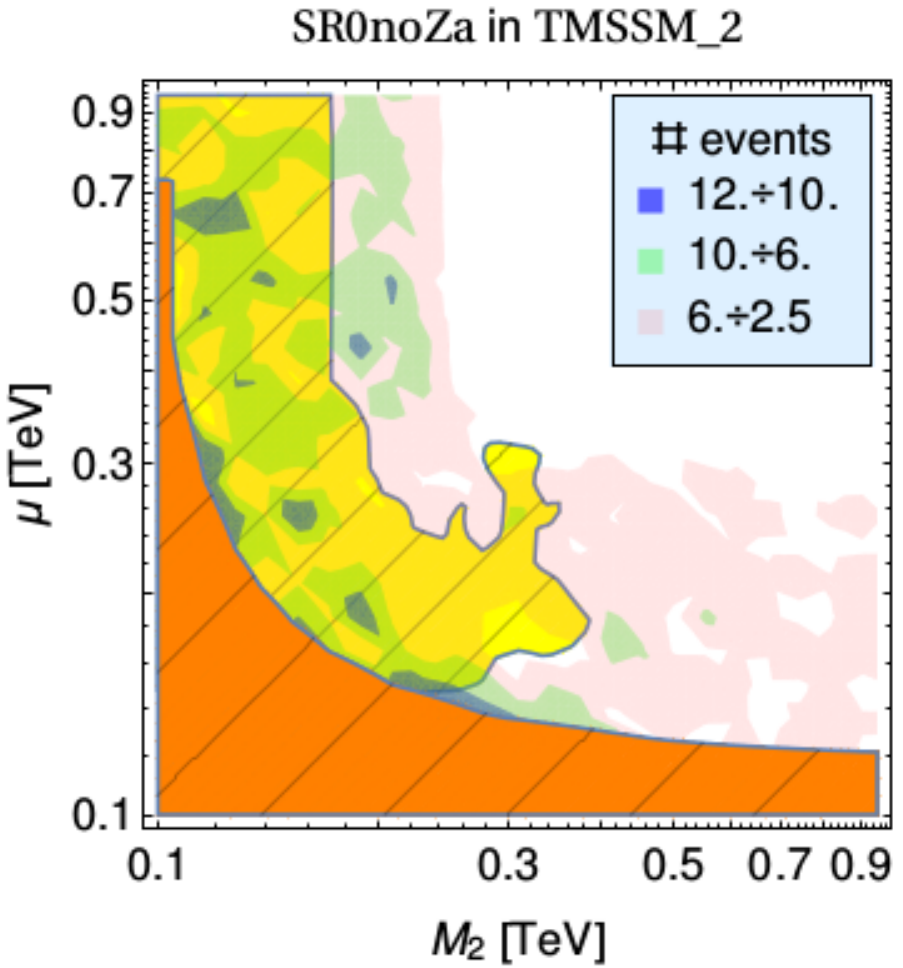}
\end{minipage}
\begin{minipage}[t]{0.32\textwidth}
\centering
\includegraphics[width=1.\columnwidth]{./figs/TMSSM65_350_13TeV-4lept-bin4.pdf}
\end{minipage}
\begin{minipage}[t]{0.32\textwidth}
\centering
\includegraphics[width=1.\columnwidth]{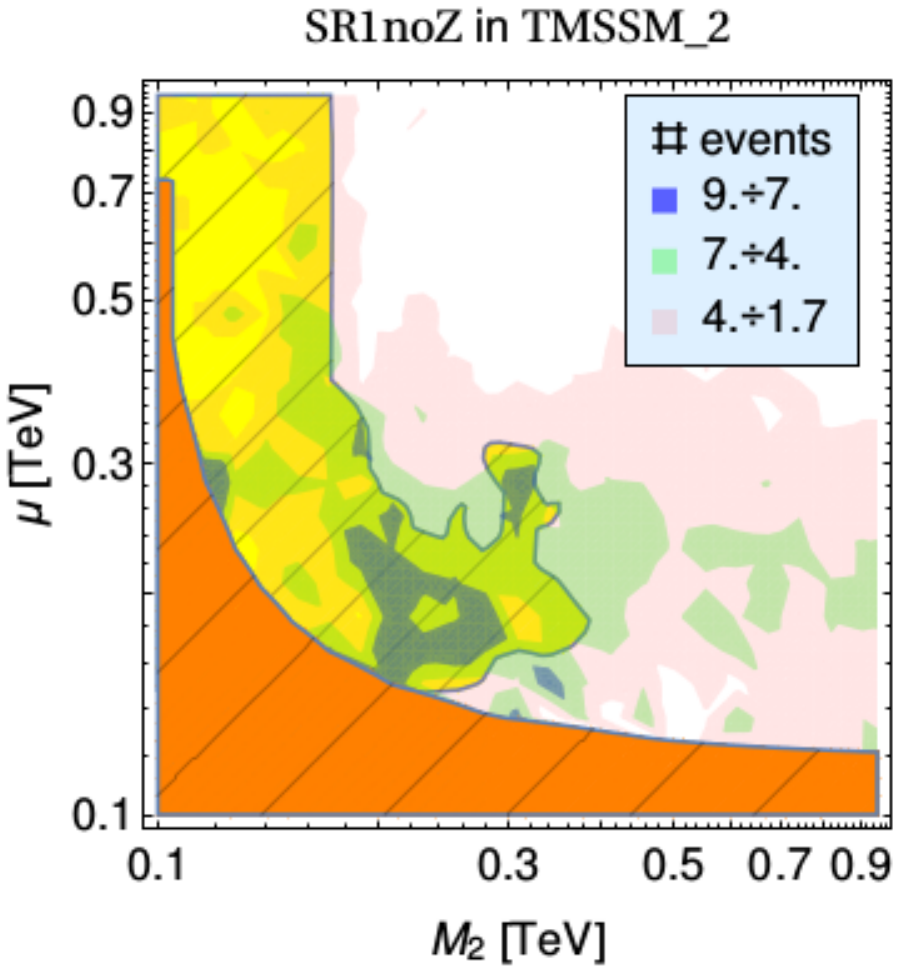}
\end{minipage}
\caption{{\bfseries Two-lepton \& four-lepton searches - TMSSM\_2 - 13 TeV:} Same as Fig.~\ref{fig:2lepton_mssm13tev_bins} for the TMSSM\_2 case ($\lambda = 0.65$ and $\mu_{\Sigma}=350$ GeV).}
\label{fig:2lepton_tmssm65350_13tev_bins}
\end{figure}

Figure~\ref{fig:2lepton_tmssm65350_13tev_bins} shows the sensitivity of the two-lepton search but now for scenario TMSSM\_2, as in Figs.~\ref{fig:2lepton_mssm13tev_bins} and~\ref{fig:2lepton_tmssm65300_13tev_bins}. This scenario behaves similarly to the case of TMSSM\_1, still featuring SRm$_{T2,90}$ as a very sensitive SR, even though the number of events is closer to the one of the MSSM scenario. Regarding the four-lepton search, we find results similar to the TMSSM\_1 case. The presence of new EWino states modify the decay chains in such a way that different EWino decays contribute to the four-lepton + MET signatures.

\begin{figure}[t]
\centering
\includegraphics[scale=0.38]{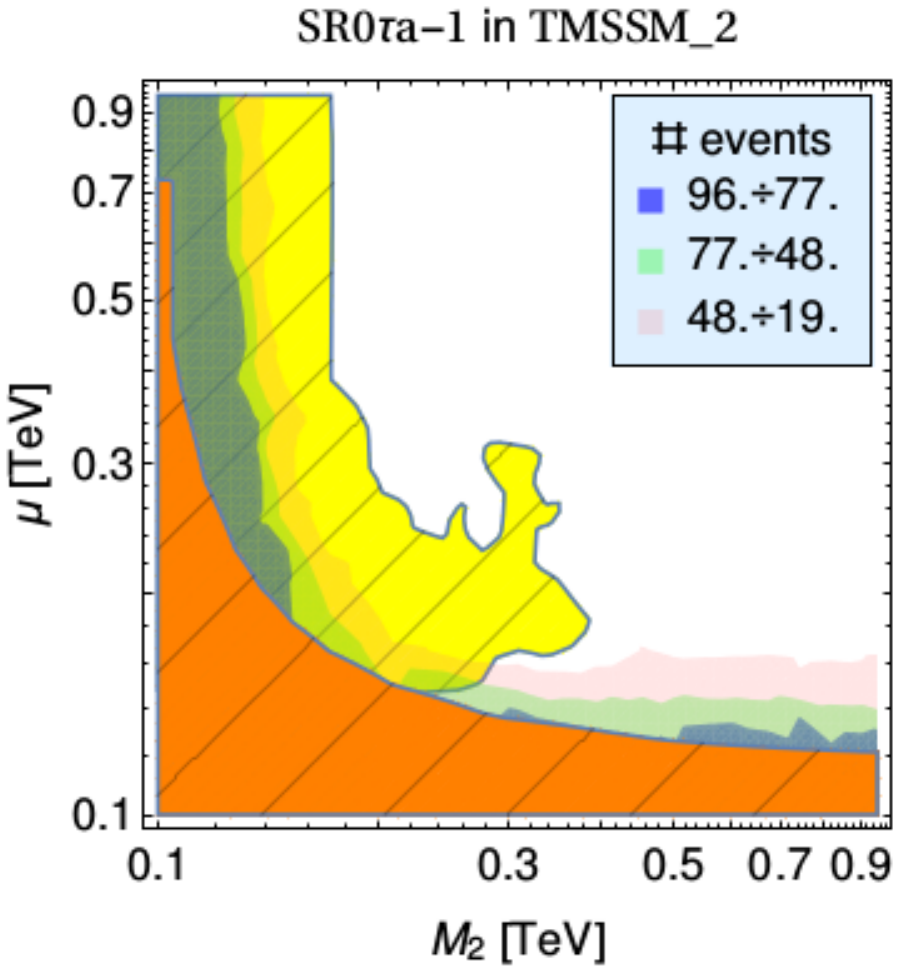}
\hspace*{0.1cm}
\includegraphics[scale=0.38]{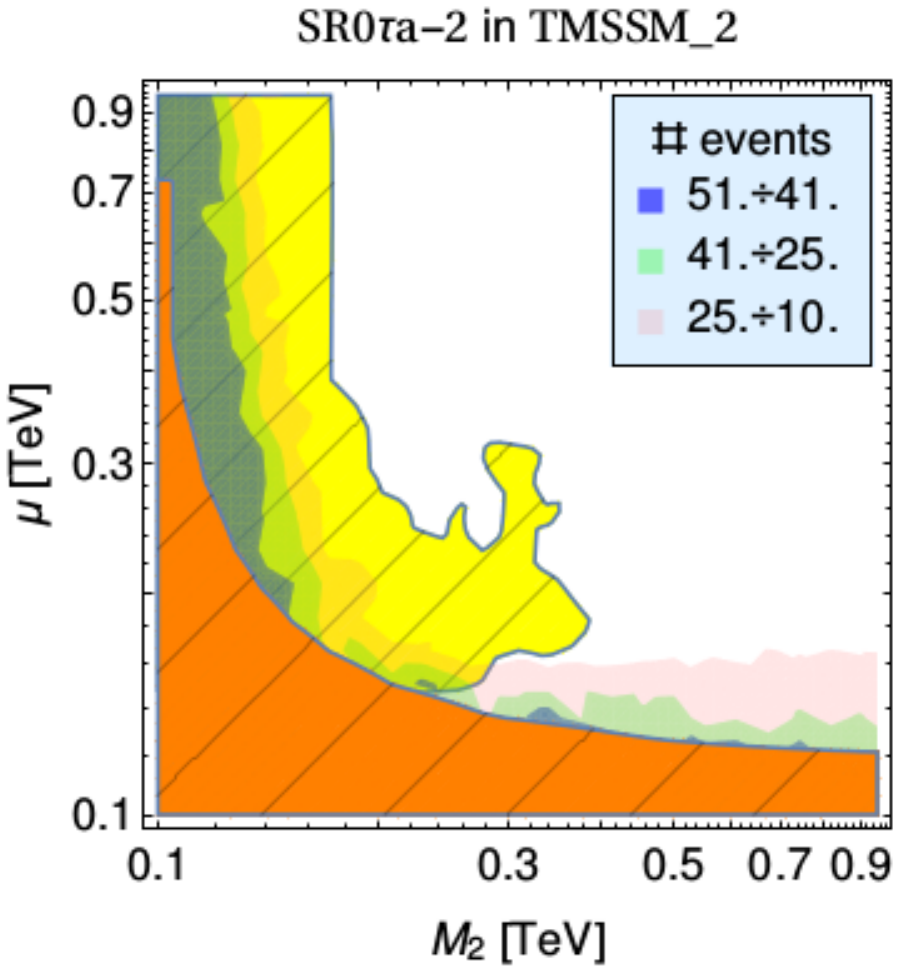}
\hspace*{0.1cm}
\includegraphics[scale=0.38]{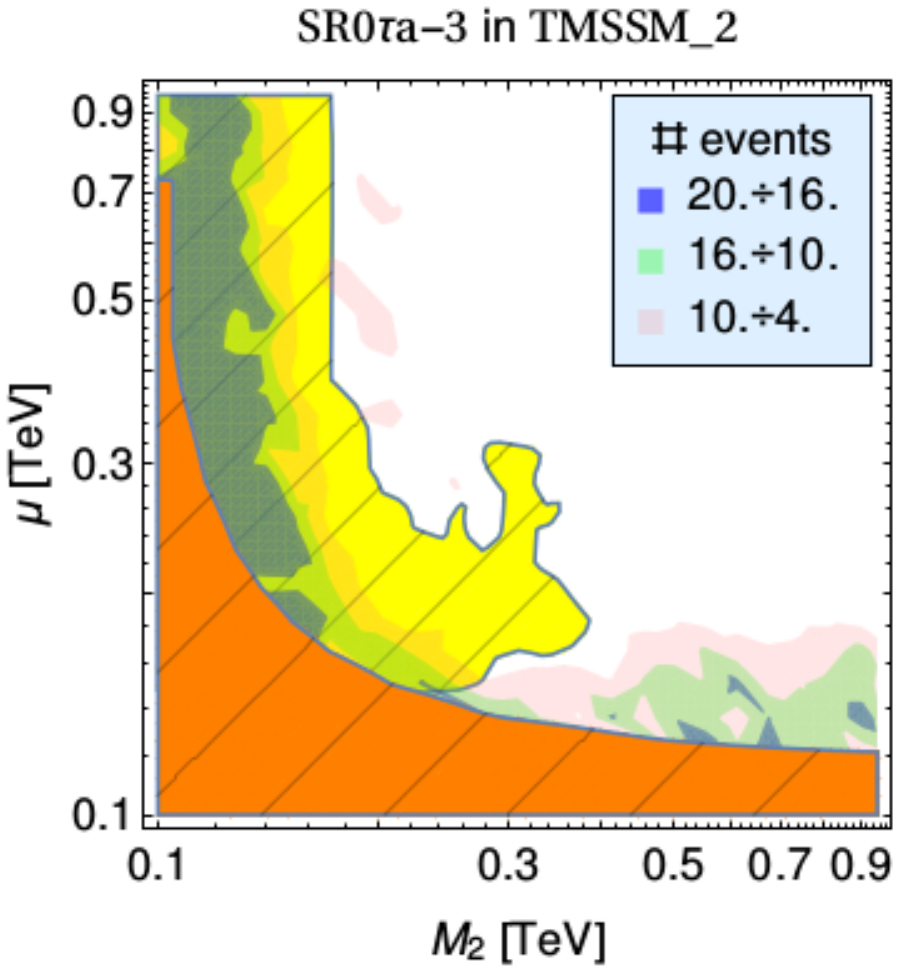}
\hspace*{0.1cm}
\includegraphics[scale=0.38]{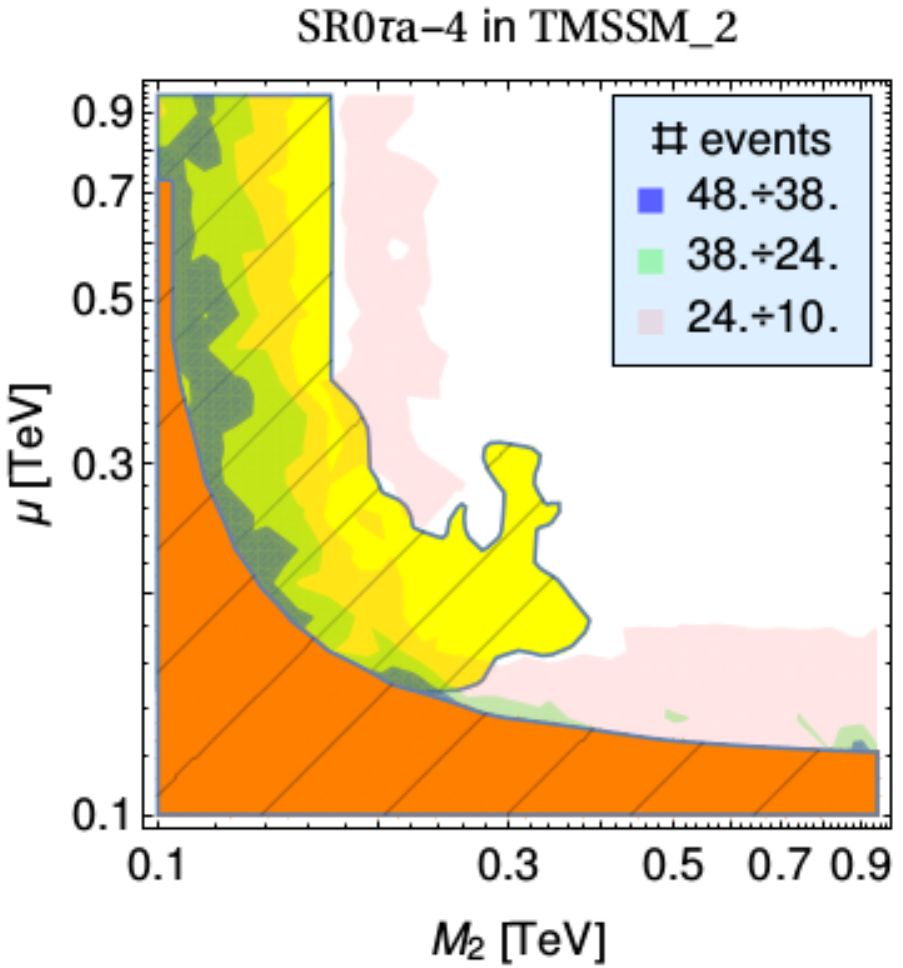}\\
\includegraphics[scale=0.38]{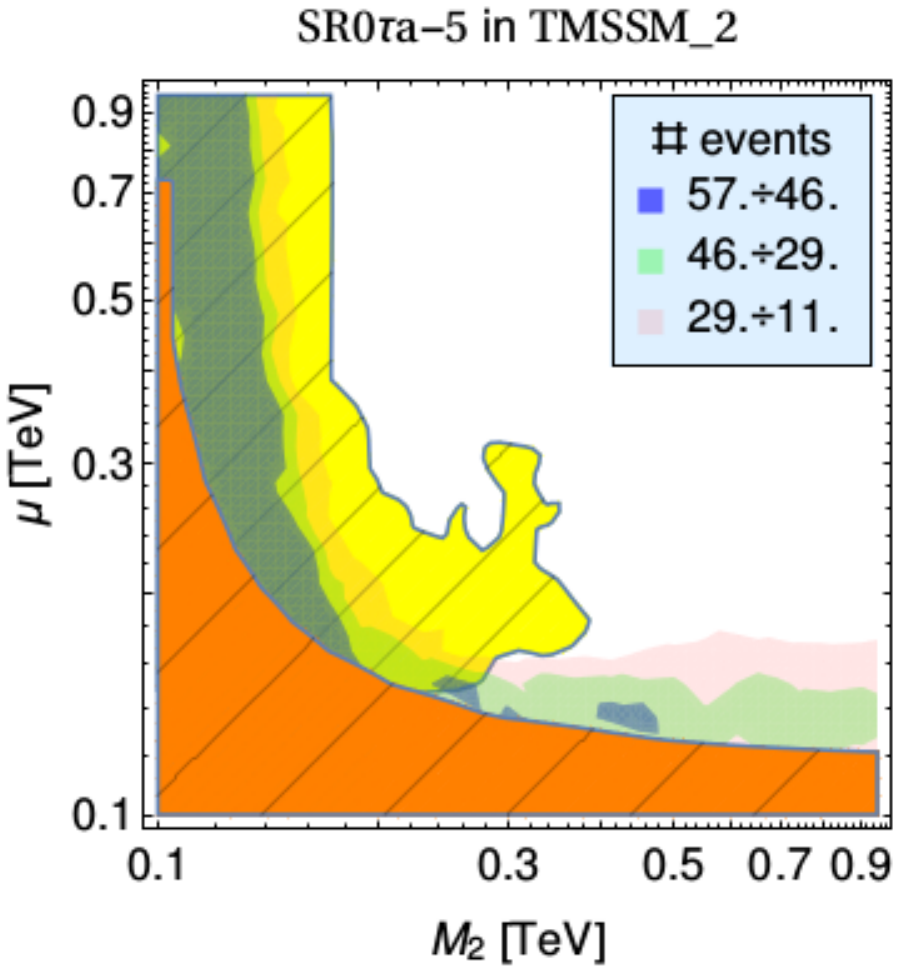}
\hspace*{0.1cm}
\includegraphics[scale=0.38]{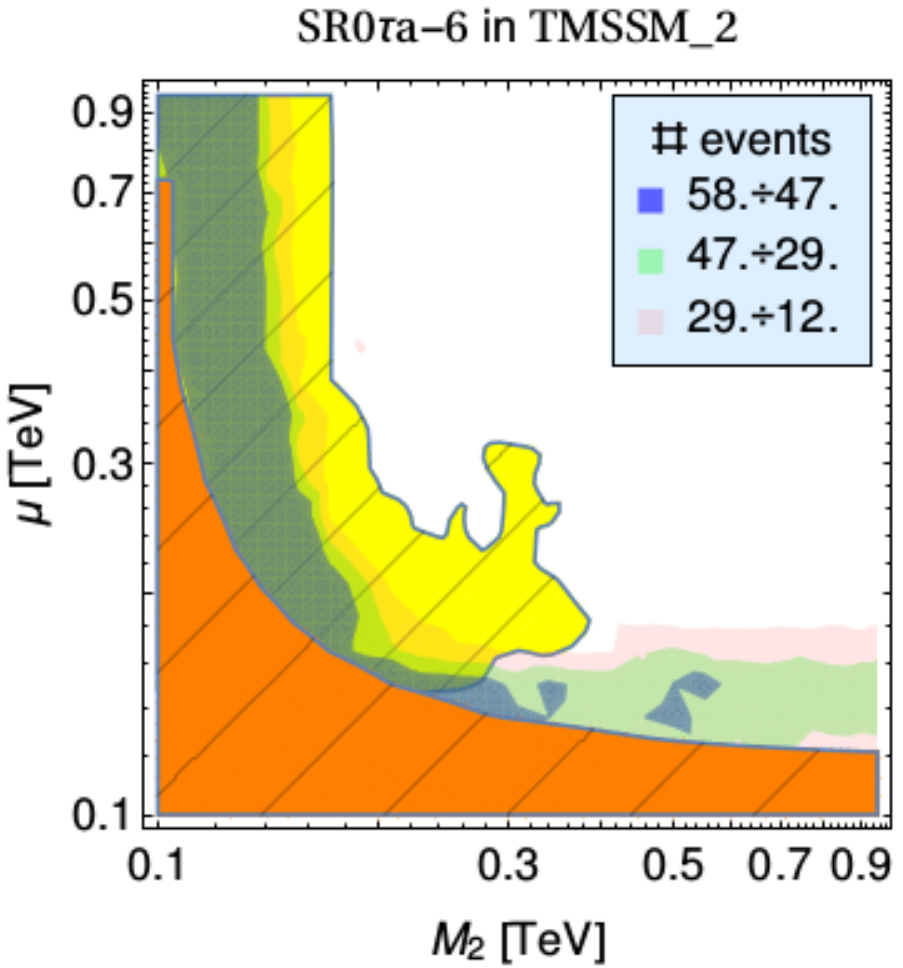}
\hspace*{0.1cm}
\includegraphics[scale=0.38]{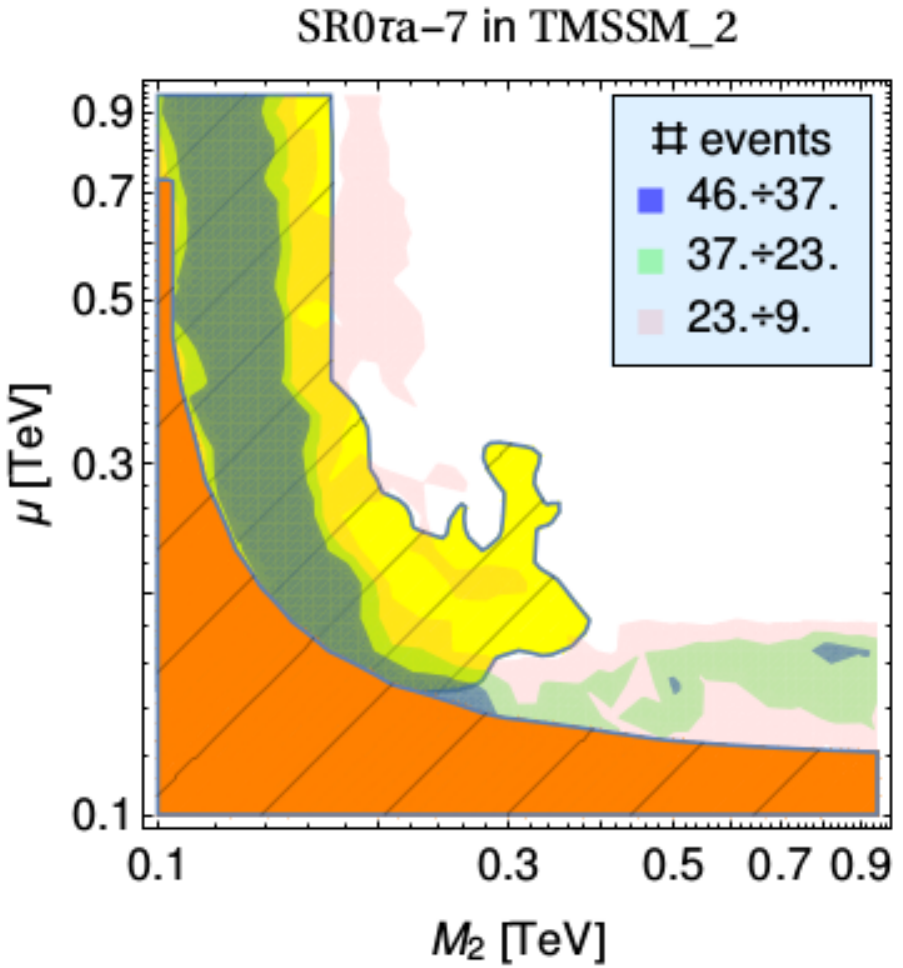}
\hspace*{0.1cm}
\includegraphics[scale=0.38]{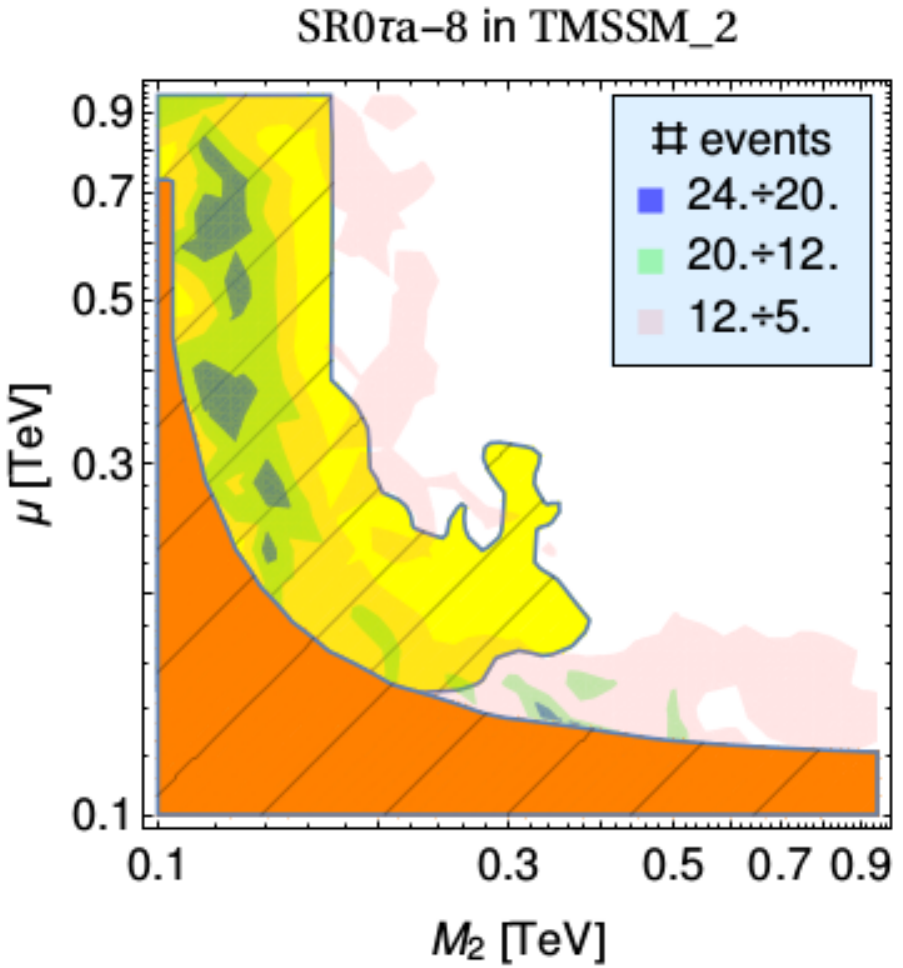}\\
\includegraphics[scale=0.38]{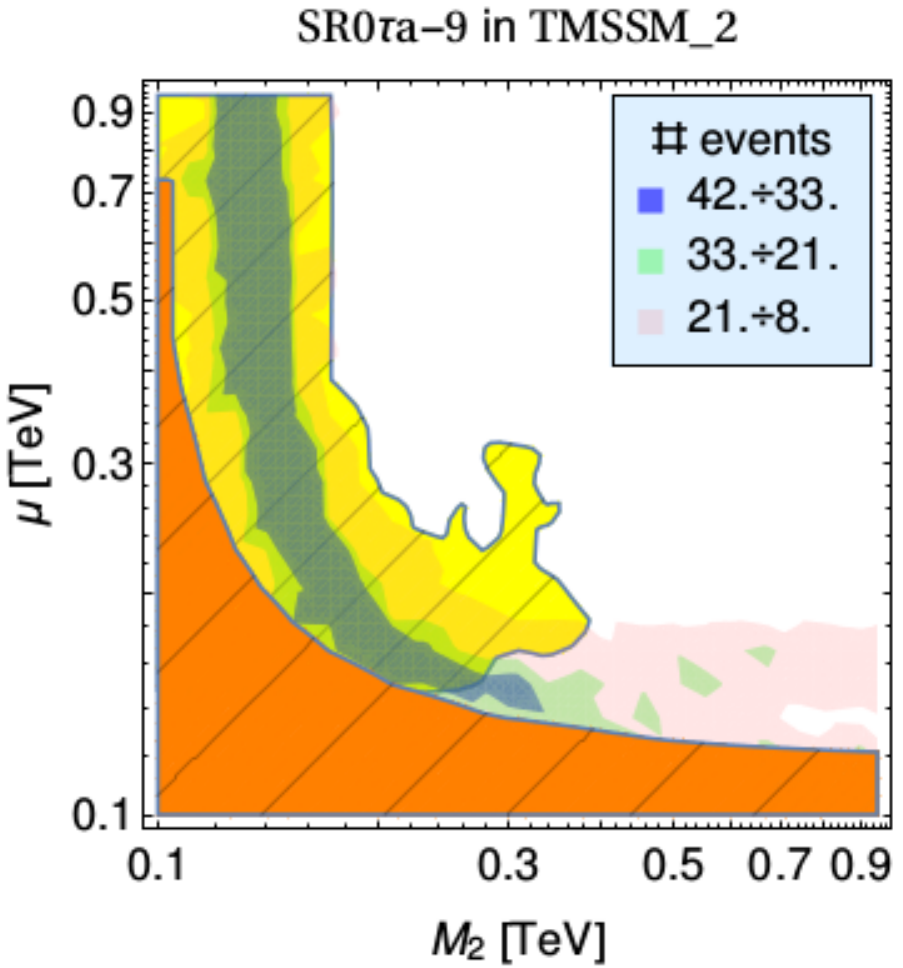}
\hspace*{0.1cm}
\includegraphics[scale=0.38]{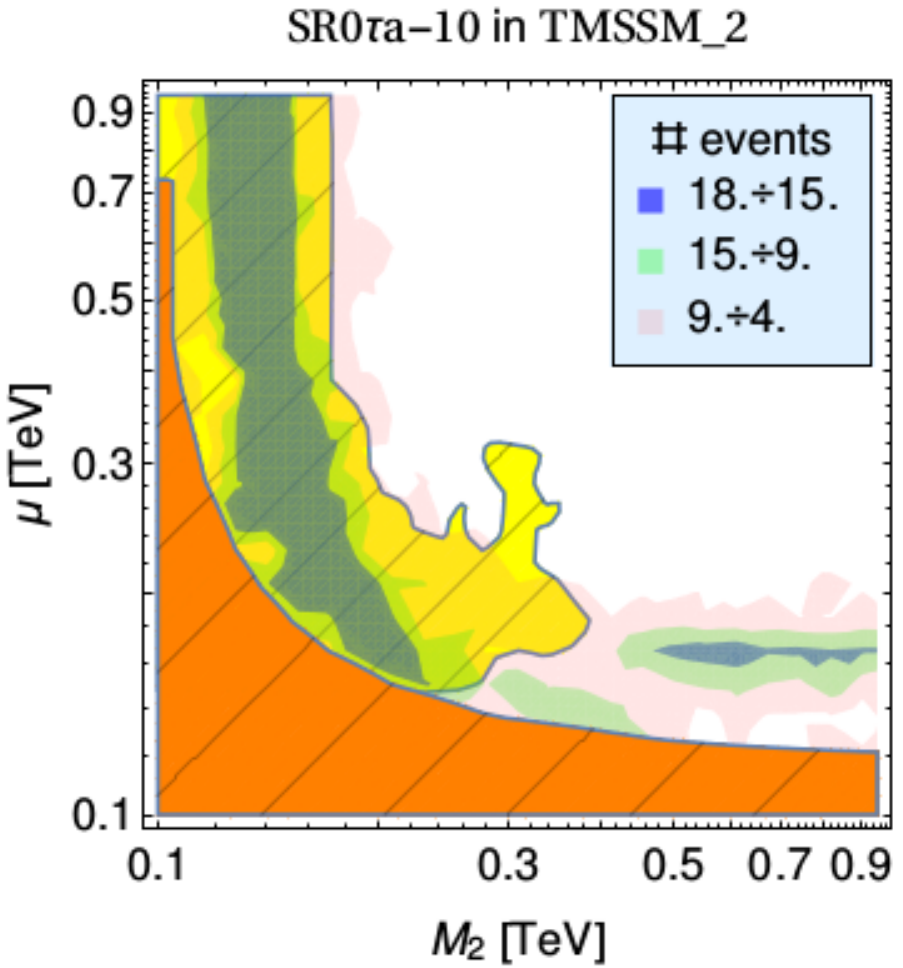}
\hspace*{0.1cm}
\includegraphics[scale=0.38]{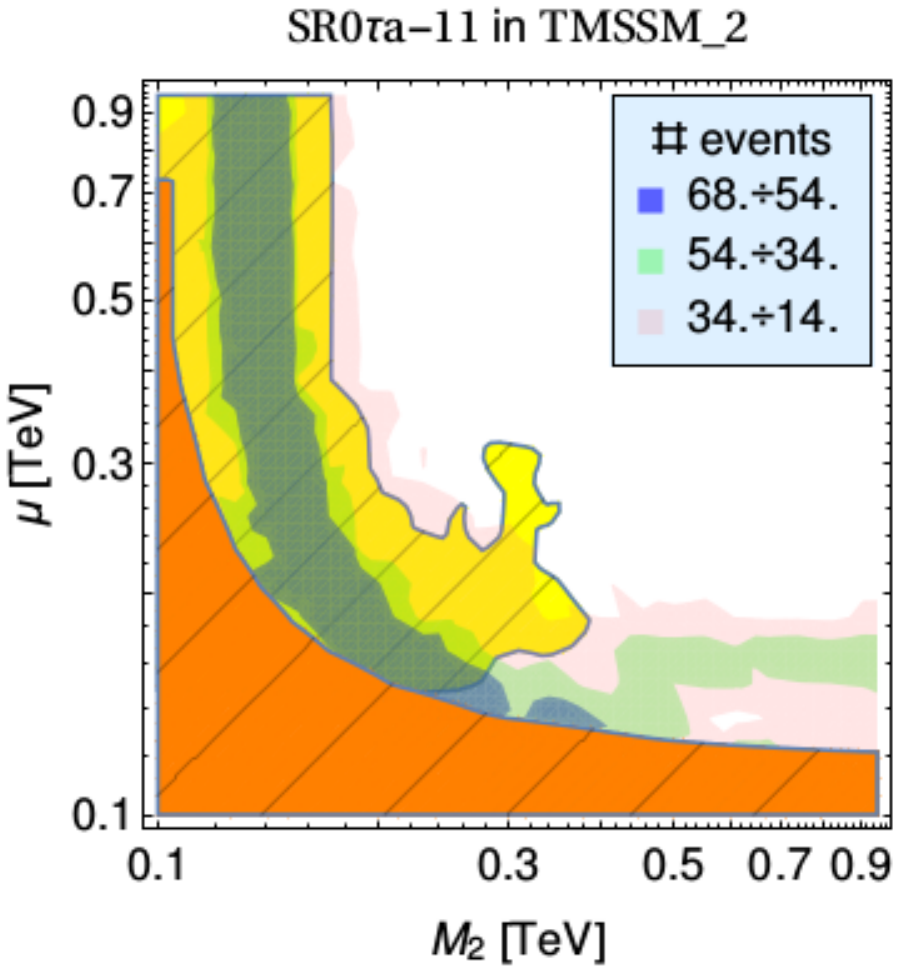}
\hspace*{0.1cm}
\includegraphics[scale=0.38]{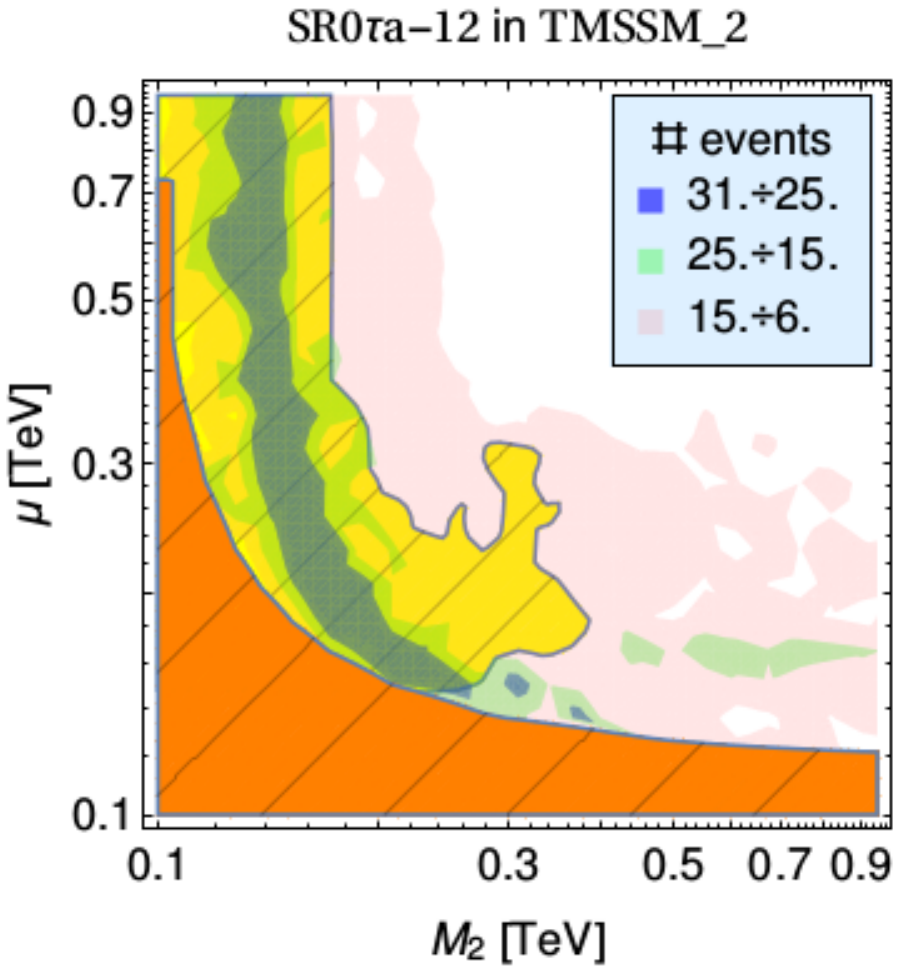}\\
\includegraphics[scale=0.38]{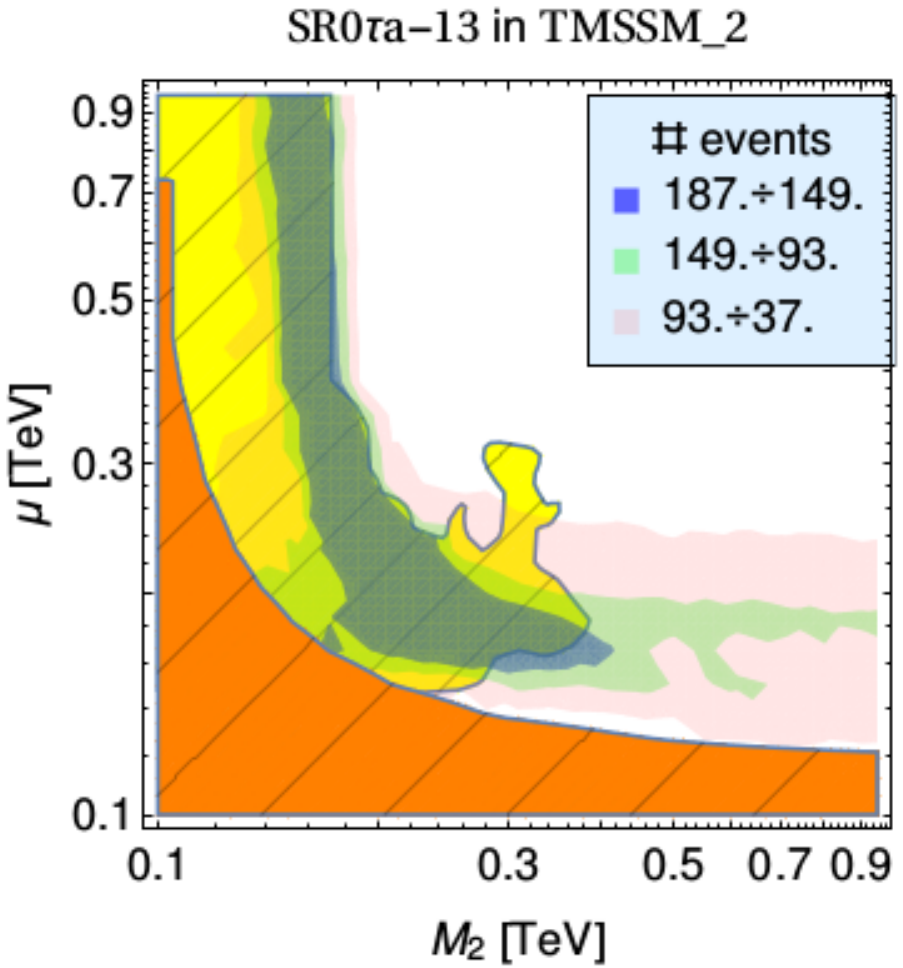}
\hspace*{0.1cm}
\includegraphics[scale=0.38]{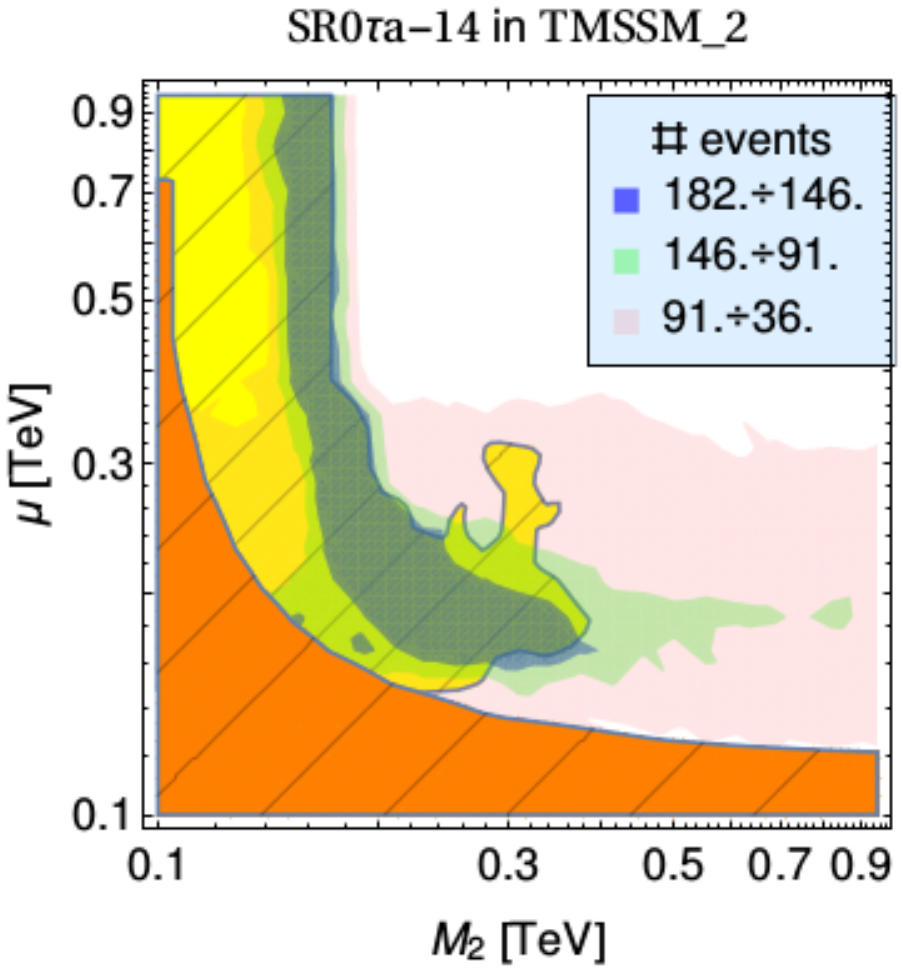}
\hspace*{0.1cm}
\includegraphics[scale=0.38]{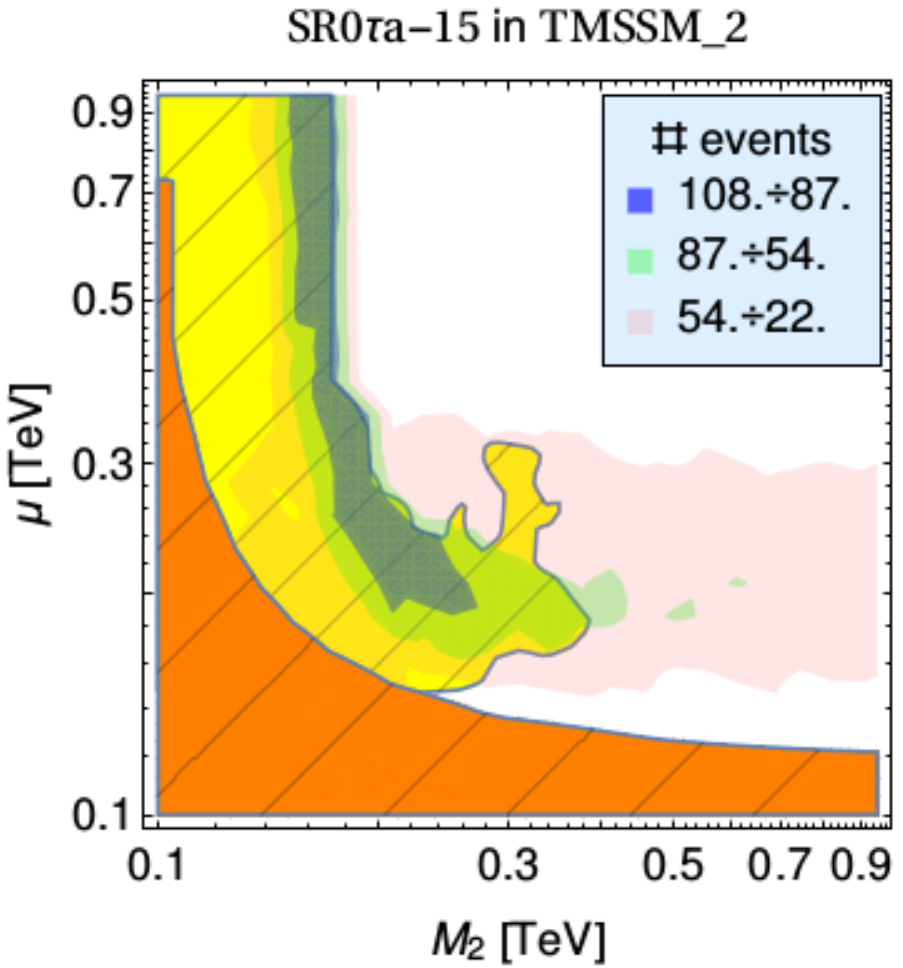}
\hspace*{0.1cm}
\includegraphics[scale=0.38]{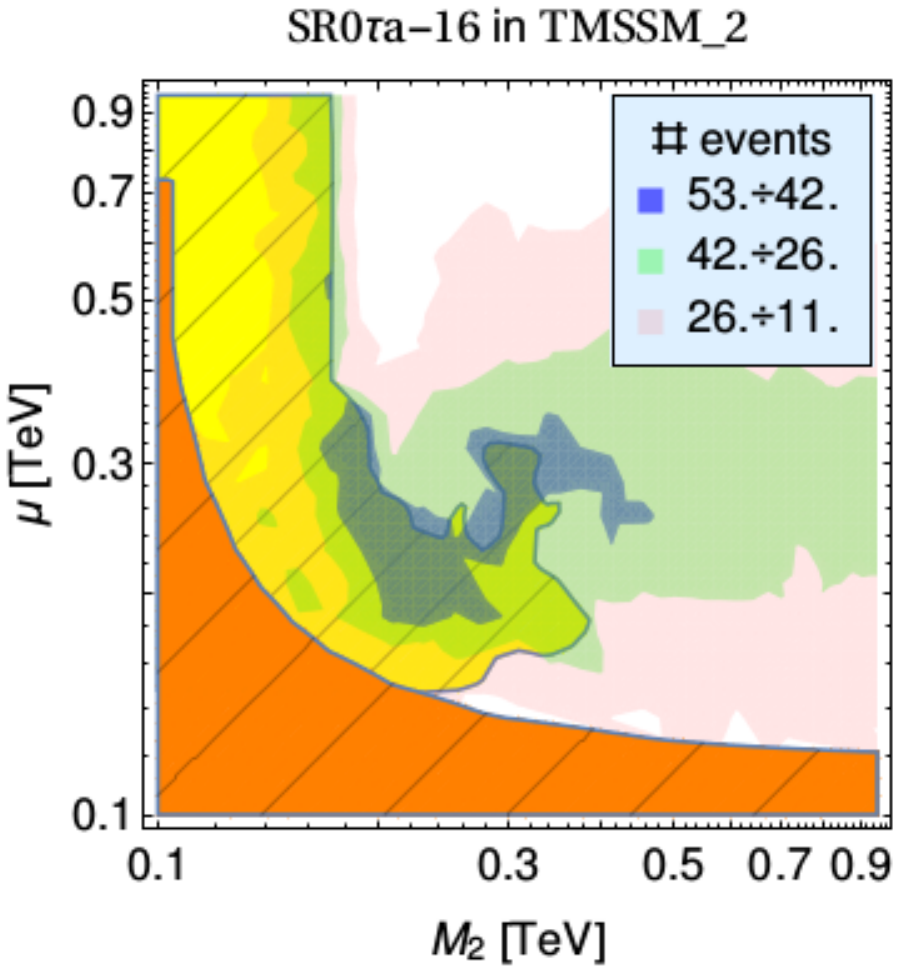}\\
\includegraphics[scale=0.38]{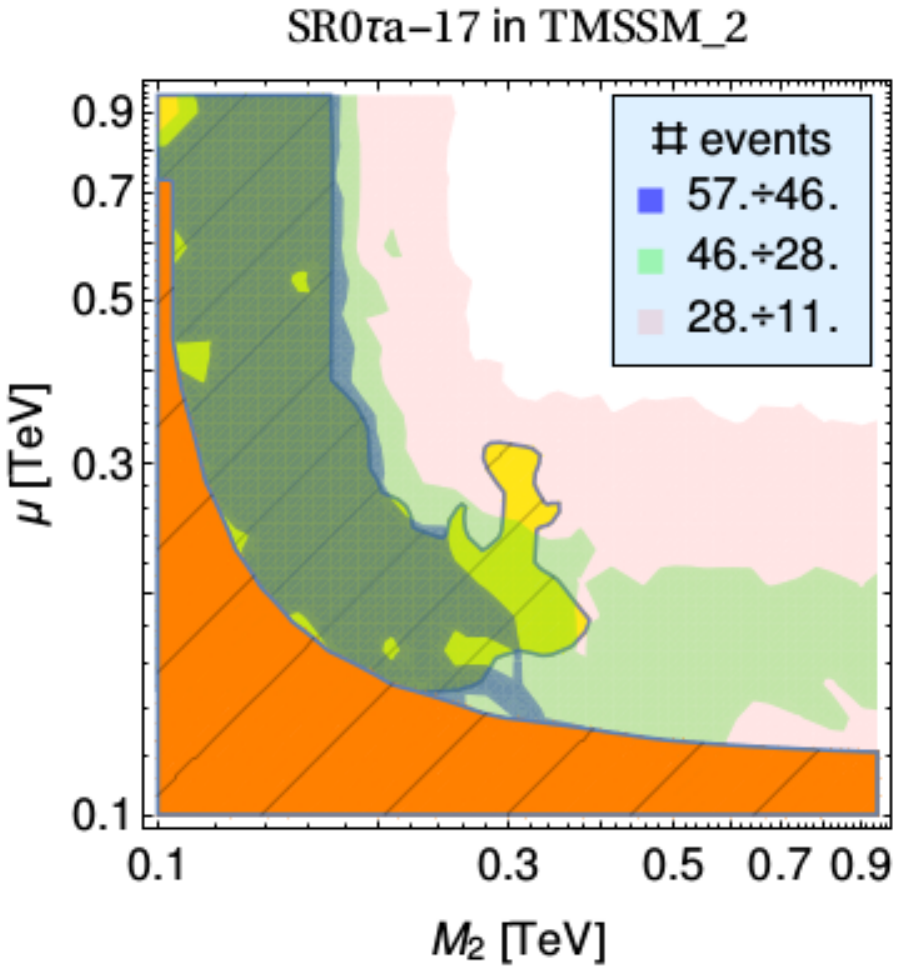}
\hspace*{0.1cm}
\includegraphics[scale=0.38]{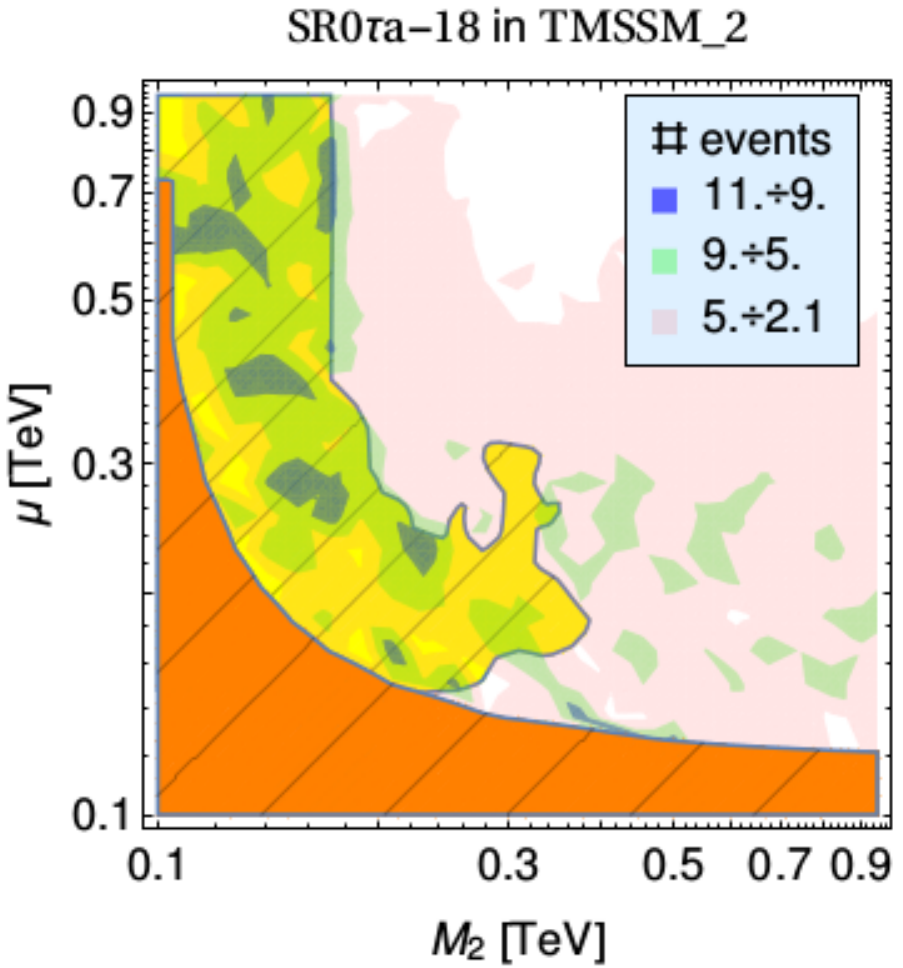}
\hspace*{0.1cm}
\includegraphics[scale=0.38]{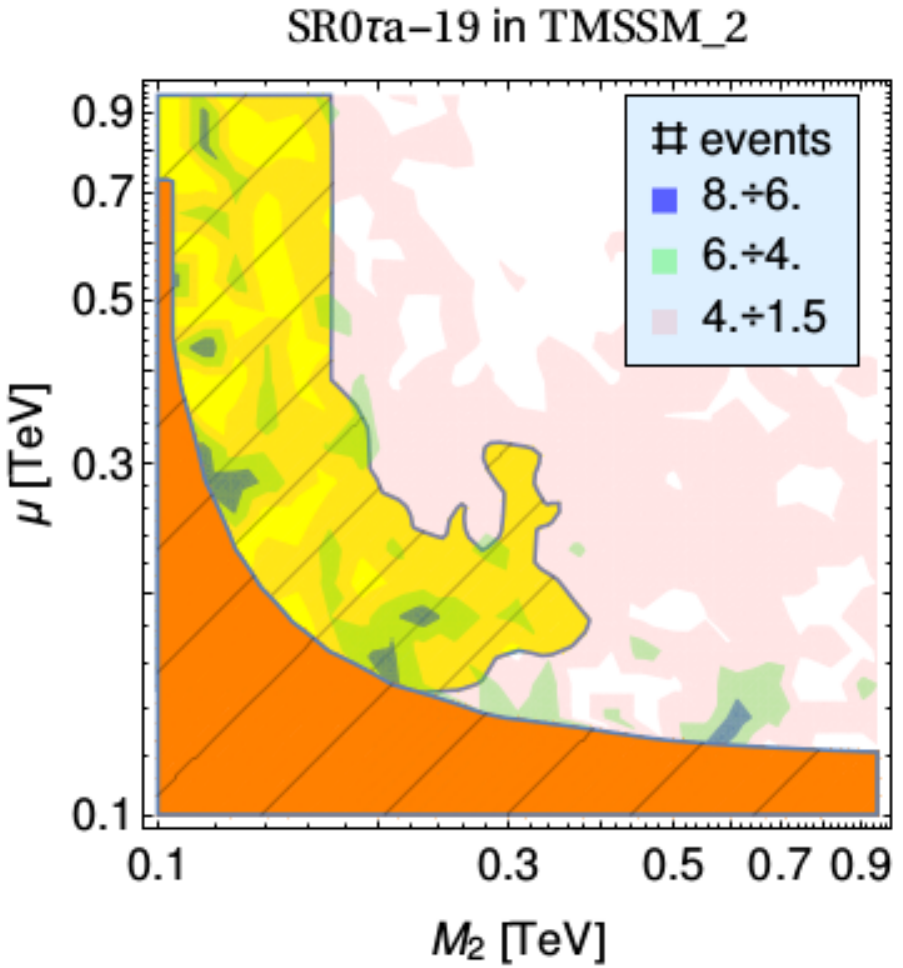}
\hspace*{0.1cm}
\includegraphics[scale=0.38]{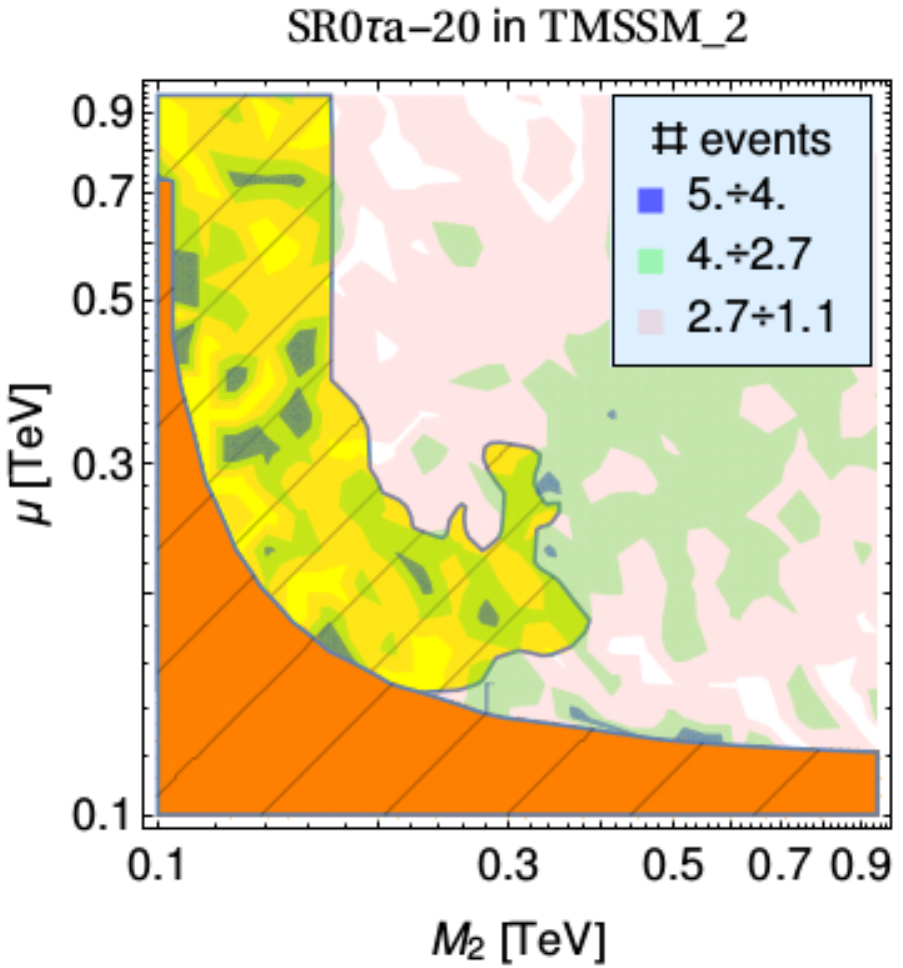}
\caption{{\bfseries Three-lepton search - TMSSM\_2 - 13 TeV:} Same as Fig.~\ref{fig:3lepton_mssm_13tev_bins} for the TMSSM\_1 case ($\lambda = 0.65$ and $\mu_{\Sigma}=350$ GeV).}
\label{fig:3lepton_tmssm65350_13tev_bins}
\end{figure}

The bins of SR0$\tau$a of the three-lepton search for the scenario TMSSM\_2 are shown in Fig.~\ref{fig:3lepton_tmssm65350_13tev_bins}. The conclusions that can be extracted from this scenario are similar to the ones of the scenario TMSSM\_1. The bin 11 is the one with the largest number of signal events, while the bin 16 seems to cover better the parameter region that is not excluded by the LHC Run 1 and it is not reachable in the MSSM scenario. Hence in the three-lepton analysis there are several bins that can be used to look for new physics in general coming from SUSY, irrespective on the details of the model, while there are others, such as bin 16, that are relevant to disentangle among different SUSY models.

\clearpage

\bibliography{biblio}
\bibliographystyle{JHEP}

\end{document}